\documentclass[a4paper,10pt]{article}
\pdfoutput=1 

\usepackage{jheppub} 

\usepackage[T1]{fontenc} 
\usepackage{lmodern} 

\usepackage{xcolor}
\usepackage{mathrsfs}
\usepackage[nointegrals]{wasysym}
\usepackage{booktabs}
\usepackage{slashed}
\usepackage{cancel}
\usepackage[font=small,labelfont=bf,labelsep=period,tableposition=top]{caption}
\usepackage[above]{placeins}
\usepackage{afterpage}
\usepackage{tikz}
\usetikzlibrary{bayesnet}

\newcommand{\pt}{p_{\rm t}}

\newcommand{\MSbar}{\overline{\text{MS}}}
\newcommand{\sss}{\scriptscriptstyle}

\newcommand{\St}   {\Sigma_{\rm true}}
\newcommand{\Sp}   {\Sigma_{\rm pert}}

\newcommand{\kas} {k_{\rm asympt}}

\newcommand{\al}{\alpha}
\newcommand{\as}{\alpha_s}
\newcommand{\az}{\alpha_0}

\newcommand{\Ord}{\mathcal{O}}

\newcommand{\mh}{m_{H}}

\newcommand{\mz}{m_{Z}}

\newcommand{\muf}{\mu_{\sss\rm F}}

\newcommand{\lr} {\ell}

\newcommand{\abs}[1]{\left| #1 \right|}

\newcommand{\sign}{{\rm sign}\,}

\let\originalleft\left
\let\originalright\right
\renewcommand{\left}{\mathopen{}\mathclose\bgroup\originalleft}
\renewcommand{\right}{\aftergroup\egroup\originalright}

\newcommand{\av}[1]{\langle #1\rangle}

\def\beq{\begin{equation}}  
\def\eeq{\end{equation}}
\def\({\left(}
\def\){\right)}
\def\[{\left[}
\def\]{\right]}

\let\oldsubsection\subsection
\renewcommand\subsection[2][\subsectiontoc]{%
  \def\subsectiontoc{#2}%
  \oldsubsection[#1]{\boldmath #2}%
}

\let\oldsubsubsection\subsubsection
\renewcommand\subsubsection[2][\subsubsectiontoc]{%
  \def\subsubsectiontoc{#2}%
  \oldsubsubsection[#1]{\boldmath #2}%
}

\allowdisplaybreaks

\numberwithin{figure}{section}

\title{\boldmath Probabilistic definition of the perturbative theoretical uncertainty from missing higher orders}

\author[]{Marco Bonvini}
\affiliation[]{INFN, Sezione di Roma 1,\\ Piazzale Aldo Moro~5, 00185 Roma, Italy}
\preprint{}
\arxivnumber{2006.16293}

\emailAdd{marco.bonvini@roma1.infn.it}

\abstract{%
We consider the problem of quantifying the uncertainty on theoretical predictions based on perturbation theory
due to missing higher orders.
The most widely used approach, scale variation, is largely arbitrary and it has no probabilistic foundation,
making it not suitable for robust data analysis.
In 2011, Cacciari and Houdeau proposed a model based on a Bayesian approach to provide a
probabilistic definition of the theory uncertainty from missing higher orders.
In this work, we propose an improved version of the Cacciari-Houdeau model, that overcomes some limitations.
In particular, it performs much better
in case of perturbative expansions with large high-order contributions (as it often happens in QCD).
In addition, we propose an alternative model based on the same idea of scale variation, which overcomes
some of the shortcomings of the canonical approach, on top of providing a probabilistically-sound result.
Moreover, we address the problem of the dependence of theoretical predictions on unphysical scales
(such as the renormalization scale), and propose a solution to obtain a scale-independent result within the probabilistic framework.
We validate these methods on expansions with known sums, and apply them to a number of physical observables in particle physics.
We also investigate some variations, improvements and combinations of the models.
We believe that these methods provide a powerful tool to reliably estimate theory uncertainty from missing higher orders
that can be used in any physics analysis.
The results of this work are easily accessible through a public code named \texttt{THunc}.
}

\begin{document}

\maketitle

\section{Introduction}

Computing exactly physical observables in a generic quantum field theory (QFT) is still an open challenge.
The closest approximation to such a computation is achieved through numerical simulations of the theory on a discretized space-time (lattice).
This approach, however, cannot be used universally.
For instance, for some observables (e.g.\ involving large momenta) the required computing power would be out of reach.
A complementary approach, applicable when the coupling of the theory is sufficiently small,
is the so-called perturbative approach, or perturbation theory.
Namely, physical observables are expressed as a power series in the coupling.
Computing the coefficients of this power series can be done systematically, e.g.\ using Feynman diagram techniques,
but the complication of the calculation grows considerably with the perturbative order.
Therefore, in practice, only a limited number of coefficients is achievable for a given observable.

This poses an important question: how far is the approximate perturbative result from the unknown exact one?
The answer to this question can be cast in an uncertainty, usually called \emph{theory uncertainty from missing higher orders}.
Since the exact result is unknown, one can only quantify this uncertainty in terms of a \emph{probability distribution}.
How to determine such a distribution is the subject of this paper.

\begin{figure}[t]
  \centering
  \includegraphics[width=0.6\textwidth]{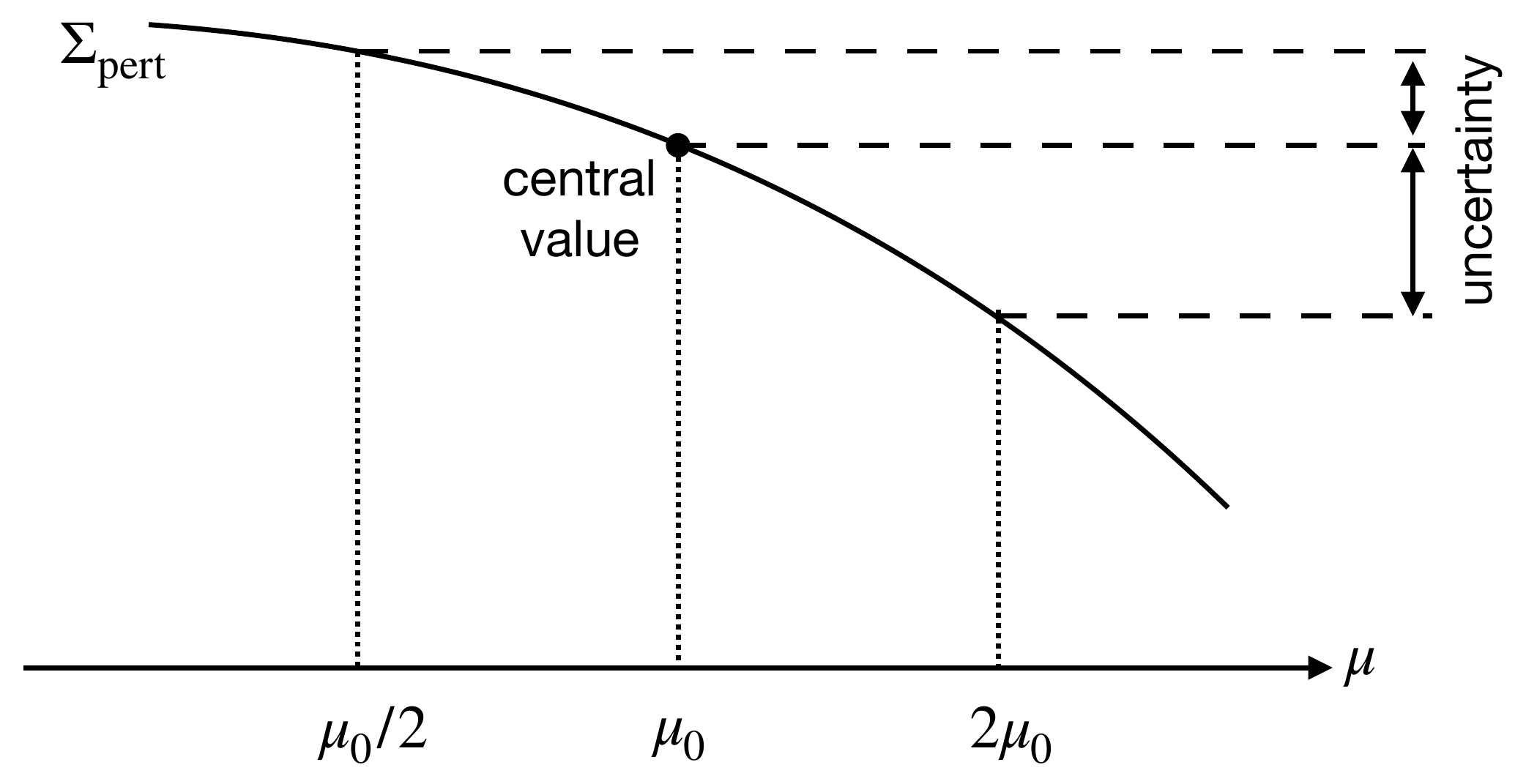}
  \caption{Schematic representation of the canonical method to estimate theory uncertainty, namely the scale variation approach.
    Since the left and right variations of the scale generally lead to different sizes of variation of the perturbative result
    (shown by the two double-headed arrows), one may either choose to keep the uncertainty asymmetric,
    or to select the largest and symmetrize it.}
  \label{fig:scalevar}
\end{figure}

The most standard and widespread approach to estimate this uncertainty is the so-called scale variation method.
This approach is based on the observation that physical observables do not depend on unphysical scales (such as the renormalization scale)
appearing in QFTs.
However, this independence is strictly valid only for the exact result.
Any approximate result computed in perturbation theory will in fact depend on unphysical scales,
the dependence being formally of higher order.
The idea is thus to estimate the theory uncertainty by varying the scale, as the result will accordingly change
by an amount that is formally of the same order as what this uncertainty wants to quantify.
While this idea is certainly valid and powerful, the canonical method used to exploit it has various caveats.
Indeed, the canonical recipe consists in varying the unphysical scale $\mu$ by a factor of two about a central value $\mu_0$ of choice,
and then using the maximal variation of the observable with respect to the value at central scale as a measure of the uncertainty.
This is depicted in Fig.~\ref{fig:scalevar}.
In formulae, the canonical way of writing a result based on a perturbative computation of a physical observable $\Sigma$ is\footnote
{In the literature there exist variants of this formulation.
  In the most widespread variant the ``error'' is left asymmetric, as shown in Fig.~\ref{fig:scalevar}.}
\begin{equation}\label{eq:CSU}
\Sigma\; \overset{\substack{\rm canonical\\ \rm scale\\ \rm variation}}{\equiv}\;
\Sigma_{\rm pert}(\mu_0) \pm \max_{\mu_0/2<\mu<2\mu_0}\Big|\Sigma_{\rm pert}(\mu)-\Sigma_{\rm pert}(\mu_0)\Big|,
\end{equation}
where $\Sigma_{\rm pert}(\mu)$ is the scale dependent perturbative result,
$\Sigma_{\rm pert}(\mu_0)$ represents the ``central value'' of the prediction,
and the scale variation is appended as a theory ``error''.
The caveats of this approach are apparent:
\begin{itemize}
\item it is largely arbitrary, in the choice of the central scale $\mu_0$ and of the interval of variation (factor of two);
\item in the vicinity of stationary points the uncertainty can become accidentally small;
\item the uncertainty has no probabilistic interpretation.
\end{itemize}
The latter point can be overcome by \emph{assigning} an interpretation
(for instance, the ``error'' could be interpreted as the standard deviation of a gaussian distribution),
but any choice would be totally arbitrary.
In addition to all this, it is well known that the scale variation uncertainty often underestimates
the size of higher order contributions.\footnote
{It has to be mentioned that at low orders not all partonic channels contribute
  to a given QCD process with hadrons in the initial state.
  Since renormalization scale variation cannot probe different channels, the opening up of new channels
  certainly contributes to the failure of scale variation in estimating the size of the higher orders.
  In proton-proton collisions, from next-to-next-to-leading order (NNLO) onwards all partonic channels are active,
  and indeed for those processes that are known beyond NNLO scale variation typically works better than at previous orders.
  A remarkable exception is given by the Drell-Yan process, for which the recently computed next-to-next-to-next-to-leading order
  correction~\cite{Duhr:2020seh,Duhr:2020sdp} lies outside the NNLO scale variation band in a wide kinematic region.}

In 2011, Cacciari and Houdeau~\cite{Cacciari:2011ze} proposed a completely different approach
to estimate the uncertainty from missing higher orders.
In their groundbreaking work they constructed a probabilistic model to define this uncertainty,
based on some assumptions on the progression of the perturbative expansion.
Roughly speaking, the model assumes that the coefficients of the power expansion in the coupling
are bounded by an unknown (hidden) parameter. The knowledge of the first few orders of the perturbative expansion
allow to perform (Bayesian) inference on the hidden parameter, whose improved knowledge can in turn be used
to make inference on the unknown subsequent perturbative coefficients.
Once the likelihood of the perturbative coefficients given the hidden parameter and the prior distribution
of the hidden parameter are defined, the model produces probability distributions for the unknown coefficients,
and thus for the physical observable itself.

This approach to theory uncertainties is unquestionably superior and more elegant than canonical scale variation,
mainly because it is probabilistically founded.
However, it also has limitations.
For instance, while it performs well for QCD observables at $e^+e^-$ colliders~\cite{Cacciari:2011ze},
it leads to less reliable results in the case of proton-proton collider observables~\cite{Bagnaschi:2014wea,Bonvini:2016frm},
that are usually affected by larger perturbative corrections.
Perhaps for this reason, in conjunction with the simplicity and the deep-rooted attitude of using the scale variation method,
the Cacciari-Houdeau (CH) approach is not very popular in the high-energy physics community,
with only few applications~\cite{Ball:2011us,
Goria:2011wa,
David:2013gaa,
Forte:2013mda,
Dulat:2015mca,
Furnstahl:2015rha,
Perez:2015ufa,
Griesshammer:2015ahu,
Bonvini:2016frm,
Perez:2016nwc,
Melendez:2017phj,
Ball:2018iqk,
Melendez:2019izc,
Ekstrom:2019hlw},
most of which in the context of effective theories.
Moreover, the CH approach does not deal with the unphysical scale dependence of the result ---
the CH prediction is computed for a given choice of the scale, so the final probability distributions
is de facto scale dependent.

\begin{figure}[t]
  \centering
  \includegraphics[width=0.6\textwidth,page=1]{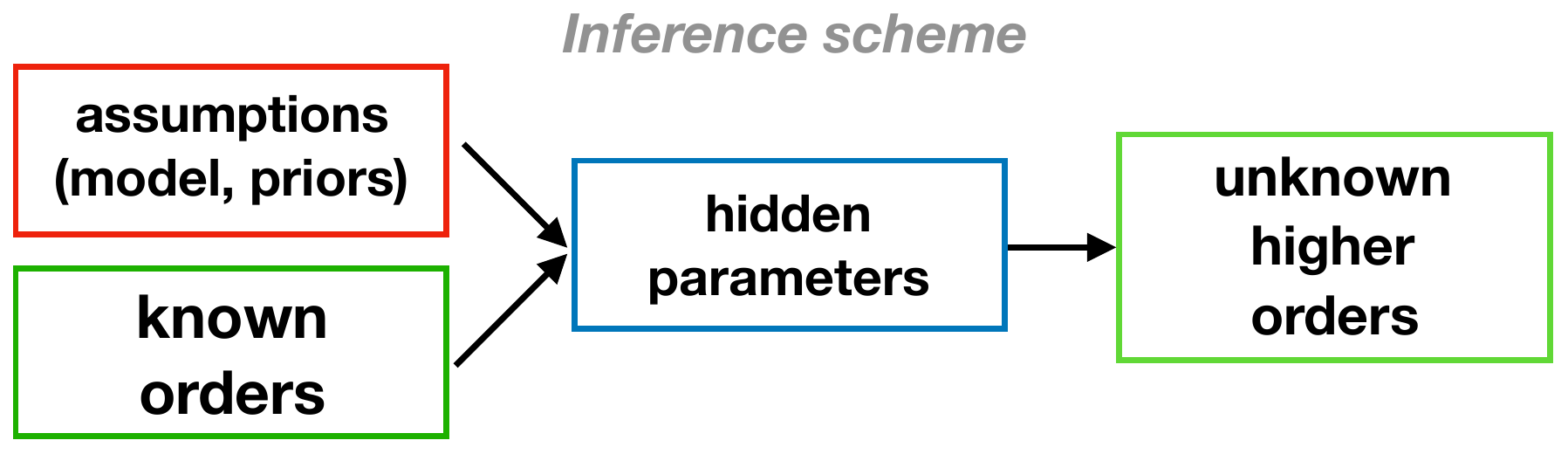}
  \caption{The general structure of the inference in any model considered in this work.}
  \label{fig:inference}
\end{figure}

In this paper, we use a Bayesian approach to build new probabilistic models, similar to the CH model,
that overcome the limitations of the previous approach.
The inference structure of any model that we will consider is depicted in full generality in Fig.~\ref{fig:inference}.
Starting from the known orders, under some model assumptions one can make inference on the
(hidden) parameters characterizing the model, to be used in turn to infer the probability distribution
of the unknown higher orders.
In formulae, we have, schematically,
\begin{equation}\label{eq:inference1}
P(\text{unknown orders}|\text{known orders}) = \int d \text{pars}\; P(\text{unknown orders}|\text{pars}) P(\text{pars}|\text{known orders})
\end{equation}
where $P(A|B)$ is the conditional probability distribution of $A$ given $B$.
Eq.~\eqref{eq:inference1} is given in terms of the posterior distribution of the hidden parameters
\begin{equation}\label{eq:inference2}
P(\text{pars}|\text{known orders}) \propto P(\text{known orders}|\text{pars}) P_0(\text{pars})
\end{equation}
which depends on the prior distribution $P_0(\text{pars})$ of the hidden parameters
and on the model assumptions through the likelihood $P(\text{orders}|\text{pars})$,
that appears also explicitly in Eq.~\eqref{eq:inference1}.
These two model-dependent ingredients are sufficient to let Bayesian inference work,
and can be used to eventually construct a probability distribution for the observable,
that contains all the information on the uncertainty from missing higher orders.

In this work, we propose two main models: one is an improved version of the CH model
that efficiently describes perturbative expansions with large perturbative corrections;
the other is a model inspired by the scale variation method, but constructed in such a way to be reliable
and probabilistically sound.
Both methods outperform the current approaches in terms of reliability.
They are also sufficiently general to be used for perturbative expansion
that are not necessarily fixed-order expansions in powers of the coupling,
but can be for instance resummed expansions or other generalized expansions.
Moreover, the first model can be applied also beyond quantum field theory,
for instance to quantum mechanics or in general to any expansion that behaves perturbatively.
We also explore variants of the methods and combinations of them.

These methods are still applied to the perturbative expansion at a given fixed value of the unphysical scale.
An important and innovative development proposed in this work is a way to ``remove'' the scale dependence of the result.
This is achieved by dealing with the scale dependence within the probabilistic framework,
and leads to a result that is to a large extent scale independent.
As a byproduct of this procedure, the central value of the prediction
(identified with the mean of the probability distribution for the observable)
does not necessarily correspond to the canonical perturbative result at a given ``central'' scale.
Thanks to this feature our method improves not only the reliability of the uncertainty,
but also that of the central prediction for the observable.

To facilitate the adoption of the results of this work, a computer code named \texttt{THunc} is publicly released.
The code is very easy to use: the user provides the perturbative expansion,
and the code outputs the probability distribution of the result, together with a number of statistical estimators
(mean, mode, median, standard deviation, degree-of-belief intervals).
It is also fast and efficient, and flexible as it allows the user to define customized models.

The structure of this paper is the following.
In Sect.~\ref{sec:prel} we give some preliminary information on the perturbative expansions
and their scale dependence, we provide some basic concepts on the probabilistic definition of the theory uncertainty
as well as a brief recap of the CH method, and we define a working example to be used in the subsequent sections.
In Sect.~\ref{sec:statmodel} we start defining some notations and describe general features
common to all models we will later consider.
We then move on presenting our two main models (at fixed scale) in Sect.~\ref{sec:GB} and Sect.~\ref{sec:ModelScaleVar}.
We propose a way to construct scale-independent results and uncertainties in Sect.~\ref{sec:scaleindep}.
We then validate our methods in Sect.~\ref{sec:applications}, where we also consider some realistic applications.
In Sect.~\ref{sec:corr}, we discuss the issue of defining correlations between theory uncertainties.
After concluding in Sect.~\ref{sec:conclusions},
we collect details on numerical implementations in App.~\ref{sec:appAlgos} and
propose a number of variants and possible improvements in App.~\ref{sec:models}.

\section{Preliminaries}
\label{sec:prel}

\subsection{Basic concepts and assumptions on perturbative expansions}
\label{sec:basics}

We consider a (renormalizable) quantum field theory (QFT) depending on a single coupling $\al$.
We will often refer to practical examples in quantum chromodynamics (QCD),
since its coupling is not very small and thus
perturbation theory produces somewhat large high-order corrections,
which is the case where reliably estimating theory uncertainties is both important and challenging.
We focus on a generic physical observable $\Sigma$.
The theory predicts a unique well-defined value for this observable, that we call $\St$.
This is the value we aim to obtain.
However, we usually cannot compute it exactly, and we thus use a perturbative approach to approximate it.

According to the perturbative hypothesis, namely that the dimensionless coupling $\al$ of the theory is sufficiently small,
it is possible to compute the observable $\Sigma$ as a power series in the coupling itself.
We then write
\begin{equation}\label{eq:pertdef}
\St \simeq \sum_{k=0}^n c_k \al^k,
\end{equation}
where $c_k$ are the coefficients of the perturbative expansion, and $n$ is some order at which we stop the expansion.
Eq.~\eqref{eq:pertdef} is not an equality.
One may be tempted to think that if $n\to\infty$, then it would become an equality.
This limit, however, does not exist, as perturbative expansions are divergent~\cite{Dyson:1952tj,tHooft:1977xjm}.
The divergence of the series is related to the fact that $\St$ is a non-analytic function of the coupling $\al$ in $\al=0$.
One may try to treat the divergent series using e.g.\ the Borel summation method.
For some known (to all orders) divergent contributions, such as those due to renormalons (see e.g.\ Ref.~\cite{Beneke:1998ui}),
one can obtain a finite result through Borel summation.\footnote
{In fact, the renormalon series is not Borel summable, but one can still define a finite sum
at the price of introducing an ambiguity proportional to a non-analytic function of the coupling.
The ambiguity term is clearly an artefact of the summation method, and it suggests that the
information contained in the perturbative expansion is not sufficient to fully reconstruct the true value of the observable.}
However, it is not guaranteed that the Borel-sum of the series captures the full result:
there may be intrinsically non-perturbative contributions that cannot be reconstructed from the perturbative expansion.\footnote
{It has been pointed out in Ref.~\cite{Serone:2017nmd} (in the context of quantum mechanics)
  that it is possible to suitably modify the action such that the (Borel-summed) perturbative result
  gives the whole exact result, without any extra non-perturbative contribution.
  It is not obvious whether this result can be extended to QFTs.}
Moreover, in order to use the Borel summation method, the series should be known to all orders,
or at least its asymptotic behaviour should be known.\footnote
{If one applies the Borel method to a finite sum, it acts as an identity.
  However, if the asymptotic behaviour is known, one can mix the Borel method with a conformal mapping,
  to obtain a ``perturbative version'' of Borel summation that is effective also on finite sums
  (see e.g.\ Refs.~\cite{Serone:2018gjo,Caprini:2019kwp}).}
However, this is usually not the case, so in practice the only information we have about an observable is
its (truncated) perturbative expansion, Eq.~\eqref{eq:pertdef}.

\begin{figure}[t]
  \centering
  \includegraphics[width=0.49\textwidth,page=1]{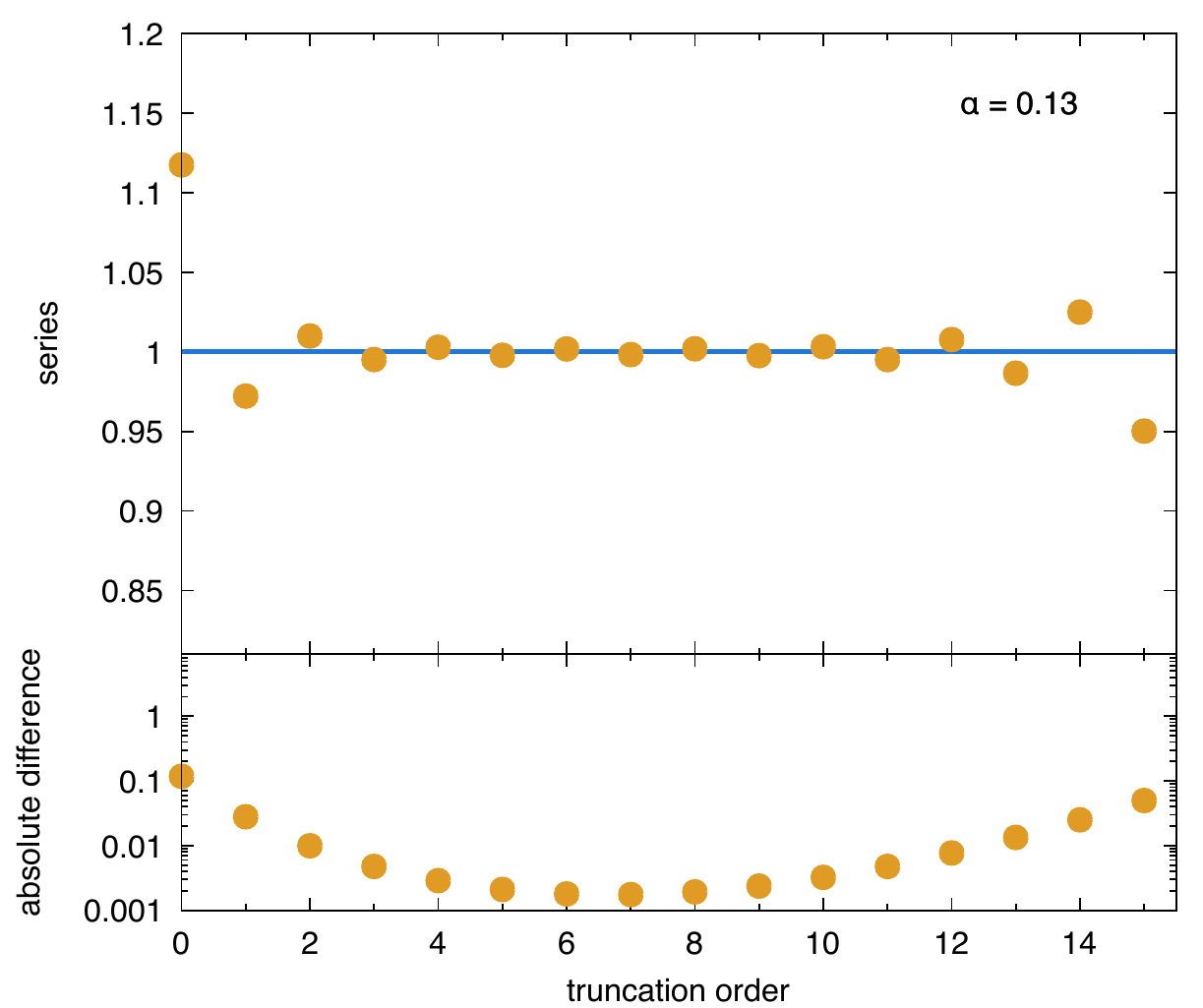}
  \hspace{\stretch{1}}
  \includegraphics[width=0.49\textwidth,page=2]{./images/divser_asympt}
  \caption{The asymptotic expansion of the function Eq.~\eqref{eq:divserasympt} truncated at various orders from 0 to 15 (orange dots),
    normalized to the value of the function itself (blue line).
    The lower panel shows the absolute difference between the truncated and exact results.
    The left plot corresponds to $\al=0.13$, the right plot corresponds to $\al=0.2$.}
  \label{fig:asympt}
\end{figure}

The fact that the series is divergent and that summation methods cannot be used may suggest that
the perturbative result is useless. However, this is in constrast with the well known fact that perturbation theory
works, namely it predicts results in decent (or even good) agreement with data, at least when the coupling is sufficiently small
(for instance, in QED it works very well).
The explanation of this fact relies on the assumption that perturbative series are asymptotic expansions of the exact result.
To our knowledge, there is no general proof of this statement, but it seems very reasonable and we take it as valid.
The asymptotic nature of the perturbative expansion implies that up to some order $\kas$
adding terms to the expansion improves the accuracy of the prediction,
but beyond $\kas$ the divergent contributions to the series dominate and the sum explodes.
A visual example of this fact is shown in Fig.~\ref{fig:asympt}, where the following
non-analytic function of $\al$ is compared with its asymptotic expansion at small $\al$:
\begin{equation}\label{eq:divserasympt}
\frac1{\al}\exp\(\frac1{\al}\)\Gamma\(0,\frac1{\al}\) \overset{\rm asympt}= \sum_{k=0}^\infty (-1)^k k! \al^k.
\end{equation}
From the figure one sees that for a sufficiently small value of $\al=0.13$ (left plot)
the first few (approximately 7) orders give a good approximation of the result,
showing an apparently converging behaviour. However, adding extra orders deteriorates the prediction,
because the factorial growth of the series wins over the power suppression.
The best prediction one can make using the asymptotic expansion is thus obtained truncating it
to $\kas\sim7$. This prediction, however, has an irreducible uncertainty due to the truncation itself,\footnote
{Note that this divergent series is in fact Borel summable,
  and its Borel sum corresponds exactly to the original function Eq.~\eqref{eq:divserasympt}.}
of the size of the last term included in the truncated expansion.
The right plot, that is obtained with a larger coupling $\al=0.2$, shows that this irreducible uncertainty
grows with the value of the coupling, and the value of $\kas$ decreases accordingly.\footnote
{Indeed, for this specific factorially divergent series, one can show that $\kas\sim1/\al$
  and the irreducible uncertainty is of order $\exp(-1/\al)$~\cite{Beneke:1998ui}.}

From these considerations we can conclude that a physical observable can be expressed as the sum
\begin{equation}\label{eq:Sigma_decomposition1}
\St = \sum_{k=0}^{\kas} c_k \al^k + \Delta_{\rm asympt} + \Delta_\text{non-pert},
\end{equation}
where the first term is perturbative expansion is truncated at $\kas$,
which is the best prediction we can make using perturbation theory,
while $\Delta_{\rm asympt}$ represents the irreducible difference between the truncated expansion and its all-order sum.
The $\Delta_\text{non-pert}$ term, instead, represents possible intrinsically non-perturbative
contributions that cannot be captured by perturbation theory.
The general expectation on the size of these contribution is
\begin{equation}\label{eq:Delta_size1}
\abs{\Delta_{\rm asympt}} \sim \abs{\Delta_\text{non-pert}} \ll \abs{\sum_{k=0}^{\kas} c_k \al^k},
\end{equation}
at least when the coupling is sufficiently small.
This expectation is again not a proof, but a consequence of the goodness of perturbation theory,
together with the fact that both $\Delta$ terms are known to be exponentially suppressed by $\exp(-a/\al)$, $a>0$,
in some established cases (namely, when $\Delta_\text{non-pert}$ contains instanton contributions
and when $\Delta_{\rm asympt}$ is dominated by factorially divergent contributions like renormalons~\cite{Beneke:1998ui}).

Unfortunately, since we do not generally know the asymptotic behaviour of the expansion,
we cannot know a priori the value of $\kas$, and we thus cannot truncate the expansion at the optimal value.
This is not a real issue,
as in most cases we know just a rather small number of orders,
typically two or three, corresponding to next-to-leading order (NLO) and next-to-next-to-leading order (NNLO) computations.
Only in very few cases we know physical quantities at N$^3$LO (four terms in the expansion) or beyond. 
We thus expect (and assume) the number $n$ of known orders to be smaller than $\kas$.\footnote
{If this is not the case, it would mean that $\kas$ is a very small number, which in turn implies
  that the expansion is badly divergent from the very first orders,
  without showing any apparently converging pattern.
  This is for instance the case for the perturbative conversion between the pole and the $\MSbar$ charm mass,
  where $\kas$ is probably 2, and the number of known orders is 4,
  see e.g.~\cite{Marquard:2015qpa,Kataev:2015gvt,Marquard:2016dcn}.
  In this condition, that may occur when the coupling is not small (e.g.\ in QCD at low energies),
  it's clear that perturbation theory is no longer predictive.}
Therefore, we can rewrite Eq.~\eqref{eq:Sigma_decomposition1} as
\begin{equation}\label{eq:Sigma_decomposition2}
\St = \sum_{k=0}^{n} c_k \al^k + \Delta_{\rm MHO}^{(n)} + \Delta_{\rm asympt} + \Delta_\text{non-pert},
\end{equation}
having defined the contribution from missing higher orders
\begin{equation}\label{eq:DeltaMHOdef}
\Delta_{\rm MHO}^{(n)} = \sum_{k=n+1}^{\kas} c_k \al^k,
\end{equation}
where $n$ is the highest known perturbative order.
Assuming that $n$ is sufficiently smaller than $\kas$,
using similar considerations to those that led to Eq.~\eqref{eq:Delta_size1}
we can conclude that in most cases
the contribution from missing higher orders is larger than the asymptotic and non-perturbative contributions, 
\begin{equation}\label{eq:Delta_size2}
\abs{\Delta_{\rm MHO}^{(n)}} \gg \abs{\Delta_{\rm asympt}} \sim \abs{\Delta_\text{non-pert}}.
\end{equation}
This implies that our knowledge of the observable $\Sigma$ is determined by the
perturbative expansion truncated at order $n$ with an uncertainty that is dominated by the
missing higher order term $\Delta_{\rm MHO}^{(n)}$.
Quantifying this term, or better determining its probability distribution, is the main task of the rest of this paper.

\subsection{Constructing a probability distribution for a physical observable}

Defining the theoretical uncertainty from missing higher orders in a probabilistic way
may sound impossible or completely arbitrary to many.
This would certainly be true in the context of the so-called frequentist approach to probability,
in which the definition of a probability requires the existence of a repeatable event,
which is typically the case for an experiment but clearly not the case for a theoretical prediction.
However, the frequentist approach to probability is not the only one --- actually,
the frequentist formulation is mathematically inconsistent (see e.g.~\cite{DAgostini:2003syq}),
and thus certainly not the best one.

The only mathematically correct formulation of a probability theory is the so-called Bayesian approach,
where the probability is defined as the ``degree of belief'' of an event, which is then intrinsically subjective.
Initially, when no information is available, the probability of an event is given by a \emph{prior} distribution,
which encodes our subjective and arbitrary prejudices.
Acquiring information on the event changes the degree of belief through statistical inference (Bayes theorem).
Therefore, any probability will depend on subjective assumptions through the prior distribution, 
but adding more and more information updates the probability making it less and less dependent on the prior.

In case of repeatable events, one can acquire information on the process by repeating them
(and, in the limit of large number of repetitions, one recovers the frequentist result).
However, repetion is not the only way of acquiring information, and thus one can use the Bayesian approach
also in cases (like ours) when the event is not repeatable.
Here ``event'' means something that can happen in different ways with different likelihoods,
which we want to describe through a probability distribution.
In our case, the event is \emph{``the observable takes the value $\Sigma$''},
and its probability distribution will be a function of $\Sigma$ ranging over all possible values.
The information on this event that we want to use is the perturbative expansion of the observable.
How this will be used in practice depends on the model and will be discussed at length in the rest of this paper.
Thanks to this information, we can then use probabilistic inference to improve the knowledge on
the observable, namely to update the distribution of $\Sigma$.

The goal of this work is thus the construction of a probability distribution of the observable $\Sigma$
given the perturbative expansion up to order $n$, namely
\begin{equation}\label{eq:PSigma}
P(\Sigma|c_0,...,c_n, H),
\end{equation}
where we have indicated with $H$ any assumption (hypothesis),
including both prior distributions and the model we want to use
(we will come back to this point later).
$P(A|B)$ indicates the probability distribution of $A$ given the information $B$.
In our case, the information is given by the first $n+1$ coefficients $c_0,...,c_n$, and $H$.
This distribution contains all the information we desire about our knowledge of the observable.
For instance, we can compute the best estimate of the observable $\Sigma$
as its expectation value according to such distribution,
\begin{equation}
\av{\Sigma} = \int d\Sigma \; \Sigma \; P(\Sigma|c_0,...,c_n, H),
\end{equation}
and its uncertainty either as the standard deviation or using degree-of-belief (DoB) intervals.
Most importantly, the probability distribution can be used directly in physical analyses,
when comparing theory predictions with data.

In the limit of ``infinite information'', namely when we know the exact result $\St$,
the probability Eq.~\eqref{eq:PSigma} should become
\begin{equation}\label{eq:Plim}
P(\Sigma|\St) = \delta(\Sigma-\St),
\end{equation}
which represents the certainty (not probability) that $\Sigma=\St$.
In this limit any a priori assumption $H$ does not matter.
Eq.~\eqref{eq:Plim} cannot be seen as the all-order limit of Eq.~\eqref{eq:PSigma},
due to the divergent nature of the series and to the non-perturbative contributions discussed in Sect.~\ref{sec:basics}.
However, it suggests that when adding information (i.e.\ when increasing the number $n$ of known orders, up to $\kas$)
the probability distribution should become narrower and more localised.
We shall consider this behaviour as a property that a good model for theory uncertainty must satisfy.

Note that knowing the probability distribution for the missing higher order term $\Delta_{\rm MHO}^{(n)}$
is practically the same as knowing the distribution for $\Sigma$.
Indeed, in the limit Eq.~\eqref{eq:Delta_size2} where we neglect the asymptotic and non-perturbative contributions,
the distributions for $\Delta_{\rm MHO}^{(n)}$ and $\Sigma$ are the same
up to a trivial shift given by the perturbative result, Eq.~\eqref{eq:Sigma_decomposition2}.
In the following, we will always deal directly with the distribution of $\Sigma$, Eq.~\eqref{eq:PSigma},
but we will compute it by estimating the missing higher orders $\Delta_{\rm MHO}^{(n)}$.

\subsection{The role of unphysical scales}
\label{sec:scales}

A general feature of renormalizable QFTs is the appearance of an unphysical scale $\mu$ as a consequence
of the regularization procedure needed to deal with ultraviolet divergences.
This is known as the renormalization scale.
Physical observables do not depend on it, as this scale is an artefact of the scheme adopted
to renormalize the theory.
However, in practical computations using perturbation theory, a scale dependence is present in each order,
in a way that it is compensated order by order.
Any finite-order truncation of the perturbative series will thus have a residual scale dependence,
which is formally of higher order.
As discussed in the introduction,
this observation is at the core of the canonical scale variation method to estimate theory uncertainties.

Because of renormalization scale dependence,
perturbative expansions do not uniquely determine a series,
but rather a family of series parametrized by the renormalization scale $\mu$.
We shall thus rewrite Eq.~\eqref{eq:Sigma_decomposition2} as
\begin{equation}\label{eq:Sigma_decomposition3}
\St = \sum_{k=0}^{n} c_k(\mu) \al^k(\mu) + \Delta_{\rm MHO}^{(n)}(\mu) + \Delta_{\rm asympt}(\mu) + \Delta_\text{non-pert}(\mu),
\end{equation}
where the left-hand side, the exact result, is scale independent:
\begin{equation}
\frac{d}{d\mu} \St = 0.
\end{equation}
Therefore, whenever we want to estimate the value of an observable using perturbation theory,
we need to face with the fact that the perturbative result is scale dependent, and also the missing higher orders are.

An immediate consequence of this fact is that also the probability distribution Eq.~\eqref{eq:PSigma}
will unavoidably depend on such the choice of scale $\mu$.
We can express this by writing the probability Eq.~\eqref{eq:PSigma} as
\begin{equation}\label{eq:PSigmamu}
P(\Sigma|c_0(\mu),...,c_n(\mu), H),
\end{equation}
where we have emphasised that each coefficient depends on the scale $\mu$,\footnote
{Note that there is a $\mu$ dependence also in $\al$, so to be rigorous we should also include $\al(\mu)$
  in the list of known parameters.}
or equivalently as
\begin{equation}\label{eq:PSigmamu2}
P(\Sigma|c_0,...,c_n, \mu, H),
\end{equation}
where the coefficients $c_k$ are intended as functions, to be computed at the value $\mu$ passed as an extra parameter.
This dependence on the scale is clearly undesired, because this probability distribution, which depends on $\mu$,
is for the true observable, which is independent of $\mu$.
In the limit of infinite knowledge, Eq.~\eqref{eq:Plim}, the distribution should tend to $\delta(\Sigma-\St)$
irrespectively of the value of $\mu$.
This implies that increasing the order, the probability distributions at different values of $\mu$
should become more and more similar.
While this feature is certainly nice, having an infinite number of different results for the same object
is obviously not ideal.
Rather, one would like to obtain a probability distribution for $\Sigma$ that does not depend on the choice of scale.
This can be achieved in two ways: either having a criterion for selecting an ``optimal value'' of the scale,
or combining in some way the results at different scales.

The first way is obviously simpler, provided such a criterion exists.
In the literature there are various approaches that aim at selecting an optimal scale,
e.g.\ the Brodsky-Lepage-Mackenzie (BLM) method~\cite{Brodsky:1982gc,Grunberg:1991ac,Brodsky:1994eh},
the principle of minimal sensitivity (PMS)~\cite{Stevenson:1981vj,Stevenson:1985dy},
the principle of maximal conformality (PMC)~\cite{Brodsky:2011ig,Brodsky:2011ta,Brodsky:2012rj,Brodsky:2013vpa}
and the recent principle of observable effective matching (POEM)~\cite{Chishtie:2020cen}.
The PMC is probably the most widespread approach.
It provides a way to select, order by order, an optimal scale that removes non-conformal $\beta$-function contributions
from the perturbative expansion. This is believed to remove the renormalons from the perturbative expansion,
thereby leading to a possibly convergent series (or at least to a less divergent one).
For our purposes, this approach could provide a way to select, among the infinitely many probability distributions for $\Sigma$,
a specific one.\footnote
{In fact, in the default PMC approach the value of the scale is different for every order,
  which implies that the PMC perturbative expansion does not correspond to any of the series with a fixed value of $\mu$ for all orders.
  A simplified variant of the PMC approach where the scale is the same at each order has also been proposed~\cite{Shen:2017pdu}.}
Note, however, that the PMC fixes the scale at each known order except the last one,
which is free and thus arbitrary.
This implies that a residual scale dependence is present also in the PMC approach,
even though it is claimed to be much milder than the canonical scale dependence.
However, it has been pointed out that a proper study of all the ambiguities in the approach
leads to larger uncertainties, comparable to the canonical scale uncertainty~\cite
{Kataev:2014jba,Kataev:2016aib,Chawdhry:2019uuv,Brambilla:2018tyu}.
We conclude that while the PMC approach is certainly intersting,
it cannot provide the full solution to our problem.

The second way to obtain a scale-independent probability distribution for $\Sigma$ is what we pursue in this work.
The treatment of the renormalization scale is addressed within the methodology for computing
the probability distribution Eq.~\eqref{eq:PSigma}.
The actual procedure to combine the probability distributions at different scales,
which represents one of the most innovative proposals of this work,
will be presented in Sect.~\ref{sec:scaleindep}.

Before moving further, another aspect of scale dependence must be discussed.
So far, we have described scale dependence as an \emph{obstacle} to obtain a unique probability distribution for the observable.
In fact, scale dependence can be also considered as a \emph{tool}.
This relies on the fact, already discussed in the introduction,
that the $\mu$-dependence of the finite-order truncation of the perturbative series
is of higher order, namely
\begin{equation}\label{eq:scale_dep_Sp}
\mu\frac{d}{d\mu} \sum_{k=0}^{n} c_k(\mu) \al^k(\mu) = \Ord\(\al^{n+1}\) = \Ord\(\Delta_{\rm MHO}^{(n)}\).
\end{equation}
This fact provides additional information on the expansion, which can be very useful as in most cases of interest
in particle physics the available information is very limited (typically $n=2$ or $3$).
In practice,
let us assume for simplicity that the $\mu$ dependence at a given order $k$ can be translated in a single number, $r_k$.
It can for instance be the canonical scale uncertainty error, or the slope of the cross section as a function of $\mu$,
or something similar (we will provide a precise definition later in Sect.~\ref{sec:notations_rk}).
We can then generalize the probability distribution as
\begin{equation}\label{eq:PSigmamu3}
P(\Sigma|c_0(\mu), r_0(\mu),...,c_n(\mu), r_n(\mu), H),
\end{equation}
where we have made explicit that also the $r_k$ numbers generally depend on the choice of $\mu$
about which the scale dependence is computed.
In other words, also in this case we get a family of distributions depending on the value of $\mu$
at which the perturbative expansion is computed, however this time we also include in each member of the family some information
on the scale dependence.
Since these parameters double the previous information\footnote
{According to the definition of the coefficients $c_k$, the first one, $c_0$, is independent of the scale,
and therefore $r_0$ does not provide any information.
However, there are cases in which the LO cross section is already scale dependent,
for instance because it starts at order $\al$.
In these cases also the first coefficient $c_0=0$,
and the first non-trivial order is $c_1$, with $r_1$ being non-zero.}
they are clearly very precious.

Note that the parameters $r_k$ represent the kind of information used in the construction of the canonical scale uncertainty.
More precisely, the canonical scale uncertainty is based only on the last one, $r_n$
(assuming $r_n$ is defined according to Eq.~\eqref{eq:CSU}),
and it does not provide a probabilistic interpretation.
In Sect.~\ref{sec:ModelScaleVar} we will instead make use of the scale variation information $r_k$
in a fully fledged probabilistic model, thereby providing a method that is,
in some sense, a more reliable and statistically sound version of canonical scale variation.

We finally stress that the renormalization scale is not the only scale appearing in perturbative computations.
For instance, in QCD processes involving hadrons in the initial or final states, the factorization
of collinear singularities introduces a (perturbative) dependence on another unphysical scale,
the so-called factorization scale.
Also, in effective field theories widely used in collider phenomenology
(e.g.\ heavy quark effective theory or soft-collinear effective theory),
other unphysical scales may appear.
The way to deal with these scales strictly depends on the scale itself.
For instance, the dependence on the factorization scale cancels between the perturbatively computable
coefficients and the non-perturbative parton distribution functions (PDFs).
Therefore, if we wish to obtain factorization scale independent probability distribution
for a physical observable, we may try to extend the procedure that we propose in Sect.~\ref{sec:scaleindep}
to this scale, with proper caveats due to the fact that PDFs are non-perturbative objects.
Instead, if we wish to include our definition of theory uncertainties in the
fits used to determine PDFs from data, the situation is completely different,
as the PDFs are not physical observables and they are thus scheme and scale dependent.
Addressing this issue is beyond the scope of this paper and it is left to future work.
Here, we only focus on the renormalization scale dependence, which is universal.

\subsection{The Cacciari-Houdeau method}
\label{sec:CH}

Before moving to our new proposals, we now present the Cacciari-Houdeau (CH) method
for estimating theory uncertainties~\cite{Cacciari:2011ze}.
Let us forget about the scale dependence (which is not dealt with in the original paper)
and consider the perturbative expansion
\begin{equation}\label{eq:SigmaPert}
\Sp = \sum_{k=0}^\infty c_k \al^k.
\end{equation}
We know that this series is divergent, but for the moment we ignore this fact.
The basic assumption made in the CH model is that all the coefficients $c_k$ are bounded in absolute value by a common number $\bar c$,
namely
\begin{equation}\label{eq:CHhyp}
\abs{c_k}\leq \bar c\qquad \forall k.
\end{equation}
The coefficient $\bar c$ is a parameter of the model, and specifically a hidden parameter,
which will disappear (through marginalization) in the final results.
Moreover, they assume that all $c_k$ are independent from each other, with the exception for the common bound,
which implies that
\begin{equation}\label{eq:CHindep}
P(c_k,c_j|\bar c) = P(c_k|\bar c) P(c_j|\bar c)\qquad \forall k,j, \quad k\neq j.
\end{equation}
The conditional probability $P(c_k|\bar c)$, that we shall call \emph{the likelihood},
encodes in a probabilistic way the assumption Eq.~\eqref{eq:CHhyp}. In the CH approach it is given by
\begin{equation}\label{eq:CHlik}
P(c_k|\bar c) = \frac1{2\bar c} \theta(\bar c-\abs{c_k}),
\end{equation}
namely the condition Eq.~\eqref{eq:CHhyp} must be strictly satisfied (the probability that the condition is violated is zero),
and within the allowed range all values are equally likely (flat distribution).
Finally, they provide a prior distribution for the hidden parameter,
\begin{equation}\label{eq:CHprior}
P_0(\bar c) \propto \frac1{\bar c}\theta(\bar c),
\end{equation}
which corresponds to a flat distribution in the logarithm of $\bar c$,
to encode the idea that the order of magnitude of $\bar c$ is a priori unknown.
Note that this prior distribution is not normalizeable, and thus it requires a regularization procedure to be used.

The ingredients above are sufficient to define the model, and using standard Bayesian inference
they allow to compute the sought probability distribution for the observable.
In practice, since the starting point is the perturbative expansion Eq.~\eqref{eq:SigmaPert},
to obtain a probability distribution for the full sum it is sufficient to have a probability distribution
for the unknown $c_k$ coefficients given the knowledge of the first $n+1$ coefficients $c_0,...,c_n$.
The key ingredient is thus $P(c_k|c_0,...,c_n)$, with $k>n$, which can be computed as
\begin{align}
P(c_k|c_0,...,c_n)
&= \frac{P(c_k,c_0,...,c_n)}{P(c_0,...,c_n)} & (k>n) \nonumber\\
&= \frac{\int d\bar c\,  P(c_k,c_0,...,c_n,\bar c)}{\int d\bar c\, P(c_0,...,c_n,\bar c)} \nonumber\\
&= \frac{\int d\bar c\,  P(c_k,c_0,...,c_n|\bar c)P_0(\bar c)}{\int d\bar c\, P(c_0,...,c_n|\bar c)P_0(\bar c)} \nonumber\\
&= \frac{\int d\bar c\,  P(c_k|\bar c) P(c_0|\bar c) \cdots P(c_n|\bar c) P_0(\bar c)}
  {\int d\bar c\, P(c_0|\bar c) \cdots P(c_n|\bar c) P_0(\bar c)}
\end{align}
where we have used the relation between the joint probability and the conditional probability $P(A,B)=P(A|B)P(B)$ in the first step,
introduced the hidden parameter in the second step, used again the definition of the conditional probability in the third step
and finally used Eq.~\eqref{eq:CHindep}.
The last line is written in terms of known functions (the likelihood and the prior), and can thus be easily computed.
This result can be easily generalized to the joint probability of more than one unknown coefficient,
\begin{equation}
P(c_{k_1},...,c_{k_m}|c_0,...,c_n)
= \frac{\int d\bar c\,  P(c_{k_1}|\bar c)\cdots P(c_{k_m}|\bar c) P(c_0|\bar c) \cdots P(c_n|\bar c) P_0(\bar c)}
  {\int d\bar c\, P(c_0|\bar c) \cdots P(c_n|\bar c) P_0(\bar c)},
\qquad k_1,...,k_m>n.
\end{equation}
At this point one can also compute, at least formally, the probability distribution for the full sum,
which is given by
\begin{equation}\label{eq:CHPSigma}
P(\Sp|c_0,...,c_n) = \int dc_{n+1}dc_{n+2}\cdots\, P(c_{n+1},c_{n+2},...|c_0,...,c_n) \delta\(\Sp-\sum_{k=0}^\infty c_k\al^k\).
\end{equation}
Since this is an infinite-dimensional integration, it is impossible to perform it numerically and too hard to compute it analytically.
Therefore, one can approximate the full sum with a truncated sum at some finite order $n+j$ to get
\begin{equation}\label{eq:CHPSigmaApprox}
P(\Sp|c_0,...,c_n) \simeq \int dc_{n+1}\cdots dc_{n+j}\, P(c_{n+1},...,c_{n+j}|c_0,...,c_n) \delta\(\Sp-\sum_{k=0}^{n+j} c_k\al^k\),
\end{equation}
which can be easily handled, at least numerically. The easiest approximation is obtained with $j=1$,
where only the first missing higher order is used to approximate the distribution,
and it leads to a simple analytical expression~\cite{Cacciari:2011ze}.

The CH approach is a breakthrough in the context of estimating the theory uncertainty from missing higher orders,
as it provides for the first time a probabilistic way to determine the uncertainty of an observable
computed in perturbation theory.
Note that this approach only considers the behaviour of the expansion,
without using any information from the scale dependence.
This is exactly the opposite of the canonical scale variation method,
which is based on the scale dependence and does not use any information on the behaviour of the expansion.

Despite the nice properties of the CH approach, there are some caveats that need to be considered.
The most obvious one is the assumption Eq.~\eqref{eq:CHhyp}, that implies that the perturbative expansion
is bounded by a convergent (geometric) series,
\begin{align}
\abs{\Sp} = \abs{\sum_{k=0}^\infty c_k \al^k}
  \leq \sum_{k=0}^\infty \abs{c_k} \al^k
  \leq \sum_{k=0}^\infty \bar c \al^k
  = \frac{\bar c}{1-\al},
\end{align}
where we have assumed $\al<1$ (which is consistent with the perturbative hypothesis).
This is in contrast with the known fact that perturbative expansions are divergent.
In a subsequent paper, Ref.~\cite{Bagnaschi:2014wea}, the CH approach has been
modified to account for the divergence of the series, by modifying the condition Eq.~\eqref{eq:CHhyp}
into
\begin{equation}\label{eq:CHmodhyp}
\abs{c_k} \leq \bar b k!\qquad \forall k,
\end{equation}
with $\bar b$ being the new hidden parameter, with the same prior as $\bar c$.
This condition on the coefficients is much less stringent and compatible with the assumption
that the divergence of the series is dominated by a factorial growth such as those due to renormalons.
However, with this choice it is no longer possible to use Eq.~\eqref{eq:CHPSigma}
to compute the full sum, as it does not exist. Therefore only the approximation
Eq.~\eqref{eq:CHPSigmaApprox} can be considered, and typically with a low value of $j$
otherwise the probability distribution becomes large due to the factorial growth.
In Ref.~\cite{Bagnaschi:2014wea} only $j=1$ is considered.

The second issue is related to the fact that Eq.~\eqref{eq:CHhyp} does not account for a possible
power growth of the coefficients.
In other words, each term of the perturbative expansion is assumed to have a power scaling given just by $\al$.
This limitation was stressed already in the first CH paper~\cite{Cacciari:2011ze},
where they propose to solve it by rescaling $\al$,
\begin{equation}
\sum_{k=0}^\infty c_k \al^k = \sum_{k=0}^\infty c_k' \(\frac{\al}{\eta}\)^k
\end{equation}
to obtain new coefficients $c_k'$ that
satisfy Eq.~\eqref{eq:CHhyp}, or Eq.~\eqref{eq:CHmodhyp} in the factorial divergent hypothesis of Ref.~\cite{Bagnaschi:2014wea}.
The trouble is how to find such a rescaling factor $\eta$.
In Ref.~\cite{Bagnaschi:2014wea}, a global survey over a quite large number of observables
is proposed to determine an optimal value of $\eta$.
In this survey they compare the uncertainty computed at the next-to-last known order with the actual (known) next order,
to quantify how reliable the uncertainty is for each given value of the rescaling factor.
Apart from the details (for which we refer the Reader to Ref.~\cite{Bagnaschi:2014wea}),
we stress that this approach assumes that the rescaling factor is the same for all the observables.\footnote
{In Ref.~\cite{Bagnaschi:2014wea} a distinction is made between two classes of observables,
  namely those involving hadrons in the initial state and those not involving hadrons in the initial state.
  For each class a value of the rescaling factor is found.}
However, this is hardly the case, as different processes and observables are characterized by different
dominant perturbative corrections.\footnote{For instance, the dominant QCD corrections to gluon jets and quark jets
are characterized by different color factors, $(C_A\as)^k$ and $(C_F\as)^k$ respectively.
Also, in differential distributions in certain kinematic conditions there are powers of kinematic
logarithms appearing with the powers of the coupling and dominating the perturbative corrections,
and depending on the kinematic conditions the logarithm can have different sizes.}
Therefore, it is more appropriate to assume that the rescaling factor $\eta$ is process and observable dependent.
A different way to obtain it has been proposed in Ref.~\cite{Forte:2013mda},
where a fitting procedure is suggested to find an optimal value of $\eta$ such that the known $c_k$
are all of the same order.
This method is observable dependent and uses only the information
on the perturbative expansion to obtain the optimal rescaling.
However, a fitting procedure to determine the rescaling factor clashes with the probabilistic nature of the rest of the procedure.

Finally, the CH approach or its modified versions do not deal with scale dependence.
The CH machinery is applied to the perturbative expansion at a given value of the scale,
and if one changes the scale the result changes accordingly.
The difference in the final probability distribution at different values of the scales
can be sizeable, see e.g.\ Ref.~\cite{Bonvini:2016frm}.

\subsection{A working example: Higgs production at the LHC}
\label{sec:ggH}

In Sect.~\ref{sec:applications} we will consider various examples of perturbative expansions,
and apply our methods to each of them.
Nevertheless, in order to be clearer when discussing our new proposals,
we think it is instructive to have a working example to immediately visualize
how the various methods work.

The observable we choose is the inclusive cross-section for Higgs production in gluon fusion at the LHC for this purpose.
This process has a number of advantages:
\begin{itemize}
\item it is known up to N$^3$LO~\cite{Anastasiou:2015ema,Anzai:2015wma,Anastasiou:2016cez,Mistlberger:2018etf}
  (four orders in the perturbative expansion)
  in the so-called large top mass effective theory
  (this is a rarity in QCD processes at LHC, most of which are only known to NLO or NNLO);
\item it is characterized by large perturbative corrections;
\item canonical scale variation underestimates the impact of (large) higher orders;
\item its factorization scale dependence is very mild, so the whole scale dependence is basically
  fully captured by its renormalization scale dependence;
\item because the process starts at $\Ord(\as^2)$, the LO is scale dependent;
\item it is a real process and not a toy example.
\end{itemize}
On top of these reasons, the process is interesting also from a phenomenological point of view (see e.g.\ Ref.~\cite{deFlorian:2016spz}),
and indeed it was subject of several investigations of theory uncertainties
from missing higher orders~\cite{David:2013gaa,Bonvini:2016frm,Bagnaschi:2014wea,Anastasiou:2016cez}.

\begin{figure}[t]
  \centering
  \includegraphics[width=0.7\textwidth,page=1]{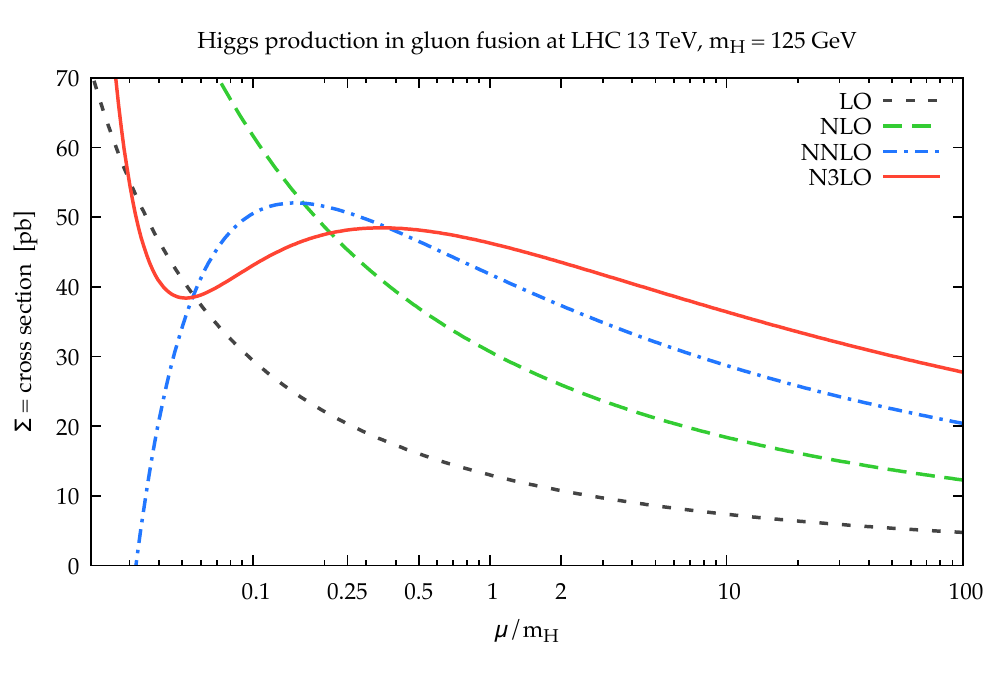}
  \caption{The Higgs production cross section in gluon fusion at LHC $\sqrt{s}=13$~TeV with $\mh=125$~GeV
    for $\muf=\mh/2$ as a function of the renormalization scale $\mu$ at the four known perturbative orders.}
  \label{fig:ggHscale}
\end{figure}
Specifically, we consider LHC at $\sqrt{s}=13$~TeV and set the Higgs mass to $\mh=125$~GeV.
We fix the factorization scale to $\muf=\mh/2$ (which is a standard choice~\cite{deFlorian:2016spz}),
even though changing this value has a negligible effect on the cross section,
in particular at high orders.
The ``raw'' result of the computation, which is the cross section as a function of the renormalization scale $\mu$,
is plotted in Fig.~\ref{fig:ggHscale}.
Note that this process depends on a single hard scale, the Higgs mass $\mh$.
Therefore, it is natural to choose $\mu$ of the order of $\mh$, in order to avoid the presence of large
unresummed logarithms of $\mu/\mh$ in the perturbative coefficients.
Nevertheless, we believe it is instructive to visualize the scale dependence for wide range
of scales: the plot covers almost four decades in $\mu/\mh$.

We see that for large values of the scale, where the QCD coupling is smaller,
the expansion is characterized by all positive contributions and it progresses very slowly, with large perturbative corrections.
Conversely, at small scales where the strong coupling blows up the expansion is highly unstable.
In a ``central'' region, where $\mu\sim\mh$, the expansion behaves in a reasonably perturbative way,
even though the perturbative corrections are rather large and it is not at all clear what could possibly be
the true cross section.
Note also the presence of a stationary point at NNLO and of two stationary points at N$^3$LO,
which could corrupt an estimate of the uncertainty based on canonical scale variation.

We stress that the full plot of Fig.~\ref{fig:ggHscale} can be constructed from just the sequence
of partial sums of the observable at the various orders at a given scale,
and the knowledge of the value of the coupling at that scale.
In this example, we have at $\mu=\mh$ (and $\muf=\mh/2$)\footnote
{These numbers have been computed with the latest version of
  \href{https://www.roma1.infn.it/~bonvini/higgs/}{\texttt{ggHiggs}}~\cite{Ball:2013bra,Bonvini:2014jma,Bonvini:2016frm,Bonvini:2018ixe,Bonvini:2018iwt}.
  Note that the N$^3$LO result differs from the one reported in Ref.~\cite{Bonvini:2016frm}
  because in the latest version of the code the N$^3$LO computation through a threshold expansion~\cite{Anastasiou:2016cez}
  has been replaced with the complete result~\cite{Mistlberger:2018etf}.}
\begin{equation}\label{eq:XSggH}
\as(\mh)=0.1126,\qquad
\Sp(\mh) = \left\{ 13.0, 30.7, 41.8, 46.3 \right\}\; \text{pb},
\end{equation}
where the values in curly brackets correspond to the cross section at
LO, NLO, NNLO and N$^3$LO respectively.
The way to reconstruct the cross section at any scale from these ingredients is discussed in Sect.~\ref{sec:appScale}.

\section{Model-independent features of the inference approach at fixed scale}
\label{sec:statmodel}

In this section we start introducing our notation and present some general features of the construction of the models.
For the time being we consider only the models at a fixed scale.
How to obtain scale-independent probability distributions will be discussed in Sect.~\ref{sec:scaleindep}.

\subsection{Basic notations}
\label{sec:notations}

Let us denote with $\Sigma_n(\mu)$ the partial sum of the perturbative series up to order $n$
depending on the scale $\mu$.
If we are considering a standard perturbative expansion in powers of $\al$ this is given by
\begin{equation}\label{eq:Sigmanal}
\Sigma_n(\mu) = \sum_{k=0}^n c_k(\mu) \al^{k+k_0}(\mu),
\end{equation}
where we have explicitly introduced an offset $k_0$ for observables starting at $\Ord\(\al^{k_0}\)$.
This notational change with respect to the previous section allows us to be sure that the
first order, $k=0$, namely the leading order (LO), is non-zero.
Note that the information contained in the coefficients $c_k$ is fully contained in the sequence
of partial sums
\begin{equation}
\Sigma_0, \Sigma_1, \Sigma_2, ...
\end{equation}
once the value of $\al$ and of $k_0$ are specified.
From now on, we shall consider the partial sums as the basic objects, and forget about the coefficients $c_k$ and even of $\al$.
In this way, the ``perturbative expansion'' is more general, as it does no longer necessarily
need to be a strict expansion in powers of $\al$.
For instance, it can be a logarithmic-ordered expansion of a resummed result,
or a \mbox{(non-)linear} transformation of the perturbative expansion.
In what follows we simply assume that $\Sigma_n$ represents the partial sum of a
non-specified expansion that behaves perturbatively, defined such that the ``LO'' $\Sigma_0$ is non-zero.

For a number of reasons that will become clear later,
it is convenient to introduce a normalized version of the expansion, where the LO is factored out,
\begin{equation}\label{eq:Sigmannorm1}
\Sigma_n(\mu) = \Sigma_0(\mu)\sum_{k=0}^n \delta_k(\mu),
\end{equation}
where we have defined the dimensionless coefficients\footnote
{A possible alternative definition of the dimensionless perturbative coefficients $\delta_k(\mu)$
is obtained dividing by $\Sigma_{k-1}(\mu)$ rather than $\Sigma_0(\mu)$.
Given the similarity to the definition used here, Eq.~\eqref{eq:deltakdef},
the models that we will propose can be straightforwardly modified using this alternative definition.
The results will in general differ, the difference being larger if perturbative corrections are large,
thus highlighting a source of arbitrariness in the procedure.
}
\begin{equation}\label{eq:deltakdef}
\delta_k(\mu) \equiv \frac{\Sigma_k(\mu)-\Sigma_{k-1}(\mu)}{\Sigma_0(\mu)}.
\end{equation}
According to this definition, $\delta_0=1$ always, so we can also write
\begin{equation}\label{eq:Sigmannorm}
\Sigma_n(\mu) = \Sigma_0(\mu)\(1+\sum_{k=1}^n \delta_k(\mu) \).
\end{equation}
The coefficients $\delta_k$ contain the information on the perturbative orders.
The fact that $\delta_0=1$ sets a common size for all perturbative expansions (useful when defining the model),
and also tells us that the LO does not contain any useful information on the behaviour of the expansion
and thus on the uncertainty due to missing higher orders
(which is obvious, as with just the LO one cannot know how large the perturbative corrections will be).
This is true even when the LO is scale dependent, because without another order to compare with
one cannot know how to reliably translate the scale dependence into an information on the size of missing higher orders.
Only when at least two orders, namely $\delta_0$ and $\delta_1$, are known one can start making inference.
The only role of the LO is to set the dimension and the size of the observable through the prefactor $\Sigma_0(\mu)$.

\subsection{Definition of the scale-dependence numbers}
\label{sec:notations_rk}

Because the expansion is scale dependent, as discussed in Sect.~\ref{sec:scales}
one can construct numbers $r_k(\mu)$ that encode this dependence at each order.
The importance of these numbers stems from the observation that scale dependence of a physical observable
is formally of higher order, Eq.~\eqref{eq:scale_dep_Sp}, namely\footnote
{We do not include $k_0$ in the power of $\al$ because we assume to define $r_k$ such that they are directly comparable
to the normalized $\delta_k$.}
\begin{equation}\label{eq:rkOrd}
r_k(\mu)=\Ord\(\al^{k+1}\).
\end{equation}
The actual construction of these numbers must be such that this equation is satisfied.
The simplest measure of the scale dependence is the slope of $\Sigma_k(\mu)$ as a function of $\mu$,
or more precisely its derivative with respect to $\log\mu$
(the logarithm is natural because the dependence on the scale is logarithmic).
In order to directly compare these numbers with the $\delta_k$ coefficients,
it is convenient to normalise the derivative to the observable itself to make them dimensionless.
The two most natural ways to do this are
\begin{equation}\label{eq:rk_options}
\frac1{\Sigma_0(\mu)}\,\mu\frac{d}{d\mu}\Sigma_k(\mu) \qquad\text{or}\qquad
\frac1{\Sigma_k(\mu)}\,\mu\frac{d}{d\mu}\Sigma_k(\mu),
\end{equation}
where in the first case we have normalized to the LO $\Sigma_0$,
while in the second case we have normalized to the observable at the same order at which the scale dependence is computed.
If the observable does not change much at different orders, the two options are equivalent.
However, in presence of large perturbative corrections there can be a substantial difference between the two.
None of them is better in an absolute sense.
However, we argue that the second option has a nicer perturbative behaviour,
with higher order $r_k$ being typically smaller than lower order ones,
as one would expect from Eq.~\eqref{eq:rkOrd}.
\begin{figure}[t]
  \centering
  \includegraphics[width=0.495\textwidth,page=2]{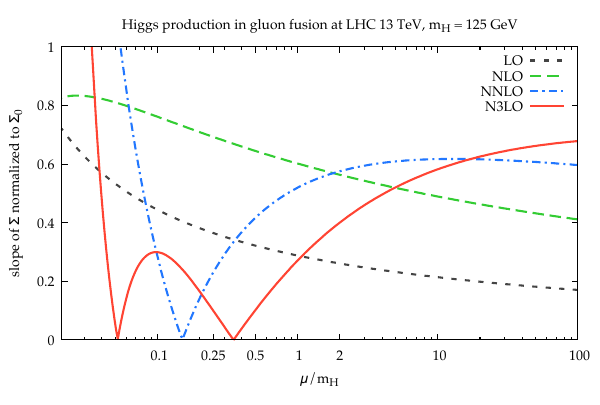}
  \includegraphics[width=0.495\textwidth,page=3]{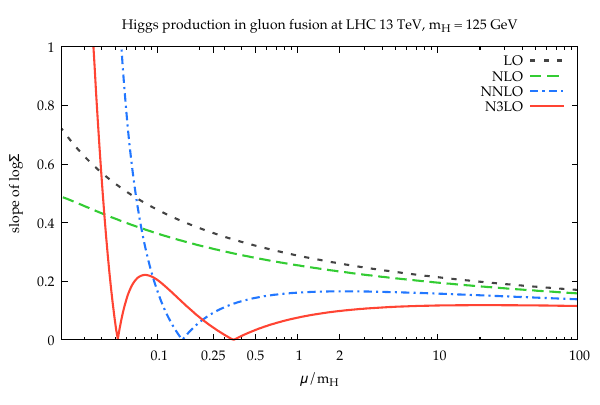}
  \caption{The absolute value of the normalized slopes defined in Eq.~\eqref{eq:rk_options},
    respectively shown in the left and right plots, for the Higgs production process.}
  \label{fig:ggHscaleslope}
\end{figure}
To prove this, we consider the example of Higgs production introduced in Sect.~\ref{sec:ggH},
and show in Fig.~\ref{fig:ggHscaleslope} the (absolute value of) the two options in Eq.~\eqref{eq:rk_options}.
From the left plot, corresponding to the left option in Eq.~\eqref{eq:rk_options},
we see that at NLO (green curve) the slope is always larger than the LO one (black curve),
and at the next orders there is a large variability without a precise hierarchy.
Conversely, in the right plot, corresponding to the right option in Eq.~\eqref{eq:rk_options},
the higher order curves are smaller than the lower order ones over a wide range of scales,
which is the expected perturbative behaviour.
So the definition of the $r_k$ coefficients that we propose is \emph{morally} given by
\begin{equation}
r_k(\mu) \simeq 
\abs{\frac1{\Sigma_k(\mu)}\,\mu\frac{d}{d\mu}\Sigma_k(\mu)}
=
\abs{\mu\frac{d}{d\mu}\log\Sigma_k(\mu)}
,
\end{equation}
where the absolute value is introduced to keep only the information on the size of the dependence but not the sign.
Note that at small scales, where the perturbative expansion is unstable (Fig.~\ref{fig:ggHscale}),
the expected hierarchy is violated. This is the case also in proximity of the stationary points.
To avoid problems arising from stationary points, we propose to define the $r_k$ numbers in a slightly different way.
Namely, we replace the derivative with a finite difference, and look for the largest finite difference
in a range around the point $\mu$.
In formulae, we have\footnote
{In the numerical implementation, the max over a continuous variable is replaced with a discretised version.}
\begin{equation}\label{eq:rkdef}
r_k(\mu) \equiv \frac1{\abs{\Sigma_k(\mu)}}\,\max_{\mu/f\leq\nu\leq f\mu}\abs{\frac{\Sigma_k(\nu) - \Sigma_k(\mu)}{\log(\nu/\mu)}},
\end{equation}
that selects the largest finite difference in the range of scales between $\mu/f$ and $f\mu$,
where $f>1$ is a factor to be fixed.
This definition is more robust than the one based on the derivative.
\begin{figure}[t]
  \centering
  \includegraphics[width=0.495\textwidth,page=5]{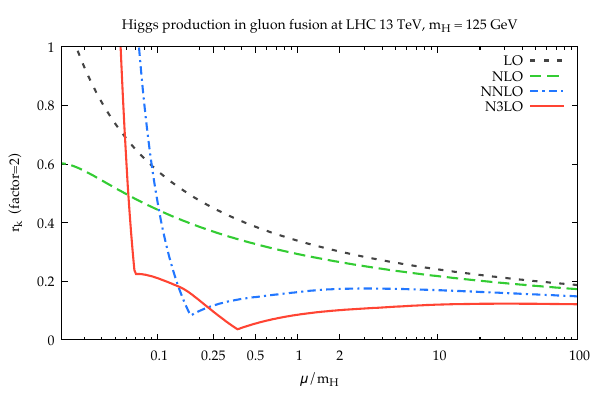}
  \includegraphics[width=0.495\textwidth,page=4]{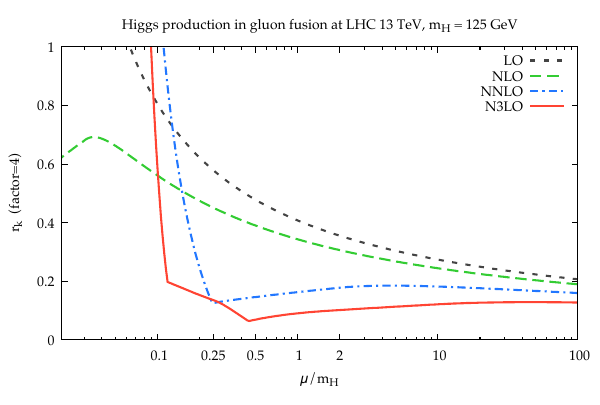}
  \caption{The coefficients $r_k(\mu)$, Eq.~\eqref{eq:rkdef}, for $f=2$ (left) and $f=4$ (right),
    for the Higgs production process.}
  \label{fig:ggHrk}
\end{figure}
We show this in Fig.~\ref{fig:ggHrk}, for $f=2$ (left plot) and $f=4$ (right plot).
Increasing $f$ allows to be less sensitive to stationary points, and indeed
in the right plot the expected hierarchy is preserved over a wide range of scales.
At low scales, the value of $r_k$ at NNLO and N$^3$LO blows up because the finite difference
probes the low-scale region where the perturbative result is unstable, Fig.~\ref{fig:ggHscale}.
For larger values of $f$, the blow-up region obviously moves to larger scales.
Therefore, the robustness gained increasing $f$ has to be balanced with the stability loss.
We believe that $f=4$ can be considered as a reasonable compromise.\footnote
{Note that the canonical scale variation method uses the scale dependence over a factor of 2 about the central scale choice.
  Our definition of $r_k$ with $f=4$ probes a larger range.
  However, while in canonical scale variation increasing the factor increases the uncertainty,
  here since the finite difference is divided by the size of the variation the dependence on the factor is much milder.
  In a hypothetical linear dependence case, $r_k$ would be independent of $f$ while the canonical scale variation
  uncertainty would change with $\log f$.}

We stress that for an efficient computation of the $r_k(\mu)$ coefficients
a fast evaluation of the scale dependence of the observable is needed.
Since the renormalization scale dependence is universal and governed by the $\beta$-function of the theory,
it can be constructed automatically from the knowledge of the
observables at the various orders at a single value of the scale.
The details are reported in Appendix~\ref{sec:appScale}.

We finally note that if the LO is scale independent, then $r_0=0$.
Given that we aim at using the $r_k$ numbers to estimate the higher orders,
a zero value would be inappropriate.
More precisely, $r_0=0$ means that the LO cannot be used to probe higher orders through scale variation.
We can fix this by assuming that, when the LO is scale independent,
we can only say that the NLO corrections may be of the order of the LO itself.
Given the definition of the $r_k$ as normalized quantities, this corresponds to assuming $r_0=\Ord(1)$.
For definiteness, we arbitrarily set $r_0(\mu)=1/2$ in these cases.
We shall see later that this assumption is rather conservative.

\subsection{General features of the models}
\label{sec:generalfeatures}

Before discussing the actual models that we propose, we want to give some general features
that are common to all of them.
Let us recall that the goal of this work is to construct a probability distribution for $\Sigma$
given the first known $n+1$ orders $\Sigma_0,\Sigma_1,...,\Sigma_n$, or equivalently
$\Sigma_0,\delta_1,...,\delta_n$ (using our new notation).
This was given in Eq.~\eqref{eq:PSigma}, or, when working at fixed scale, in Eq.~\eqref{eq:PSigmamu2}.

According to our assumption Eq.~\eqref{eq:Delta_size2} that the uncertainty of a theoretical prediction
based on perturbation theory is dominated by the missing higher orders,
we shall compute the distribution for $\Sigma$ through these missing higher orders,
ignoring the other contributions from the asymptotic expansion truncation and the non-perturbative part.
In other words, we shall approximate the observable $\Sigma$ as
\begin{equation}\label{eq:Sigmaapprox}
\Sigma\simeq \Sigma_{k}(\mu), \qquad k\leq\kas,
\end{equation}
where the best approximation is obtained using $k=\kas$, namely including all the missing higher orders
up to the point in which the asymptotic expansion starts to grow.
Since we typically do not know a priori the value of $\kas$, the best we can do is to include a few
extra orders beyond the known ones, without exaggerating in order not to risk to go beyond $\kas$.

The inference on the observable $\Sigma$ can be obtained in terms of the more fundamental inference of the higher orders
from the known ones. This was already done in Sect.~\ref{sec:CH}, and we repeat that derivation here in more generality
and with our new notation.
Assuming we know the coefficients of the expansion up to order $n$ and we approximate the observable
with its expansion up to order $n+j$ with $j>0$, $n+j\leq\kas$, the probability distribution is given by
(suppressing for ease of notation the implicit dependence on the hypothesis $H$)
\begin{align}\label{eq:PSigmaderiv}
P(\Sigma|\delta_n,...,\delta_1,\Sigma_0,\mu)
  &= \frac{P(\Sigma,\delta_{n},...,\delta_1,\Sigma_0|\mu)}{P(\delta_n,...,\delta_1,\Sigma_0|\mu)} \nonumber\\
  &= \frac{\int d\delta_{n+j}\cdots d\delta_{n+1}\,
    P(\Sigma,\delta_{n+j},...,\delta_1,\Sigma_0|\mu)}{P(\delta_n,...,\delta_1,\Sigma_0|\mu)} \nonumber\\
  &= \frac{\int d\delta_{n+j}\cdots d\delta_{n+1}\,
    P(\Sigma|\delta_{n+j},...,\delta_1,\Sigma_0,\mu)P(\delta_{n+j},...,\delta_1,\Sigma_0|\mu)}{P(\delta_n,...,\delta_1,\Sigma_0|\mu)} \nonumber\\
  &= \int d\delta_{n+j}\cdots d\delta_{n+1}\,
    P(\Sigma|\delta_{n+j},...,\delta_1,\Sigma_0,\mu) P(\delta_{n+j},...,\delta_{n+1}|\delta_n,...,\delta_1,\Sigma_0,\mu) \nonumber\\
  &\simeq \int d\delta_{n+j}\cdots d\delta_{n+1}\,
 \delta\(\Sigma-\Sigma_{n+j}(\mu)\) P(\delta_{n+j},...,\delta_{n+1}|\delta_n,...,\delta_1,\Sigma_0,\mu),
\end{align}
where we have used in the last step the $(n+j)$-th order approximation
\begin{equation}\label{eq:PSigmaapprox}
P(\Sigma|\delta_{n+j},...,\delta_1,\Sigma_0,\mu) \simeq \delta\(\Sigma-\Sigma_{n+j}(\mu)\),
\end{equation}
which is a direct consequence of Eq.~\eqref{eq:Sigmaapprox},
and the definition of $\Sigma_n(\mu)$, Eq.~\eqref{eq:Sigmannorm}.\footnote
{Note that Eq.~\eqref{eq:PSigmaapprox} is a rather crude approximation.
To be conservative, one could replace the delta function with a smooth distribution,
for instance a gaussian. However, in this case one would need some parameters,
like the width of the distribution, which would then be arbitrary.}
If $\Sigma$ is a positive definite observable, like a cross section, one could also impose
a positivity constraint through a factor $\theta(\Sigma)$, which however requires the computation
of a normalization factor as in general the distribution is no longer normalized.
The result Eq.~\eqref{eq:PSigmaderiv} is written in terms of
$P(\delta_{n+j},...,\delta_{n+1}|\delta_n,...,\delta_1,\Sigma_0,\mu)$,
which represents the probability of the higher orders given the known ones,
and depends on the model under consideration.
Note that we have also included $\Sigma_0$ among the known information,
even though it is just a prefactor: indeed $\Sigma_0$
is required to compute the $r_k$ numbers introduced in Sect.~\ref{sec:notations_rk} 
that are needed in models that use information on the scale dependence.

The delta function in Eq.~\eqref{eq:PSigmaderiv} can be used to perform the integral
over one of the higher orders, say $\delta_{n+j}$, which gives
\begin{align}\label{eq:PSigmaj}
P(\Sigma|\delta_n,...,\delta_1,\Sigma_0,\mu)
  &\simeq \frac1{\Sigma_0(\mu)} \int d\delta_{n+j-1}\cdots d\delta_{n+1} \nonumber\\
  &\qquad\qquad\times
P\(\delta_{n+j}=\frac{\Sigma-\Sigma_{n+j-1}(\mu)}{\Sigma_0(\mu)}, \delta_{n+j-1},...,\delta_{n+1}|\delta_n,...,\delta_1,\Sigma_0,\mu\).
\end{align}
The other integrations are more complicated, and should be performed numerically (unless the model is particularly simple).
The simplest result is obtained when considering $j=1$, namely when approximating
the observable using just the first unknown order,
\begin{align}\label{eq:PSigmaj=1}
P(\Sigma|\delta_n,...,\delta_1,\Sigma_0,\mu)
  &\overset{j=1}\simeq \frac1{\Sigma_0(\mu)} P\(\delta_{n+1}=\frac{\Sigma-\Sigma_{n}(\mu)}{\Sigma_0(\mu)}|\delta_n,...,\delta_1,\Sigma_0,\mu\).
\end{align}
In practice, since the order $n+j$ that approximates best the observable is not known a priori,
a convenient approach to choose properly $j$ is the following.
We start considering the simplest approximation, $j=1$, Eq.~\eqref{eq:PSigmaj=1}.
Then, we include the next order, namely we use $j=2$.
If the distribution changes visibly,
then we further increase $j$ by a unity, and so on until the distribution changes only mildly,
up to a tolerance decided by the user.

We now turn our attention to the probability $P(\delta_{n+j},...,\delta_{n+1}|\delta_n,...,\delta_1,\Sigma_0,\mu)$.
As we said, this is model dependent. However, we can further write it in terms of more fundamental
probabilities, using the relation
\begin{equation}\label{eq:Pdeltadecomp}
P(\delta_{n+j},...,\delta_{n+1}|\delta_n,...,\delta_1,\Sigma_0,\mu)
=\frac{P(\delta_{n+j},...,\delta_{n+1},\delta_n,...,\delta_1,\Sigma_0|\mu)}{P(\delta_n,...,\delta_1,\Sigma_0|\mu)}.
\end{equation}
The numerator and the denominator are the same object,
simply with a different number of $\delta_k$ terms.
In the numerator only some of them are known while the others are unknown,
but mathematically this does not make any difference.
The joint distribution $P(\delta_m,...,\delta_1,\Sigma_0|\mu)$ at fixed scale is the basic object of the model
that is needed to make inference on the higher orders.
In all the models we will consider, we will assume that there is a number of hidden parameters characterizing the model.
Denoting with $\vec p$ the vector of such parameters, the joint distribution can be written as
\begin{align}\label{eq:Pdeltamdelta1}
  P(\delta_m,...,\delta_1,\Sigma_0|\mu)
  &= \int d\vec p\; P(\delta_m,...,\delta_1,\Sigma_0,\vec p|\mu) \nonumber\\
  &= \int d\vec p\; P(\delta_m,...,\delta_1,\Sigma_0|\vec p,\mu) P_0(\vec p|\mu),
\end{align}
where in the second line we have written explicitly the prior of these parameters,
as they are hidden and thus can never be known exactly.

\begin{figure}[t]
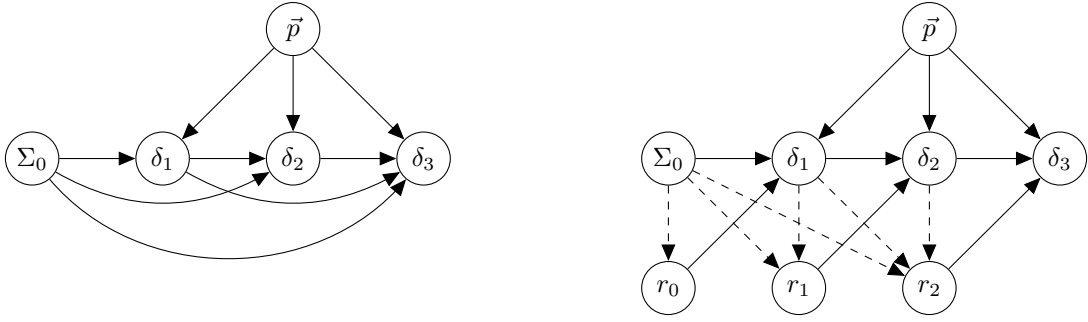

\centering
  \tikz[baseline=(current bounding box.north)]{ %
    \node[latent] (d0) {$\Sigma_0$} ; %
    \node[latent, right=of d0] (d1) {$\delta_1$} ; %
    \node[latent, right=of d1] (d2) {$\delta_2$} ; %
    \node[latent, right=of d2] (d3) {$\delta_3$} ; %
    \node[latent, above=of d2] (pars) {$\vec p$} ; %
    \edge {pars} {d1,d2,d3} ; %
    \edge {d0} {d1} ; %
    \edge {d1} {d2} ; %
    \edge {d2} {d3} ; %
    \path (d0) edge [->, >={triangle 45}, bend right=30] (d2) ;
    \path (d0) edge [->, >={triangle 45}, bend right=50] (d3) ;
    \path (d1) edge [->, >={triangle 45}, bend right=30] (d3) ;
  }
  \hspace{15ex}
  \tikz[baseline=(current bounding box.north)]{ %
    \node[latent] (d0) {$\Sigma_0$} ; %
    \node[latent, right=of d0] (d1) {$\delta_1$} ; %
    \node[latent, right=of d1] (d2) {$\delta_2$} ; %
    \node[latent, right=of d2] (d3) {$\delta_3$} ; %
    \node[latent, above=of d2] (pars) {$\vec p$} ; %
    \node[latent, below=of d0] (r0) {$r_0$} ; %
    \node[latent, right=of r0] (r1) {$r_1$} ; %
    \node[latent, right=of r1] (r2) {$r_2$} ; %
    \edge[dashed] {d0} {r0} ; %
    \edge[dashed] {d0,d1} {r1} ; %
    \edge[dashed] {d0,d1,d2} {r2} ; %
    \edge {pars} {d1,d2,d3} ; %
    \edge {r0} {d1} ; %
    \edge {r1} {d2} ; %
    \edge {r2} {d3} ; %
    \edge {d0} {d1} ; %
    \edge {d1} {d2} ; %
    \edge {d2} {d3} ; %
  }
  \caption{The most general Bayesian network of models of inference of missing higher orders (left),
    and a more specific one, using explicitly the scale variation numbers $r_k$,
    that covers all the cases considered in this work (except the one of Sect.~\ref{sec:SCv2}).}
  \label{fig:net}
\end{figure}
To compute the joint distribution, it is sometimes useful to visualize the relation between the different objects using
a Bayesian network.
The most general network for any model of theory uncertainties is rather simple,
and it is depicted in Fig.~\ref{fig:net} (left) showing only the first four orders
(the generalization to more than four orders is obvious).
The various orders depend in general on all previous orders
(but not on future ones, otherwise the model cannot be predictive),
and on the hidden parameters $\vec p$.
Since $\Sigma_0$ is the first order and it only sets the size of the observable, it does not depend on anything.
In addition to this general structure, we want to consider more explicitly the role of the scale dependence numbers $r_k$.
Since these are functions of the $\delta_k$ coefficients and $\Sigma_0$, there is no need to specify them in the network.
However, it may be useful to introduce them explicitly in order to better appreciate their role.
Therefore, in the same Fig.~\ref{fig:net} (right) we also show explicitly a network depending on these $r_k$.
The dashed arrows represent \emph{deterministic links}, namely analytic relations rather than probabilistic ones,
and mean that the $r_k$ numbers are computable analytically from the various orders.
Note that this network is not as general as the previous one. Indeed now we have made the assumption
that each $\delta_k$ depend only on the previous $\delta_{k-1}$ and $r_{k-1}$ (and on $\vec p$),
but not on all the previous orders.
This simplification is not necessary, but it will be adopted in all our models
(except the one of Sect.~\ref{sec:SCv2}).

\section{Model 1: geometric behaviour model}
\label{sec:GB}

We now present our first model, that uses only information on the behaviour of the expansion.
As such, this model is very general, and not restricted to a QFT application.

\subsection{The hypothesis of the model}

The first model that we consider is a generalization of the Cacciari-Houdeau model introduced in Sect.~\ref{sec:CH}.
The main difference is that our model accounts for a possible power growth of the coefficients of the expansion
within the probabilistic approach.
In the CH model each term of the expansion is bounded by
\begin{equation}\label{eq:CHhyp2}
\abs{c_k\al^k} \leq \bar c \al^k\qquad \forall k,
\end{equation}
which is Eq.~\eqref{eq:CHhyp} in which we have emphasised that the power behaviour of the full $\Ord(\al^k)$
term is entirely described by $\al^k$.
As we have discussed in Sect.~\ref{sec:CH}, this hypothesis is hardly satisfied,
and a variant of the CH method in which the expansion parameter $\al$ is rescaled by a factor $\eta$ is advisable.
So far, this has never been done within the context of the probabilistic model.

Now, in our new notation, Eq.~\eqref{eq:Sigmannorm}, we have lost information on the coupling $\al$,
as the whole information is contained in the $\delta_k$ coefficients Eq.~\eqref{eq:deltakdef}.
This was done on purpose and is to be considered as an advantage,
as the expansion Eq.~\eqref{eq:Sigmannorm} is more general
than a strict expansion in powers of $\al$.
If we want to translate the CH condition Eq.~\eqref{eq:CHhyp2} in the new language,
we are forced to introduce back an expansion parameter.
Since from our point of view this is a new parameter (as we have lost information about $\al$),
it is natural to consider it as a parameter of the model, rather than an external one.
As such, it is not fixed to be $\al$ or a fraction of it.
The condition that we consider is thus
\begin{equation}\label{eq:GBhyp}
\abs{\delta_k(\mu)} \leq c a^k\qquad \forall k< \kas,
\end{equation}
where both $c$ and $a$ are positive hidden parameters of the model.
This condition implies that the expansion is bounded by a geometric expansion,
and we thus call this model a \emph{geometric behaviour model}.
We have specified that this bound can only be valid for orders $k$ smaller than $\kas$,
otherwise it is certainly violated. This is anyway the only region which we are interested in,
according to the discussion in Sect.~\ref{sec:basics}.

Eq.~\eqref{eq:GBhyp}, though very similar to the CH condition Eq.~\eqref{eq:CHhyp2},
differs from it in a number of very important aspects, that we now list.
\begin{itemize}
\item The fact that $a$ is a parameter makes not only the model more general than CH,
  but it also allows to find, through inference, the most appropriate values (in a probabilistic sense)
  of the expansion parameter $a$ compatible with the behaviour of the expansion.
  In other words, $a$ can be interpreted as the rescaled expansion parameter $\al/\eta$
  introduced in Sect.~\ref{sec:CH}, but with the rescaling factor $\eta$ being determined
  through inference from the perturbative expansion itself, as opposed to the approaches of Refs.~\cite{Bagnaschi:2014wea,Forte:2013mda}.
\item The parameter $c$ is dimensionless, as opposed to $\bar c$ which has the dimension of the observable.
  This is useful as we can legitimately use a universal prior for $c$ without knowing anything about the observable.\footnote
  {In the CH approach, the prior for $\bar c$ is universal as well.
    However, in that case, one could have decided to use an observable-dependent prior.}
\item Since the condition Eq.~\eqref{eq:GBhyp} is limited to $k< \kas$,
  the fact that the perturbative expansion is typically factorially divergent does not imply
  that the geometric bound is unacceptable.
  Of course one cannot say that the entire series is bounded by a geometric series,
  but a small portion of it may well be.
  Therefore, the condition Eq.~\eqref{eq:GBhyp} limited to the first few orders can be considered
  as perfectly acceptable.\footnote
  {Whether or not it is also good has to be judged on the basis of the performance of the model.
    We will discuss this later in Sect.~\ref{sec:GBposterior}.}
\end{itemize}
The CH assumption Eq.~\eqref{eq:CHhyp} can be recovered from this new approach
by fixing $a=\al$ (or $a=\al/\eta$ in the rescaled variant) and rewriting $c=\bar c/\Sigma_0(\mu)$.
Note that the hidden parameters $c,a$ depend on the scale $\mu$.
This has not been written explicitly, because in a statistical language this information
is expressed by saying that $c$ and $a$ are correlated with $\mu$.

In order to construct a probabilistic model to estimate theory uncertainties, we need to translate
the condition Eq.~\eqref{eq:GBhyp} into a likelihood function.
We assume the simple conditional probability
\begin{equation}\label{eq:GBlik}
P(\delta_k| c, a, \mu) = \frac{1}{2ca^k}\theta\(ca^k - \abs{\delta_k(\mu)}\), \qquad k<\kas,
\end{equation}
which is the straightforward extension of the CH choice, Eq.~\eqref{eq:CHlik}.
We have considered the idea of allowing violation of the bound, by adding tails to the likelihood.
However, the fact of having two hidden parameters already makes the model much more flexible than the original CH model,
so adding a violation of the bound would not lead to any substantial improvement to the model stability.

\begin{figure}[t]
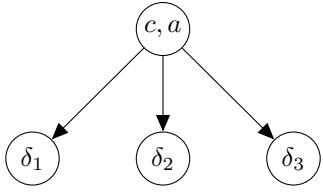

\centering
  \tikz[baseline=(current bounding box.north)]{ %
    \node[latent, right=of d0] (d1) {$\delta_1$} ; %
    \node[latent, right=of d1] (d2) {$\delta_2$} ; %
    \node[latent, right=of d2] (d3) {$\delta_3$} ; %
    \node[latent, above=of d2] (pars) {$c,a$} ; %
    \edge {pars} {d1,d2,d3} ; %
  }
  \caption{Bayesian network for the geometric behaviour model.
    Each $\delta_k$ depends only on the hidden parameters (and $\mu$) through the likelihood Eq.~\eqref{eq:GBlik}.}
  \label{fig:GBnet}
\end{figure}

In similarity with the CH model, we assume that all $\delta_k$ are independent of each other
at fixed $c$, $a$ and $\mu$, namely
\begin{equation}\label{eq:GBindep}
P(\delta_k,\delta_j| c, a, \mu) = P(\delta_k| c, a, \mu) P(\delta_j| c, a, \mu) \qquad \forall k,j, \quad k\neq j,
\end{equation}
which generalizes to any set of $\delta_k$ coefficients.
These conditions, together with the prior distributions for the hidden parameters that we are going to discuss,
are sufficient to fully define the model.
The Bayesian network of this model is a simplified version of the general one introduced in Sect.~\ref{sec:generalfeatures},
and is depicted in Fig.~\ref{fig:GBnet}.

\subsection{The choice of priors}
\label{sec:GBprior}

The two hidden parameters $c,a$ need a prior distribution.
We assume that a priori the two variables are uncorrelated
\begin{equation}\label{eq:priorfact}
P_0(c,a|\mu) = P_0(c|\mu) P_0(a|\mu),
\end{equation}
where we have emphasised that in principle the prior can depend on the scale $\mu$,
considered to be externally given for the moment (this will change in Sect.~\ref{sec:scaleindep}).

Before proposing a functional form for the prior, let us comment on the first step of the inference.
The information from the LO is encoded in $\Sigma_0$,
that appears as a prefactor in our normalized expansion, Eq.~\eqref{eq:Sigmannorm}.
Since the likelihood Eq.~\eqref{eq:GBlik} looks at the $\delta_k$ and not at $\Sigma_0$,
the knowledge of the LO does not change our prior: $P(c,a|\Sigma_0,\mu)=P_0(c,a|\mu)$.
While this is strictly speaking correct according to our notation, it misses a conceptual point.
Indeed, it exists also a $\delta_k$ for the LO, which is $\delta_0$.
The fact that $\delta_0=1$ makes it a trivial variable, in the sense that it carries no information,
which is the reason why it does not appear in our functions.
However, from a mathematical viewpoint, it would play a role if we \emph{assume}
that the likelihood Eq.~\eqref{eq:GBlik} is also valid for the LO.
Indeed, at order zero, the likelihood becomes
\begin{equation}
P(\delta_0|c,a,\mu) = \frac1{2c}\theta\(c-\abs{\delta_0}\) = \frac1{2c}\theta\(c-1\).
\end{equation}
This equation implies that the distribution for $a$ is unmodified by the knowledge of the LO,
while the distribution for $c$ changes as
\begin{equation}\label{eq:posteriorc}
P(c|\delta_0,\mu) \propto \int da\,P(\delta_0|c,a,\mu) P_{0}(c,a|\mu) \propto \frac{P_{0}(c|\mu)}{c} \theta(c-1).
\end{equation}
This result shows that the requirement that the likelihood Eq.~\eqref{eq:GBlik}
applies also at LO implies the constraint $c\geq1$.
Since $\delta_0$ is not explicitly part of our parameters,
we will not perform the inference in Eq.~\eqref{eq:posteriorc} in our model.
In other words, our prior Eq.~\eqref{eq:priorfact} is to be considered
as the posterior after the (trivial and universal) knowledge of $\delta_0=1$.
We will keep track of this result by constructing the prior such that the condition $c\geq1$ is satisfied.

At this point we are free to choose the functional form of our prior.
Note that it is convenient to use simple functional forms, such that analytic computations can be performed.
Let us start with the prior for $c$.
Since we do not have any a priori knowledge on the expected size of $c$,
only a monotonic prior is acceptable.
We find it reasonable to assume a power law function
\begin{equation}\label{eq:GBpriorc}
P_0(c|\mu) =\frac{\epsilon}{c^{1+\epsilon}} \theta(c-1), \qquad \epsilon>0,
\end{equation}
where we have included the $\theta(c-1)$ for the reason explained above.
Note that we do not include any dependence on the scale, namely for any value of $\mu$ we use the same prior distribution.
The parameter $\epsilon$ is an arbitrary parameter, and can be chosen at will
(we will discuss our favourite choices later in Sect.~\ref{sec:GBposterior}).
We note that for $\epsilon=1$ we obtain the form $\theta(c-1)/c^2$,
which results from a ``pre-prior'' (without knowledge of $\delta_0$)
proportional to $\theta(c)/c$, as obvious from Eq.~\eqref{eq:posteriorc}.
This is the prior used for $\bar c$ in the CH model, Eq.~\eqref{eq:CHprior}.
Since $\theta(c)/c$ is an improper distribution, a regularization procedure is needed in CH to perform practical computation.
Rather, when including the trivial information $\delta_0=1$ within the model,
the prior is a proper distribution, and the computations are simplified.
This is one advantage of using from the start the universal
information $\delta_0=1$, which is in turn a consequence of using the normalized version of the expansion,
Eq.~\eqref{eq:Sigmannorm}.

We stress that the choice $\theta(c)/c$ for the ``pre-prior'' corresponds to a flat (and thus non-informative)
distribution for the variable $\log c$, justified in the CH work by the argument that the order of magnitude of
the hidden parameter is unknown.
The other natural non-informative ``pre-prior'' is given simply by $\theta(c)$, namely a flat distribution in the hidden parameter $c$,
which is again improper.
This choice corresponds to the value $\epsilon=0$ in Eq.~\eqref{eq:GBpriorc}.
With $\epsilon=0$ also the prior Eq.~\eqref{eq:GBpriorc} is improper,
and indeed in that equation we have assumed $\epsilon$ to be strictly greater than zero.
In fact, computations can be easily performed also in the $\epsilon\to0$ limit with just a little care.
We find however that this complication is not necessary: if we wish to mimic the effect of a flat ``pre-prior'' in $c$,
we can just use a very small positive value for $\epsilon$. This will indeed be our favourite choice, see Sect.~\ref{sec:GBposterior}.
Variations of this parameter will be explored in Sect.~\ref{sec:GBpriorscan}.

As far as the parameter $a$ is concerned, we have to make an initial choice about the expected behaviour of the expansion.
Indeed, the geometric bound is convergent (namely, at finite order, decreasing with the order) only for $a<1$.
In principle, we could allow $a\geq1$, which would describe a (power) divergent behaviour of the expansion.\footnote
{The ``divergent region'' $a\geq1$ would predict large higher orders, which in turn would enlarge
the probability distribution for the observable.
The impact of this region can be limited by a strongly suppressed prior for $a\geq1$.}
However, allowing $a\geq1$ is in contrast with the asymptotic nature of the expansion that we are assuming, see Sect.~\ref{sec:basics}.
Therefore, we suggest to limit our interest to the region $a<1$. We thus propose the functional form
\begin{equation}\label{eq:GBpriora}
P_0(a|\mu) = (1+\omega)(1-a)^\omega\theta(a)\theta(1-a), \qquad \omega\geq0.
\end{equation}
Once again, we assume that this prior is independent of the scale $\mu$.
For $\omega=0$, we obtain a flat distribution in the allowed region $0\leq a\leq1$ (in this case, the extreme value $a=1$ is included),
while for $\omega>0$ we suppress the region $a\to1$ to favour small values of $a$.
The actual value of $\omega$ that we recommend will be discussed in Sect.~\ref{sec:GBposterior},
and its variations will be considered in Sect.~\ref{sec:GBpriorscan}.

\subsection{Inference on the unknown higher orders}
\label{sec:GBinf}

We have now all the ingredients to perform the inference in this model.
The basic probability that we need is the conditional probability of
unknown higher orders given the first $n$ known non-trivial orders $\delta_1,...,\delta_n$.
Note that because of our assumption Eq.~\eqref{eq:GBhyp}, only a limited number of higher orders,
up to order $\kas$, can be predicted within this model.
Therefore, the most generic probability distribution we need to consider, according to Eq.~\eqref{eq:Pdeltadecomp}, is
\begin{equation}\label{eq:GBPdnew}
P(\delta_{n+j},...,\delta_{n+1}|\delta_n,...,\delta_1,\mu)
=\frac{P(\delta_{n+j},...,\delta_{n+1},\delta_n,...,\delta_1|\mu)}{P(\delta_n,...,\delta_1|\mu)},
\qquad j\geq1, \quad n+j<\kas,
\end{equation}
namely the probability of the unknown orders $n+1,...,n+j$ given the known orders $1,...,n$.
In contrast with Eq.~\eqref{eq:Pdeltadecomp}, we have removed here the explicit dependence on $\Sigma_0$,
as it does not appear in the likelihood and thus it does not play any role in the inference procedure.

The numerator and the denominator are the same object,
therefore we can focus on the distribution (at fixed scale) for a generic number $m$ of consecutive coefficients,
which is given by
\begin{align}\label{eq:GBPd1dm}
P(\delta_m,...,\delta_1|\mu)
&=\int dc\int da\, P(\delta_m,...,\delta_1|c,a,\mu) P_0(c,a|\mu) \nonumber\\
&=\int dc\int da\, P(\delta_m|c,a,\mu)\cdots P(\delta_1|c,a,\mu) P_0(c,a|\mu)\nonumber\\
&= \frac1{2^m} \int \frac{da}{a^{\frac{m(m+1)}2}}\, P_0(a|\mu)
\int \frac{dc}{c^m}\, P_0(c|\mu)\, \theta\(c-\max\[\frac{\abs{\delta_1(\mu)}}{a},..., \frac{\abs{\delta_m(\mu)}}{a^m}\]\)
\end{align}
which corresponds to Eq.~\eqref{eq:Pdeltamdelta1} specialized to our case.
In the first line we have introduced the hidden parameters,
in the second line we have used the independence of the coefficients, Eq.~\eqref{eq:GBindep},
and in the third line we have explicitly written the likelihood Eq.~\eqref{eq:GBlik}
and used the prior independence of the hidden parameters Eq.~\eqref{eq:priorfact}.
The integral in Eq.~\eqref{eq:GBPd1dm} can be computed analytically for our choice
of priors Eq.~\eqref{eq:GBpriorc} and Eq.~\eqref{eq:GBpriora}.
The inner integral is given by
\begin{equation}
\int \frac{dc}{c^m}\, P_0(c|\mu)\, \theta\(c-\max\[\frac{\abs{\delta_1(\mu)}}{a},..., \frac{\abs{\delta_m(\mu)}}{a^m}\]\) =
\frac{\epsilon}{m+\epsilon}\max\[1,\frac{\abs{\delta_1(\mu)}}{a},..., \frac{\abs{\delta_m(\mu)}}{a^m}\]^{-m-\epsilon}.
\end{equation}
Depending on the value of $a$, the max function selects a different term with a different $a$ dependence.
In order to compute the $a$ integral analytically, it is therefore convenient to partition
the integration region $0\leq a<\infty$ into a finite number of intervals,
in each of which the max function returns one of its arguments.
Since the arguments of the max function contain powers of $a$ that grow with $k$,
the intervals are ordered with $k$.
More precisely, smaller values of $a$ will select larger powers $k$, and viceversa.
We can thus introduce consecutive decreasing numbers $a_k$, representing the boundaries of these consecutive intervals,
defined such that
\begin{equation}\label{eq:GBakdef}
a_{k+1}< a< a_k
\qquad\Leftrightarrow\qquad
\max\[1,\frac{\abs{\delta_1(\mu)}}{a},...,\frac{\abs{\delta_m(\mu)}}{a^m}\] = \frac{\abs{\delta_k(\mu)}}{a^k}
\end{equation}
and assuming $a_0\equiv\infty$ and $a_{m+1}\equiv0$.
An algorithm for extracting the various $a_k$'s from the knowledge of the $\delta_k$'s is described in App.~\ref{sec:appAlgo}.
The $a$ integral is then given by
\begin{align}
P(\delta_m,...,\delta_1|\mu)
&= \frac1{2^m} \int \frac{da}{a^{\frac{m(m+1)}2}}\,P_0(a|\mu)\frac{\epsilon}{m+\epsilon}
  \max\[1,\frac{\abs{\delta_1(\mu)}}{a},..., \frac{\abs{\delta_m(\mu)}}{a^m}\]^{-m-\epsilon} \nonumber\\
&=\frac1{2^m} \sum_{k=0}^{m} \int_{a_{k+1}}^{a_k} \frac{da}{a^{\frac{m(m+1)}{2}}}\, P_0(a|\mu) \frac{\epsilon}{m+\epsilon}
  \(\frac{\abs{\delta_k(\mu)}}{a^k}\)^{-m-\epsilon} \nonumber\\
&=\frac{\epsilon(1+\omega)}{2^m(m+\epsilon)}\sum_{k=0}^{m} \frac1{\abs{\delta_k(\mu)}^{m+\epsilon}}
  \int_{\min(1,a_{k+1})}^{\min(1,a_k)} da\, a^{(m+\epsilon)k-\frac{m(m+1)}{2}} (1-a)^\omega,
\end{align}
where in the last line we have used the explicit form of the prior for $a$, Eq.~\eqref{eq:GBpriora},
that further restricts the integration region to $a\leq1$.
The general result of this integral can be written in terms of the incomplete Beta function.
However, a simpler form is obtained if $\omega$ is an integer.
Indeed in this case
\begin{align}\label{eq:GBPd1dm3}
P(\delta_m,...,\delta_1|\mu)
&=\frac{\epsilon(1+\omega)}{2^m(m+\epsilon)}\sum_{k=0}^{m} \frac1{\abs{\delta_k(\mu)}^{m+\epsilon}}
  \sum_{j=0}^\omega(-1)^j\binom{\omega}{j} \int_{\min(1,a_{k+1})}^{\min(1,a_k)} da\, a^{(m+\epsilon)k-\frac{m(m+1)}{2}+j} \nonumber\\
&=\frac{\epsilon(1+\omega)}{2^m(m+\epsilon)}\sum_{k=0}^{m} \frac1{\abs{\delta_k(\mu)}^{m+\epsilon}}
  \sum_{j=0}^\omega(-1)^j\binom{\omega}{j}\times\nonumber\\
&\quad\times 
  \begin{cases}
\log\frac{\min(1,a_k)}{\min(1,a_{k+1})} \qquad \qquad \qquad \qquad \qquad \text{if }(m+\epsilon)k-\frac{m(m+1)}{2}+j+1=0\\
    \frac{\min(1,a_k)^{(m+\epsilon)k-\frac{m(m+1)}{2}+j+1}-\min(1,a_{k+1})^{(m+\epsilon)k-\frac{m(m+1)}{2}+j+1}}{(m+\epsilon)k-\frac{m(m+1)}{2}+j+1}\qquad \text{elsewhere}.
  \end{cases}
\end{align}
The advantage of having such a simple analytic form is that the numerical evaluation is very fast.
However, nothing prevents one from making more complicated choices for the prior distributions,
paying the price that the numerical integration will typically slow down the computation of the distribution.

Eq.~\eqref{eq:GBPd1dm3} can be used directly in Eq.~\eqref{eq:GBPdnew} to obtain the probability distribution
of the unknown higher orders.
Following the derivation of Sect.~\ref{sec:generalfeatures} we can then construct the distribution for
the observable $\Sigma$, Eq.~\eqref{eq:PSigmaj}.
A useful property of the result is that the tails of such distributions are dominated by the first
missing higher order (this is a consequence of the hierarchy in the arguments of the max function, Eq.~\eqref{eq:GBakdef}).
Therefore, the asymptotic behaviour of the distribution is given by
\begin{equation}\label{eq:GBasympt}
P(\Sigma|\delta_n,...,\delta_1,\Sigma_0,\mu) \sim \frac1{\abs{\Sigma-\Sigma_n}^{n+1+\epsilon}}.
\end{equation}

\subsection{The posterior of the hidden parameters}
\label{sec:GBposterior}

\begin{figure}[t]
  \centering
  \includegraphics[width=0.495\textwidth,page=1]{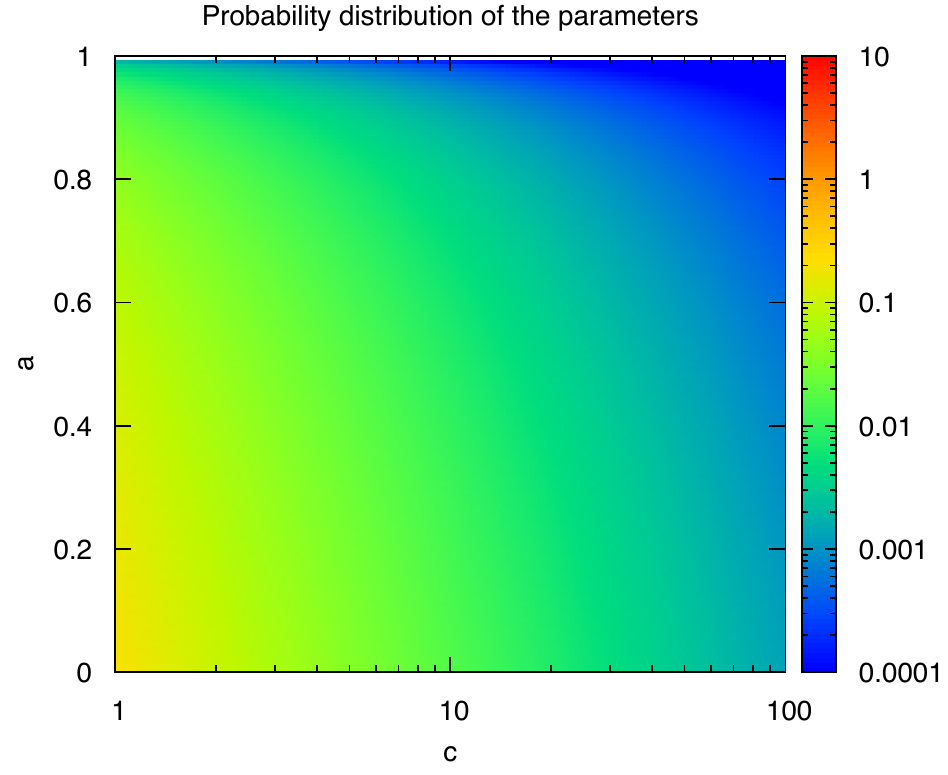}
  \includegraphics[width=0.495\textwidth,page=2]{./images/plot_Posterior_c_a_ggH_paper.pdf}\\
  \includegraphics[width=0.495\textwidth,page=3]{./images/plot_Posterior_c_a_ggH_paper.pdf}
  \includegraphics[width=0.495\textwidth,page=4]{./images/plot_Posterior_c_a_ggH_paper.pdf}
  \caption{The plots show the probability distribution of the parameters $c$ and $a$.
    The first plot (upper left) is the prior, the second (upper right) is the posterior after the knowledge of $\delta_1$,
    the third (bottom left) adds the knowledge of $\delta_2$ and the last (bottom right) the one of $\delta_3$.
    The observable under consideration is the inclusive Higgs cross section, Sect.~\ref{sec:ggH}, for fixed scale $\mu=\mh/2$.
    Each line represents the contour $ca^k=\abs{\delta_k}$, for increasing values of $k$ from 1 to 3,
    to clarify the role of the theta function.
    Note that the $c$ axis is shown in log scale, but the probability distribution is for $c$ and not $\log c$.
    For this plot, we used $\omega=1$ and $\epsilon=0.1$ for the prior parameters.}
  \label{fig:GBposterior}
\end{figure}

Even if the hidden parameters of the model are never part of the final distribution,
it is instructive to understand how their distribution changes with the knowledge of the first few orders.
The posterior distribution of $c$ and $a$ can be easily computed as
\begin{align}
P(c,a|\delta_n,...,\delta_1,\mu)
&=\frac{P(\delta_n,...,\delta_1,c,a|\mu)}{P(\delta_n,...,\delta_1|\mu)} \nonumber\\
&=\frac{P(\delta_n,...,\delta_1|c,a,\mu)P_0(c,a|\mu)}{P(\delta_n,...,\delta_1|\mu)} \nonumber\\
&=\frac
  {P(\delta_n|c,a,\mu)\cdots P(\delta_1|c,a,\mu)P_0(c,a|\mu)}{P(\delta_n,...,\delta_1|\mu)},
\end{align}
where the denominator is given in Eq.~\eqref{eq:GBPd1dm3}, and also corresponds to the
integral over $c$ and $a$ of the numerator.
It's clear that even if in our prior $c$ and $a$ were uncorrelated,
correlations arise from the model after inference takes place.
Using the explicit form of our likelihood, Eq.~\eqref{eq:GBlik},
the posterior becomes
\begin{align}\label{eq:GBposterior}
P(c,a|\delta_n,...,\delta_1,\mu)
&=\frac{P_0(c,a|\mu)}{P(\delta_n,...,\delta_1|\mu)}\,
\frac1{a^{\frac{n(n+1)}2}}\, \frac1{c^n}\, \theta\(c-\max\[\frac{\abs{\delta_1}}{a},..., \frac{\abs{\delta_n}}{a^n}\]\)\nonumber\\
&=\frac{\epsilon(1+\omega)\theta(a)\theta(1-a)}{P(\delta_n,...,\delta_1|\mu)}\,
\frac{(1-a)^\omega}{a^{\frac{n(n+1)}2}}\, \frac1{c^{n+1+\epsilon}}\, \theta\(c-\max\[1,\frac{\abs{\delta_1}}{a},..., \frac{\abs{\delta_n}}{a^n}\]\),
\end{align}
where in the second line we have also used the explicit form of our prior, Eq.~\eqref{eq:GBpriorc} and Eq.~\eqref{eq:GBpriora}.
The theta function cuts out the region of small $c$ and small $a$, with a boundary given by a sequence of
contours identified by $ca^k=\abs{\delta_k}$ in the region $a_{k+1}<a<a_k$ selected by the max function, see Eq.~\eqref{eq:GBakdef}.
On the other hand, the growing negative power of both $c$ and $a$ tend to favour small values of $c,a$, thus close to this boundary.
Since the power of $a$ grows quadratically with $n$ while that of $c$ only linearly,
inference tends to favour smaller $a$ at the price of having somewhat larger $c$.
A visual example of this behaviour is given in Fig.~\ref{fig:GBposterior}.

This preference is a nice outcome of the model: inference favours small values of the parameter $a$ that lead to
a better behaviour of the expansion, with smaller higher orders.
The prediction of the observable will then be more \emph{precise}, namely subject to a smaller uncertainty.
Of course one also (and more importantly) wants the prediction to be \emph{accurate},
namely with a reliable uncertainty that does not underestimate the missing higher orders.
Judging whether the outcome of the model is reliable is not immediate,
and requires explicit examples to verify it. We will come back later to this point
in Sect.~\ref{sec:GBresults} and Sect.~\ref{sec:applications}.

Note that all the considerations so far are independent of the prior.
The prior has the only role of changing the ``starting point'' of the inference procedure.
With sufficiently many known orders, our choice of the prior will not matter.
However, since the number of known orders is typically limited,
choosing wisely is important.
Since we like the ``direction'' selected by the inference procedure 
(it's better to have larger $c$ and smaller $a$ than the opposite)
it seems convenient to choose a prior distribution that already favours the same region of parameter space.
This is achieved in our Eq.~\eqref{eq:GBpriorc} and Eq.~\eqref{eq:GBpriora}
choosing a small value for $\epsilon$ and a large value for $\omega$.
Note however that a large $\omega$ suppresses the region $a\sim1$,
while in the inference there is no such suppression,
simply the small $a$ region is enhanced by a negative power of $a$.
Therefore, using a large value of $\omega$ may introduce a significant bias.
A good compromise, that we advocate as the best choice, is $\omega=1$.
In this way there is a preference for smaller $a$ with only a mild suppression for $a\sim1$.
On the other hand, for $\epsilon$ we can choose an arbitrarily small (positive) value,
for instance $\epsilon=0.1$ or $\epsilon=0.01$.
The difference between either choice is relevant only at very low orders:
in Fig.~\ref{fig:GBposterior} only the first plot (the prior) would change visibly.
We thus use $\epsilon=0.1$ in the rest of this work,
with the exception of Sect.~\ref{sec:GBpriorscan} where we will consider variations of the prior parameters.

\subsection{Representative results}
\label{sec:GBresults}

Before moving further, we now present some representative results of this method.
We use our working example of Higgs production to examine the distribution for the cross section.
We fix the renormalization scale $\mu=\mh/2$, which is the most widely used choice for this process.
Of course the result of this model depends on the choice of scale made,
and in addition it does not know anything about the scale dependence.
So the input for this model are just 4 numbers, the values of the cross section
at the chosen scale at LO, NLO, NNLO and N$^3$LO.
How to deal with scale dependence in this model will be discussed in Sect.~\ref{sec:scaleindep}.

\begin{figure}[t]
  \centering
  \includegraphics[width=0.9\textwidth,page=2]{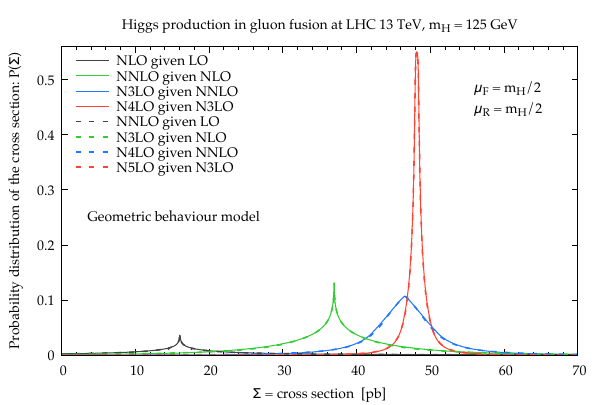}
  \caption{Probability distributions of the Higgs cross section with different states of knowledge.}
  \label{fig:GBdistr}
\end{figure}

In Fig.~\ref{fig:GBdistr} we plot the distribution for the observable $\Sigma$ (the Higgs cross section),
Eq.~\eqref{eq:PSigmaj}, using only the first missing higher order (solid lines)
or the first two missing higher orders (dashed lines).
The four colors correspond to different knowledge: LO (black),
NLO (green), NNLO (blue) and N$^3$LO (red).

We immediately notice that the solid and dashed curves are basically identical.
This implies that, for any given status of knowledge,
the uncertainty in this model is dominated by the first missing higher order.
This is a consequence of having $a<1$, that implies that higher and higher orders are smaller and smaller in this model.
Moreover, given that the distribution of $c$ and $a$ favours small values of $a$, Fig.~\ref{fig:GBposterior},
the impact of the next higher orders is significantly smaller than the first missing higher order.
The reason why the curves are almost identical also depends on the fact that the tails of these
distributions behave as a negative power, Eq.~\eqref{eq:GBasympt}, and are thus very ``long'':
therefore, the effect by the next higher orders of ``enlarging'' the distribution is almost invisible
as the tails already cover large deviations of the observable.
We thus conclude that for this model it is sufficient to consider only the first missing higher order,
so we can directly use Eq.~\eqref{eq:PSigmaj=1} for all practical applications.
This makes the implementation of this model very fast, as no numerical integration is needed.

Let us now comment on the predictions of this model.
Of course, every distribution is centered on the cross section at the known order,
and it is symmetric, because in our assumption Eq.~\eqref{eq:GBhyp}
the sign of the missing higher order is treated agnostically
(we will consider a possible way of taking the sign into account in Sect.~\ref{sec:sign}).
We consider the four states of knowledge in turn.
\begin{itemize}
\item When only the LO is known, the shape of the curves is fully determined by the prior
  (no inference took place yet), so the black curve is not particularly useful.
  Indeed, for our choice of prior, the distribution is very broad, so it is basically not predictive.
  Note also that the distribution is barely normalizable, because it asymptotically behaves as $\abs{\Sigma-\Sigma_0}^{-1-\epsilon}$,
  Eq.~\eqref{eq:GBasympt}, with $\epsilon$ small.
  Another consequence of this behaviour is that the variance of this distribution is infinite.
  Also, the distribution allows with a high probability ($\sim40\%$) unphysical negative values of the cross section.
  These can be avoided imposing a positivity constraint on $\Sigma$,
  but it is instructive to see how much information is needed to obtain a distribution sufficiently narrow to
  have negligible probability of negative cross section.
\item The knowledge of the NLO allows to perform the first step of the interference,
  so the green curve provides the first non-trivial prediction.
  However, it is still an ``immature'' prediction, because only one piece of information has been used.
  Indeed, this distribution is still very broad, with tails asymptotically behaving as $\abs{\Sigma-\Sigma_0}^{-2-\epsilon}$,
  and a cusp that favours a single point but not really a region.
  For $0<\epsilon<1$, as we advocate, also this distribution has infinite variance.
  Also in this case, there is a significant probability ($\sim4\%$) of negative cross section.
  Therefore, the NLO alone does not provide sufficient information in this model to make a precise prediction.
\item Once the NNLO is known, the situation changes. On top of having tails that decrease more rapidly,
  the blue curve clearly identifies a more probable region where the distribution has some sort of bump.
  The prediction is still rather uncertain, but at least the procedure seems to converge.
  The probability of negative cross section is reduced to less than $0.3\%$.
\item As a confirmation of this, we see that the knowledge of the N$^3$LO improves the situation even more.
  Now the red distribution is well localized, with a clear bump, which is also well compatible
  with the prediction at the previous order.
  The uncertainty is clearly reduced, and the probability of negative cross section becomes negligible (less than $0.01\%$)
\end{itemize}
We conclude that this method works well, but requires a sufficient number of known orders to be predictive.
Two orders (NLO) is the absolute minimum, but three orders (NNLO) are probably necessary to achieve a decent precision.
Beyond NNLO (four or more orders) it should work very well.
Note also that the distributions do not look like a gaussian,
and therefore they \emph{cannot} be approximated with a gaussian distribution in applications.

\begin{figure}[t]
  \centering
  \includegraphics[width=0.495\textwidth,page=1]{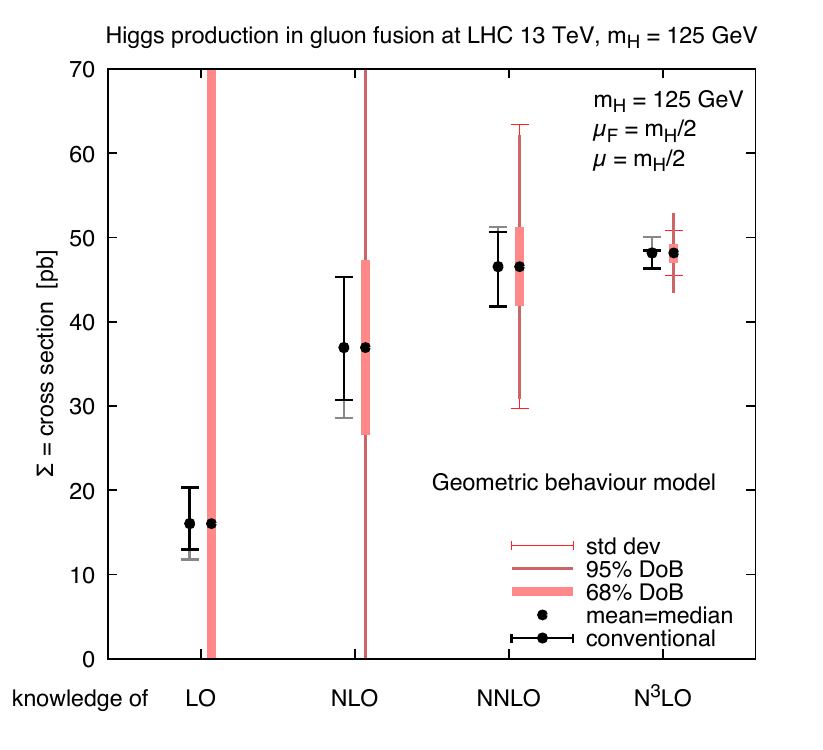}
  \includegraphics[width=0.495\textwidth,page=2]{./images/ggH_uncertainty_summary_paper.pdf}
  \caption{Summary of the distributions of Fig.~\ref{fig:GBdistr} at the various orders given
    in terms of mean, standard deviation and degree of belief intervals.
    The canonical scale variation uncertainty bands are also shown.
    The left plot is for the standard scale $\mu=\mh/2$, while the right plot is for the ``natural'' scale $\mu=\mh$.}
  \label{fig:GBsummary}
\end{figure}

To further appreciate the results of the approach, we compute from the distributions a number of
quantifiers, namely mean (which equals mode and median given the symmetry of the distributions),
standard deviation, and degrees of belief (DoB) intervals.
We summarize them in Fig.~\ref{fig:GBsummary}.
In this summary, we consider both the scale choice used previously, namely $\mu=\mh/2$,
and also another one, namely $\mu=\mh$, to emphasise the scale dependence of this approach at this level.
A number of comments are in order.
\begin{itemize}
\item The uncertainty is clearly reduced visibly and substantially when adding information from perturbative orders.
  All the uncertainty quantifiers (standard deviation, 68\% and 95\% DoB intervals) shrink with increasing the knowledge.
\item Because of the high tails, the standard deviation is quite large: infinite in the first two cases (knowledge of LO and NLO),
  and larger than the 68\% DoB interval in the other two cases.
  For the very same reason, the 95\% DoB interval is always much larger than the 68\% DoB interval.
\item In the shrinking of the uncertainty when adding information the results are always well compatible with the previous orders.
  All 95\% DoB intervals are contained in those of the previous orders. The same is true for the standard deviation.
  The 68\% DoB intervals, instead, are contained in the same interval of the previous order only in some cases,
  but this is perfectly acceptable, as there is large probability (32\%) that the true result is outside that interval.
\end{itemize}
We conclude once again that the knowledge of the NLO in this method is not sufficient to achieve a decent precision,
while when at least the NNLO is known the method works very well.
Note that the reliability is manifest, as the uncertainties always cover nicely the next orders.
This is achieved at a price, namely having large uncertainties with a poor state of knowledge.
But this is perfectly meaningful, because with too few information on the perturbative expansion
it is impossible to predict with precision the value of the observable.\footnote
{This is the reason for which using additional information on the expansion would be very precious.
The information on the scale dependence, that we will exploit in the next section, can also be combined
with this model, as we shall consider in Sect.~\ref{sec:mix}.}
We note that, despite the large uncertainties at low orders, once the N$^3$LO is known
the prediction is very precise, at least for some ``good'' scale choices (we will come back to this point in Sect.~\ref{sec:scaleindep}).

In the same plot also the conventional scale variation uncertainty is shown for comparison,
both its asymmetric version (in black) and its symmetrized version (in grey).
We stress once again that these conventional uncertainty bands have no probabilistic interpretation,
but they just represent the ``error'' usually assigned to perturbative results.
Since the probability distributions are symmetric, their mean coincides with the ``central value''
of the conventional scale variation approach.
There is little to say about these ``error bars'', given the absence of a probabilistic meaning.
We observe that, accidentally, the width of these ``error bars'' is similar in size to the 68\% DoB intervals,
with the exception of that of the LO.

\subsection{Dependence on the prior}
\label{sec:GBpriorscan}

\begin{figure}[t]
  \centering
  \includegraphics[width=0.59\textwidth,page=1]{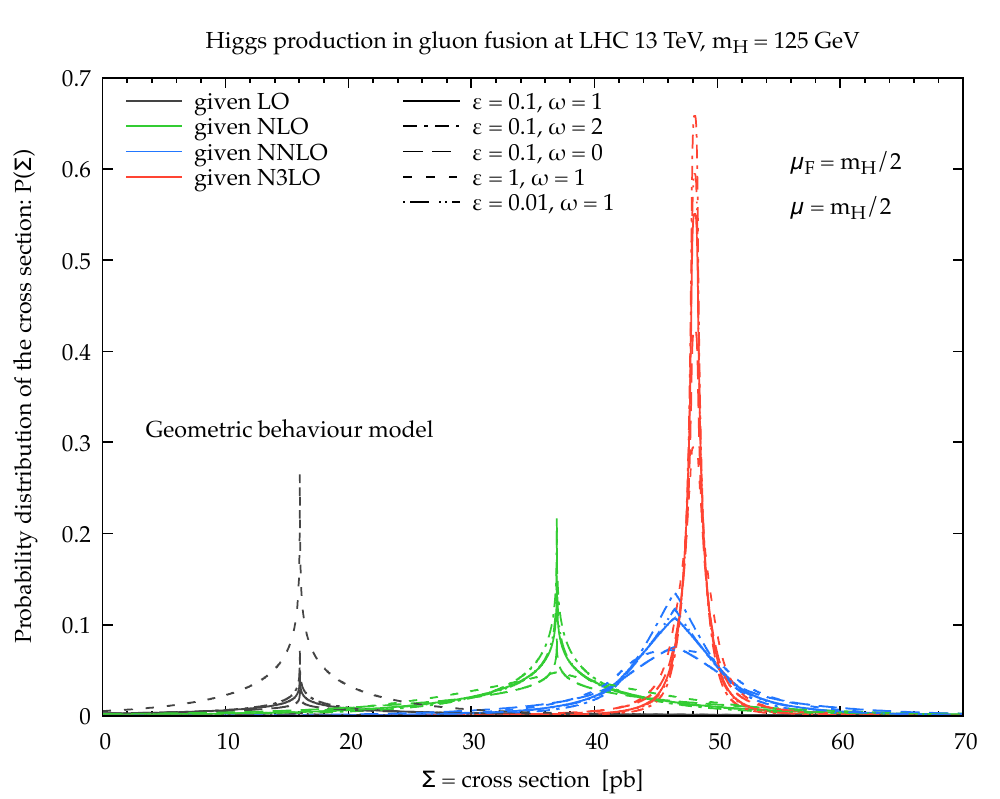}
  \includegraphics[width=0.4\textwidth,page=1]{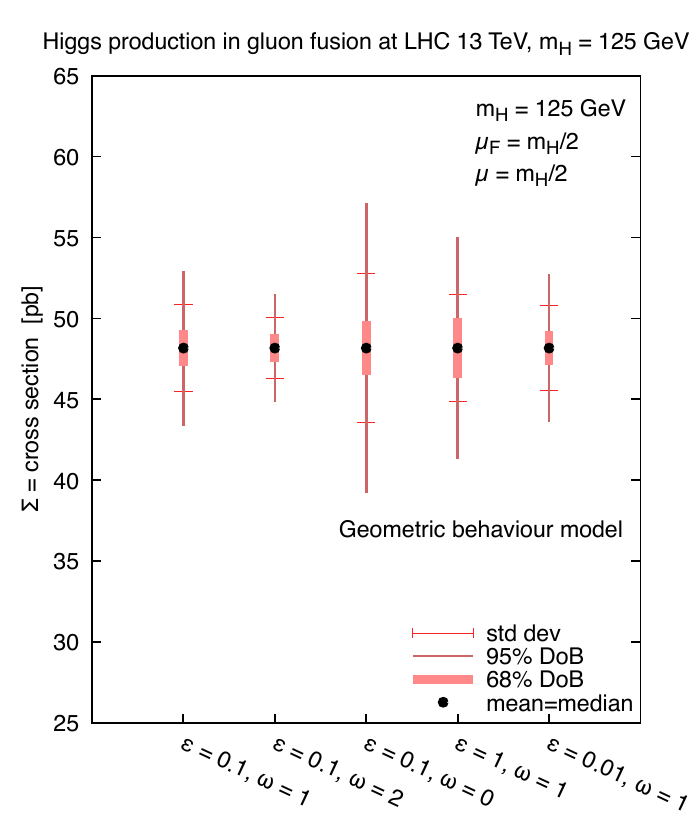}
  \caption{Dependence of the probability distribution of the observable at various orders on the parameters of the prior (left plot),
    and statistical estimators for the distribution at N$^3$LO for the same choices of parameters.}
  \label{fig:GBpriorscan}
\end{figure}

Before concluding the section we present a study on the impact of the choice of the prior
in the final probability distribution for the observable.
To do this, we consider variations of the prior parameters, Sect.~\ref{sec:GBprior}, and see how the final distributions change.
For definiteness, starting from the default choice of parameters $\epsilon=0.1$ and $\omega=1$,
we vary one of the two keeping the other at its default value.
For the prior for $c$, we consider $\epsilon=0.01$ and $\epsilon=1$,
while for $a$ we take $\omega=0$ and $\omega=2$.
The results of these variations are shown in Fig.~\ref{fig:GBpriorscan},
both at the level of the distributions for the cross section, and in terms of
the statistical estimators (considering in this case only the most accurate results with the knowlege of the N$^3$LO).

At the distribution level, we notice that the shape of the distribution is not particularly affected
by the variation of parameters, as they all look qualitatively the same.
However, they vary quantitatively, as for instance the ``width'' and the ``height'' of each distribution
changes with the prior parameters.
It has to be noted that, as expected, these differences are more marked at low orders
(less information, thus more dependence on the prior),
and become less and less relevant increasing the order (more information).

To better appreciate the impact of the different priors in the right plot of Fig.~\ref{fig:GBpriorscan}
the statistical estimators (standard deviation and DoB intervals) are shown for the distribution
at N$^3$LO obtained with each choice of the prior parameters.
We see that the difference are somewhat marked.
In particular, the largest uncertainties are obtained with $\omega=0$,
which is obvious as it allows larger values of the expansion parameter $a$
with larger probability (see discussion in Sect.~\ref{sec:GBposterior}).
Similarly, larger $\epsilon$ also leads to a larger uncertainty.
This is due to the fact that, if smaller values of $c$ are favoured,
the constraints from the theta function in Eq.~\eqref{eq:GBposterior}
forces $a$ to be larger (see also Fig.~\ref{fig:GBposterior}).
In agreement with these observations, the result with $\omega=2$ has smaller uncertainty
due to the stronger suppression of the $a\to1$ region,
and also the result with $\epsilon=0.01$ is more precise, even though here the difference
is very mild due to the tiny difference in the parameter value.
Note that in all cases the largest difference in the 95\% DoB interval,
which is dominated by the tails of the distributions that are mostly affected by the choice of prior.
The 68\% DoB interval, dominated by the peak region, is more stable.

We conclude that there certainly is a non-negligible bias due to the choice of the prior.
Nevertheless, we believe that our suggested choice of parameters represents a good compromise
between accuracy and precision.
One can obtain more conservative results with larger $\epsilon$ and/or smaller $\omega$,
but if the suggested values already provide accurate results
(as we shall verify with other examples in Sect.~\ref{sec:applications})
varying them does not seem convenient.
Conversely, one could obtain more precise results with a more aggressive prior for $a$, namely with larger $\omega$,
but this may lead to a loss of accuracy, which is to be avoided.

\section{Model 2: a new approach using scale variation information}
\label{sec:ModelScaleVar}

We now move to our second model, that uses information on the scale dependence of the perturbative expansion.
This model is specific for a QFT application.

\subsection{The hypothesis of the model}
\label{sec:SChyp}

We now consider another model that rather than looking at the behaviour of the perturbative expansion
uses only the information coming from the scale variation to infer the size of the missing higher orders,
similarly to what is done in the canonical scale variation approach.
The foundation of this method relies on the fact already stressed several times
that the scale dependence of a perturbative expansion truncated at order $n$
is formally of order $n+1$, namely of the same order as the missing higher orders, Eq.~\eqref{eq:scale_dep_Sp}.
In Sect.~\ref{sec:notations_rk} we have introduced the estimators $r_k(\mu)$, Eq.~\eqref{eq:rkdef}
to quantify the scale dependence of the perturbative expansion at order $k$,
which are thus objects of order $k+1$, Eq.~\eqref{eq:rkOrd}.
We shall thus expect
\begin{equation}\label{eq:SCidea}
\abs{\delta_k(\mu)} \sim r_{k-1}(\mu),
\end{equation}
namely, the term of the perturbative expansion at order $k$ should be
similar in size to the scale variation estimator of the previous order $k-1$.
From this vague statement we can now propose the hypothesis of our model, which is
\begin{equation}\label{eq:SChyp}
\abs{\delta_k(\mu)} \leq \lambda r_{k-1}(\mu) \qquad k<\kas,
\end{equation}
where $\lambda$ is a hidden parameter of the model, again dependent on (correlated with) $\mu$.
This condition does not tell us anything about the behaviour of the expansion,
but only relies on the goodness of the scale variation numbers $r_k$ as estimators of the next order,
up to a factor $\lambda$.

Note that if the scale variation estimator $r_k$ is accidentally small,
$\lambda$ is forced to be large to accomodate the condition Eq.~\eqref{eq:SChyp}.
This is definitely undesirable, as a large value of $\lambda$ due to an accident
of the scale dependence would enlarge the uncertainty making the model less predictive.
This would be the case if $r_k$ was constructed as the derivative of the observable:
close to stationary points, the derivative is accidentally small,
see Fig.~\ref{fig:ggHscaleslope} in comparison with Fig.~\ref{fig:ggHscale}.
Our definition of $r_k$, Eq.~\eqref{eq:rkdef}, is indeed designed to avoid
such problematic behaviours, as one can see from Fig.~\ref{fig:ggHrk}.
Since it is not unusual that the leading order is scale independent,
one has to redefine $r_0$, because in this cases it would be identically zero.
As anticipated in Sect.~\ref{sec:notations_rk}, in these cases we simply set
$r_0=1/2$, which represents a rather conservative choice
(larger values can also be considered to be even more conservative).
Note that in these cases the first non-trivial information comes from the NNLO,
so the probability distributions obtained from the knowledge of the NLO only have to be
considered biased by the arbitrary choice of $r_0$.

It is instructive to understand the connection between our hypothesis Eq.~\eqref{eq:SChyp}
and the canonical scale variation approach.
The relation between the two is very simple in the case in which the scale dependence
is linear in $\log\mu$.
In such a case, $r_k(\mu)$ coincides with the derivative of the observable with respect to $\log\mu$.
The canonical scale variation Eq.~\eqref{eq:CSU} in this limit would predict the ``error'' to be
$\log 2$ times the same derivative.\footnote
{This comparison is not exact because in our definition $r_k$ is the slope of the observable normalized to the observable
at order $k$, while the $\delta_k$ coefficients are normalized to the LO.}
Therefore, if the canonical scale variation ``error'' is thought as an absolute limit on the next order,
it would coincide with Eq.~\eqref{eq:SChyp} with $\lambda=\log 2$ \emph{fixed}.
This new method can thus be viewed as an improved version of canonical scale variation
in which the variation factor is not fixed to be 2, rather it is inferred
from the perturbative expansion itself to be such that the uncertainty estimate is reliable.

The condition Eq.~\eqref{eq:SChyp} must be translated into a probability distribution
for the coefficients $\delta_k$ given $\lambda$ and $r_{k-1}$.
We assume that the condition is strictly satisfied, and within the allowed range all values are equally likely.
This leads to the likelihood
\begin{equation}\label{eq:SClik}
P(\delta_k| r_{k-1},\lambda,\mu) = \frac{1}{2\lambda r_{k-1}(\mu)}\theta\(\lambda r_{k-1}(\mu) - \abs{\delta_k(\mu)}\),\qquad k>0, \quad k<\kas,
\end{equation}
where we have stressed that this conditional probability makes sense only for $k>0$,
because at LO there is no previous order to be used to compute scale dependence
(in other words, $r_{-1}$ does not exist).
This is once again in line with the fact that the LO alone does not bring any information on the behaviour
of the expansion.
Since the likelihood depends on a single hidden parameter,
the resulting model is not very ``flexible'', and will for example lead to non-smooth distributions for the observable.
To add some flexibility, one may allow violations of the bound Eq.~\eqref{eq:SChyp}.
This possibility will be explored later in Sect.~\ref{sec:SCviol}.

\begin{figure}[t]
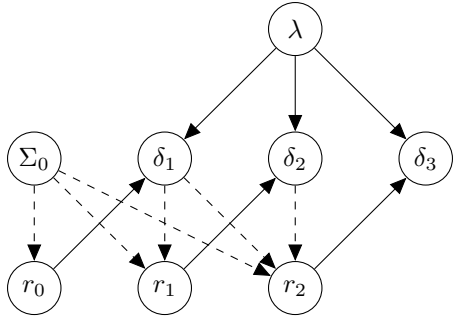

\centering
  \tikz[baseline=(current bounding box.north)]{ %
    \node[latent] (d0) {$\Sigma_0$} ; %
    \node[latent, right=of d0] (d1) {$\delta_1$} ; %
    \node[latent, right=of d1] (d2) {$\delta_2$} ; %
    \node[latent, right=of d2] (d3) {$\delta_3$} ; %
    \node[latent, above=of d2] (lam) {$\lambda$} ; %
    \node[latent, below=of d0] (r0) {$r_0$} ; %
    \node[latent, right=of r0] (r1) {$r_1$} ; %
    \node[latent, right=of r1] (r2) {$r_2$} ; %
    \edge[dashed] {d0} {r0} ; %
    \edge[dashed] {d0,d1} {r1} ; %
    \edge[dashed] {d0,d1,d2} {r2} ; %
    \edge {lam} {d1,d2,d3} ; %
    \edge {r0} {d1} ; %
    \edge {r1} {d2} ; %
    \edge {r2} {d3} ; %
  }
  \caption{Bayesian network for the scale variation model.}
  \label{fig:SCnet}
\end{figure}

The $r_{k-1}$ coefficient is assumed to be a given information in the likelihood Eq.~\eqref{eq:SClik}.
In fact, $r_{k-1}$ can be computed from the knowledge of all the previous $\delta_{k-1}(\mu),...,\delta_1 (\mu)$ and $\Sigma_0 (\mu)$.
This notation must thus be interpreted as a shorthand notation for the complete expression
\begin{equation}\label{eq:SClikdef}
P(\delta_k| r_{k-1},\lambda,\mu) \equiv P(\delta_k| \delta_{k-1},...,\delta_1,\Sigma_0,\lambda,\mu).
\end{equation}
Note that, differently from the geometric behaviour model of Sect.~\ref{sec:GB},
here there is an explicit dependence on $\Sigma_0(\mu)$, which is needed to compute the scale variation estimators $r_k$.
Eq.~\eqref{eq:SClikdef} implies that the $\delta_k$ coefficients are not independent
(as they were in the geometric behaviour model).
The dependence, however, in always only on the previous ones.
Thanks to this, the inference procedure is straightforward.
The Bayesian network of this model, using explicitly the $r_k$ parameters,
is depicted in Fig.~\ref{fig:SCnet}.

\subsection{The choice of prior}
\label{sec:SCprior}

This model depends on a single parameter $\lambda$.
In principle, it can take any positive value.
However, Eq.~\eqref{eq:SCidea} suggests that it should be of $\Ord(1)$.
For this reason, we find it convenient to choose its prior distribution such that large values are strongly suppressed.
We thus suggest an exponential behaviour
\begin{equation}\label{eq:SCprior}
P_0(\lambda|\mu) = \frac1{\Gamma(1+\gamma)}\,\lambda^\gamma\, e^{-\lambda}\, \theta(\lambda),\qquad \gamma\geq0,
\end{equation}
where $\gamma$ is a parameter that changes the shape of the distribution in the region of small $\lambda$.
The mode of the distribution is in
\begin{equation}
\lambda_{\rm mode} = \gamma.
\end{equation}
Therefore, it makes sense to choose either $\gamma=1$,
so that our expectation $\lambda=\Ord(1)$ is contained in the prior,
or $\gamma=0$, that represents a more agnostic (less informative) choice.
The dependence on $\gamma$ will be explored in Sect.~\ref{sec:SCposterior} and Sect.~\ref{sec:SCpriorscan}.
Similarly to the previous model, the prior Eq.~\eqref{eq:SCprior} is chosen to be independent of $\mu$.

\subsection{Inference on the unknown higher orders}
\label{sec:SCinf}

The probability distribution of the missing higher orders given the known ones, Eq.~\eqref{eq:Pdeltadecomp}, is
\begin{equation}\label{eq:SCPdnew}
P(\delta_{n+j},...,\delta_{n+1}|\delta_n,...,\delta_1,\Sigma_0,\mu)
=\frac{P(\delta_{n+j},...,\delta_{n+1},\delta_n,...,\delta_1|\Sigma_0,\mu)}{P(\delta_n,...,\delta_1|\Sigma_0,\mu)},
\qquad j\geq1, \quad n+j<\kas,
\end{equation}
where on the right-hand side we have included $\Sigma_0(\mu)$ as part of the information
as it is only needed to compute the scale dependence.
Both numerator and denominator can be directly computed from
\begin{align}\label{eq:SCPd1dm}
P(\delta_m,...,\delta_1|\Sigma_0,\mu)
&=\int d\lambda\, P(\delta_m,...,\delta_1,\lambda|\Sigma_0,\mu) \nonumber\\
&=\int d\lambda\, P(\delta_m|\delta_{m-1},...,\delta_1,\lambda, \Sigma_0,\mu) P(\delta_{m-1},...,\delta_1,\lambda|\Sigma_0,\mu) \nonumber\\
&=\int d\lambda\, P(\delta_m|\delta_{m-1},...,\delta_1,\lambda, \Sigma_0,\mu) \cdots P(\delta_1|\lambda, \Sigma_0,\mu) P_0(\lambda|\Sigma_0,\mu) \nonumber\\
&\equiv\int d\lambda\, P(\delta_m|r_{m-1},\lambda,\mu)\cdots P(\delta_1|r_0,\lambda,\mu) P_0(\lambda|\mu)\\
&= \frac1{2^m\, r_0(\mu)\cdots r_{m-1}(\mu)} \int \frac{d\lambda}{\lambda^m}\, P_0(\lambda|\mu)\, \theta\(\lambda-\max\[\frac{\abs{\delta_1(\mu)}}{r_0(\mu)},..., \frac{\abs{\delta_m(\mu)}}{r_{m-1}(\mu)}\]\),\nonumber
\end{align}
where in the first step we have introduced the hidden parameter,
then in the next two steps we have recursively written the probability of each $\delta_k$ in terms of the previous ones,
in the fourth step we introduced the shorthand notation Eq.~\eqref{eq:SClikdef} and removed $\Sigma_0$
from the prior which does not depend on it, and finally we have written explicitly the likelihoods Eq.~\eqref{eq:SClik}.
Note that the arguments of the max function depend only on the perturbative coefficients and not on the parameter.
Therefore, the max function itself is just a number, so we can conveniently define it
\begin{equation}\label{eq:lambdamdef}
\lambda_m(\mu) \equiv \max\[\frac{\abs{\delta_1(\mu)}}{r_0(\mu)},..., \frac{\abs{\delta_m(\mu)}}{r_{m-1}(\mu)}\]
\end{equation}
to simplify the expressions.
With our choice of prior, Eq.~\eqref{eq:SCprior}, it is easy to compute the integral
\begin{align}\label{eq:SCPd1dm2}
P(\delta_m,...,\delta_1|\Sigma_0,\mu)
&= \frac1{2^m\, r_0(\mu)\cdots r_{m-1}(\mu)\Gamma(1+\gamma)} \int_{\lambda_m(\mu)}^\infty d\lambda\,\lambda^{\gamma-m}e^{-\lambda}\nonumber\\
&= \frac{\Gamma\(1+\gamma-m,\lambda_m(\mu)\)}{2^m\, r_0(\mu)\cdots r_{m-1}(\mu)\Gamma(1+\gamma)},
\end{align}
where $\Gamma(a,b)$ is the incomplete Gamma function.
This result is very simple, thanks to the simplicity of the model (just one parameter)
and of the choice of likelihood and prior.
The distribution Eq.~\eqref{eq:SCPdnew} has an equally simple form
\begin{equation}\label{eq:SCPdnew2}
P(\delta_{n+j},...,\delta_{n+1}|\delta_n,...,\delta_1,\Sigma_0,\mu)
= \frac{\Gamma\(1+\gamma-n-j,\lambda_{n+j}(\mu)\)}{2^j\, r_n(\mu)\cdots r_{n+j-1}(\mu)\Gamma\(1+\gamma-n,\lambda_n(\mu)\)}.
\end{equation}
The derivation of the distribution for the observable $\Sigma$ is then straightforward
according to the procedure of Sect.~\ref{sec:generalfeatures}.
There is, however, an important difference with respect to the geometric behaviour model.
To make inference on the first missing higher order $n+1$ the knowledge
of the scale dependence $r_n$ of the order $n$ is required.
This is fine because the order $n$ is known.
However, to make inference on the next higher order, it is necessary to have
the subsequent coefficient $r_{n+1}$. This is not known.
In principle, this is not a real problem, as $r_{n+1}$ can be computed
from $\delta_{n+1}$.\footnote{
Note that in practice computing $r_{n+1}$ from $\delta_{n+1}$ is time consuming because
$\delta_{n+1}$ is not fixed but rather it is integrated over in the
probability of the observable Eq.~\eqref{eq:PSigmaj}.}
However, to compute the scale dependence at order $m$, the $m$-loop $\beta$-function is required.
Therefore, the scale dependence can only be computed up to the order at which we know
the $\beta$-function of the theory.
In the case of QCD, the $\beta$-function is presently known up to 5 loops~\cite{Baikov:2016tgj},
which implies that the largest coefficient which can be inferred in this model is $\delta_7$
for observables that are scale independent at LO and $\delta_6$ otherwise
(see Sect.~\ref{sec:appScale} for further detail).

\subsection{The posterior of the hidden parameter}
\label{sec:SCposterior}

We now turn our attention to the posterior distribution of the hidden parameter,
that as we have seen in the previous model brings interesting information.
The posterior for $\lambda$ is given by
\begin{align}
P(\lambda|\delta_n, ..., \delta_1, \Sigma_0,\mu)
&=\frac{P(\delta_n,...,\delta_1,\lambda|\Sigma_0,\mu)}{P(\delta_n,...,\delta_1|\Sigma_0,\mu)} \nonumber\\
&=\frac{P(\delta_n,...,\delta_1|\lambda)P_0(\lambda)}{P(\delta_n,...,\delta_1)} \nonumber\\
&=\frac{P(\delta_n|r_{n-1},\lambda,\mu)\cdots P(\delta_1|r_0,\lambda,\mu) P_0(\lambda|\mu)}{P(\delta_n,...,\delta_1)} \nonumber\\
&= \frac1{\Gamma\(1+\gamma-n,\lambda_n(\mu)\)}\,\lambda^{\gamma-n}\,e^{-\lambda}\,\theta\(\lambda-\lambda_n(\mu)\),
\end{align}
where in the last line we have already used the specific expressions for our choice of likelihood and prior,
with the definition of $\lambda_n$ Eq.~\eqref{eq:lambdamdef}.
This distribution is very simple: it has the same functional form of the prior,
but the power of $\lambda$ is reduced by a unity for each non-trivial known order,
and the region of small $\lambda$ is cut out by the theta function.

\begin{figure}[t]
  \centering
  \includegraphics[width=0.495\textwidth,page=1]{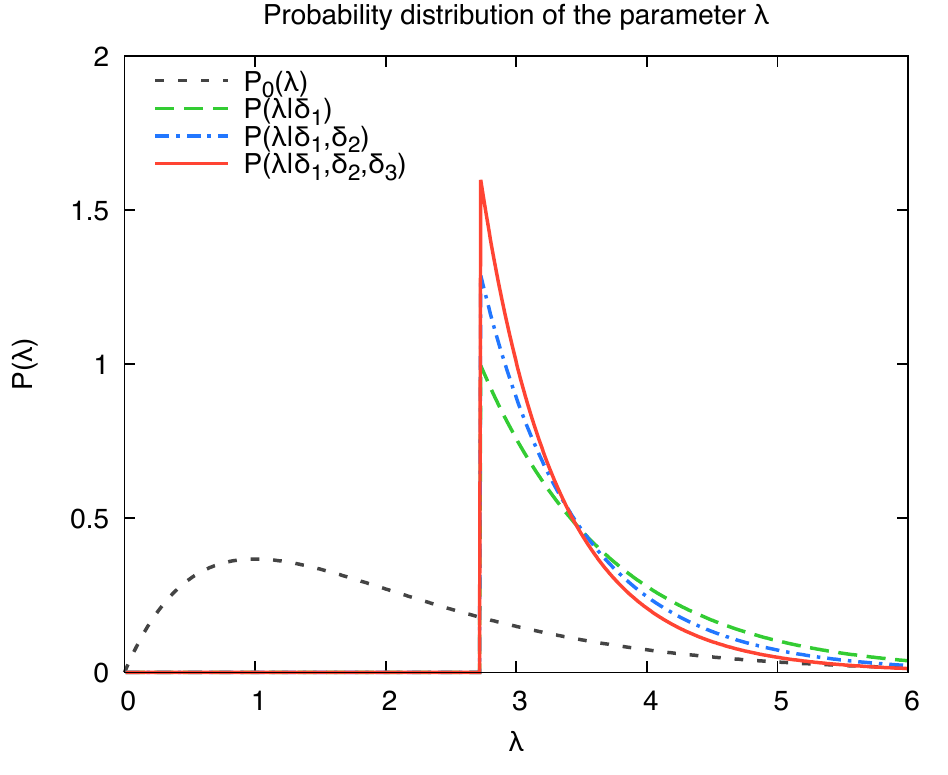}
  \includegraphics[width=0.495\textwidth,page=2]{./images/plot_Posterior_lambda_ggH_paper.pdf}
  \caption{The plots show the probability distribution of the parameter $\lambda$,
    for the prior with $\gamma=1$ (left plot) and $\gamma=0$ (right plot).
    The observable under consideration is the inclusive Higgs cross section, Sect.~\ref{sec:ggH}, for fixed scale $\mu=\mh/2$.}
  \label{fig:SCposterior}
\end{figure}
In Fig.~\ref{fig:SCposterior} we show the posterior distribution for our example of Higgs production.
In the left plot we use $\gamma=1$, and in the right plot $\gamma=0$.
We see that the difference is obviously marked for the prior (black dashed curve),
but becomes less and less relevant when adding information.
Most importantly, the lower limit on $\lambda$ imposed by the theta function,
which plays an important role, is independent of the parameter $\gamma$.
The figure also shows a striking effect of the inference:
the lower limit on $\lambda$, imposed by $\theta(\lambda-\lambda_n(\mu))$,
is fully determined by the first non-trivial order, $\delta_1$.
Indeed, it's clear that $\lambda_n$ Eq.~\eqref{eq:lambdamdef} coincides with
$\abs{\delta_1}/r_0=2.72$ for all values of $n=1,2,3$ (for $\mu=\mh/2$).
Indeed, the next orders give $\abs{\delta_2}/r_1=1.53$ and $\abs{\delta_3}/r_2=0.69$,
both smaller than the first (and decreasing).
This means that, in this case,
the actual next order is smaller than what estimated by the scale variation parameter $r_k$
times the smallest allowed value of $\lambda$.
This is the case despite our definition of $r_k$, which is normalized to $\Sigma_k$
rather than $\Sigma_0$ exactly with the purpose of giving a smaller number.
This fact can imply two things:
\begin{itemize}
\item either this behaviour is just an accident of the observable under consideration, or
\item the model itself is based on a wrong (or at least non-optimal) assumption.
\end{itemize}
We will see later that this behaviour is shared by other observables,
which seems to suggest that it is indeed the model assumption to be problematic.
However, the actual pattern of $\abs{\delta_k}/r_{k-1}$ also depends on the scale $\mu$
at which they are computed. For instance, for Higgs production,
we have verified that for $\mu>5\mh$ they all become of the same order, in agreement with the model assumption.
It is thus difficult to draw a sharp conclusion on the goodness of the model based just on these observations.

It is interesting to note that, since for some scales the $\abs{\delta_k}/r_{k-1}$ values are decreasing,
it could seem convenient to modify the assumption by adding a parameter $\tilde a$ behaving like a power,
namely
\begin{equation}\label{eq:SChyp?}
\abs{\delta_k(\mu)} \leq \lambda \tilde a^k r_{k-1}(\mu) \qquad k<\kas.
\end{equation}
This model would better describe the behaviour of the known orders, but seems unjustified:
where should such a power come from?
An alternative interpretation could be that the original assumption Eq.~\eqref{eq:SChyp} is meaningful,
but it takes a few orders before the bound is homogeneously satisfied,
with the first orders being more ``unstable''.
In this interpretation, it would make sense to allow a violation of the bound.
This option will be explored in Sect.~\ref{sec:SCviol}.

We stress that using this model as it is can be regarded as a conservative method:
indeed the allowed values of $\lambda$ are larger than actually needed,
thus predicting a larger uncertainty.
So for the moment we keep using it, but having in mind this conservative interpretation.

\subsection{Representative results}
\label{sec:SCresults}

\begin{figure}[t]
  \centering
  \includegraphics[width=0.9\textwidth,page=4]{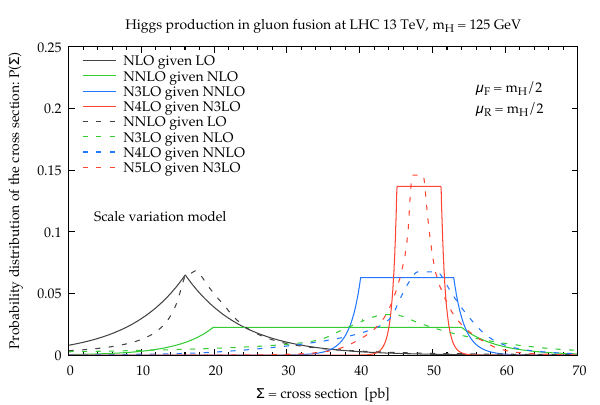}
  \caption{Same as Fig.~\ref{fig:GBdistr}, but for the scale variation model.}
  \label{fig:SCdistr}
\end{figure}

We now present some representative results of this method, using Higgs production
as an example, as we did in Sect.~\ref{sec:GBresults}.
In Fig.~\ref{fig:SCdistr} we plot the distributions for the observable given by Eq.~\eqref{eq:PSigmaj},
using a single higher order (solid lines) or two higher orders (dashed lines) to approximate the true cross section,
with different status of knowledge: LO (black), NLO (green), NNLO (blue), N$^3$LO (red).

The first striking difference with respect to the geometric behaviour model is the fact that the distribution
obtained using two unknown higher orders to approximate the cross section is very different from that
obtained using only the first unknown higher order.
According to the discussion after Eq.~\eqref{eq:PSigmaj}, we should then keep adding unknown orders in our approximation
until the distribution stops changing visibly.
Unfortunately, this will likely never happen, and the distribution will likely get broader and broader with higher and larger tails.
The reason for this comes from the way the model works.
When both $\delta_{n+1}$ and $\delta_{n+2}$ are used in the approximation for $\Sigma$,
the model needs $r_{n+1}$ to obtain the distribution for $\delta_{n+2}$. But $r_{n+1}$
depends in turn on $\delta_{n+1}$, which is not fixed, but varies among all possible values,
in principle from minus infinity to plus infinity.
The trouble is that the models relies on the assumption that the scale variation numbers $r_k$
are good estimators of the higher orders, which in turn relies on the fact that they behave in a perturbative way too.
For many of the possible values of $\delta_{n+1}$, however, the corresponding $r_{n+1}$
will violate this assumption and will be much larger than expected,
thus predicting large values of $\delta_{n+2}$.
These large values unavoidably contaminate the distribution for $\Sigma$,
making it broader with higher tails. This pattern can only get worse using more orders in the approximation of $\Sigma$.

One possible solution to this problem would be to impose some constraint also on $r_k$,
requiring that they behave in a perturbative way, and use this constraint in the inference
for the unknown $\delta_k$.
This approach is certainly interesting and potentially very powerful, but it is much more complicated to implement.
We will discuss it in Sect.~\ref{sec:SCv2}.
For the time being we stick to another possible solution, namely restricting our attention to the uncertainty
coming from the first unknown higher order only.
This approach is acceptable, provided the result is interpreted for what it is:
not a probability for the true observable $\Sigma$, but just the probability
for the observable at the next order, $\Sigma_{n+1}$.

With this limitation in mind, we now comment the shapes of the distributions in Fig.~\ref{fig:SCdistr},
focussing on those using only the first unknown higher order (solid lines).\footnote
{The only comment on the dashed curves that we make is that they become asymmetric,
  and interestingly the asymmetry tends to favour regions in the direction where the actual next orders indeed are.
  The price to pay is the higher tails.
  The model of Sect.~\ref{sec:SCv2} will preserve this nice property while
  removing the issue of the high tails.}
With the exception of the first one (black) based only on the knowledge of the LO,
they all feature a plateau, surrounded by exponentially decreasing tails.\footnote
{Because the tails die exponentially, the probability of negative cross section
is much lower than in the geometric behaviour model. Only at LO it is significant ($\sim6\%$),
while at NLO it is already below $0.3\%$, becoming totally negligible ($<10^{-5}$) at NNLO and beyond.}
The plateau is a direct consequence of the likelihood being a theta function of the absolute value of the order,
and of the model being dependent on a single hidden parameter $\lambda$.
Indeed the inference on $\lambda$ sets a lower limit, Fig.~\ref{fig:SCposterior},
which in turn implies that all values of the next order lower (in absolute value)
than the lower limit of $\lambda$ times the previous $r_k$ are equally probable.
We also see that the distributions shrink nicely by adding information.
Because also the tails die faster and faster increasing the number of known orders,
it's clear that these distributions tend to a uniform distribution.
Therefore, one can interpret this model as a sort of provider of an ``absolute error'',
a region where one is almost certain that the next order will lie,
but without any clue on where inside that region.

\begin{figure}[t]
  \centering
  \includegraphics[width=0.495\textwidth,page=3]{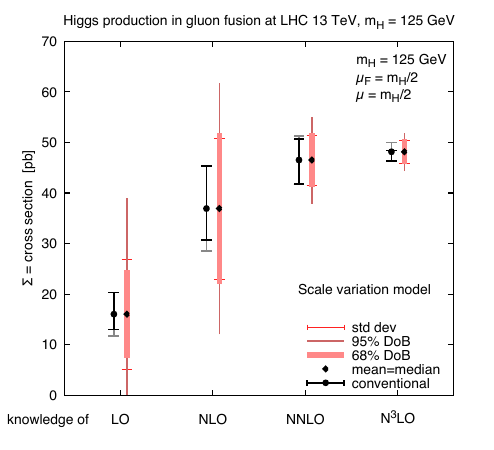}
  \includegraphics[width=0.495\textwidth,page=4]{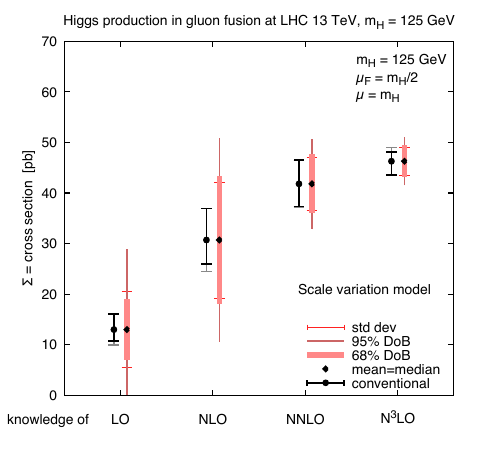}
  \caption{Same as Fig.~\ref{fig:GBsummary}, but for the scale variation model.}
  \label{fig:SCsummary}
\end{figure}

In Fig.~\ref{fig:SCsummary} we show the results using quantifiers of the distributions,
for two values of the scale, $\mu=\mh/2$ (left) and $\mu=\mh$ (right).
We see that the standard deviation is very similar to the 68\% DoB interval,
and the 95\% DoB interval is only slightly larger than the 68\% DoB interval due to the exponential suppression of the tails,
differently from what happened in the geometric behaviour model.
Also in this case the bands shrink nicely increasing the order, with a good compatibility with the previous bands
(remember however that each band now just represents the uncertainty of the next order, and not of the full result).
The only exception is the LO, which has small uncertainties not very compatible with the NLO.
Note that in this case the distribution at LO is not fully determined by the prior,
but also depends on $r_0$, the scale variation number of the LO.
Therefore the small uncertainty may also be due to the fact that $r_0$
is not a very good representative of the NLO.
Anyway, also in this case one must at least know two orders to let the inference work,
so the LO uncertainty is not very relevant.

We also compare these results with the canonical scale variation approach.
We observe that the ``error bars'' of the canonical method at NNLO and N$^3$LO are close in size
(when they are symmetrized) to the standard deviation (or the 68\% DoB interval) of our model.
This is purely accidental.

\subsection{Dependence on the prior}
\label{sec:SCpriorscan}

\begin{figure}[t]
  \centering
  \includegraphics[width=0.59\textwidth,page=2]{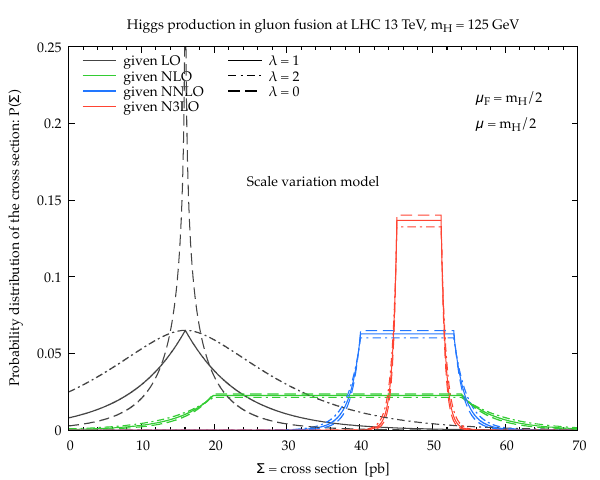}
  \includegraphics[width=0.4\textwidth,page=2]{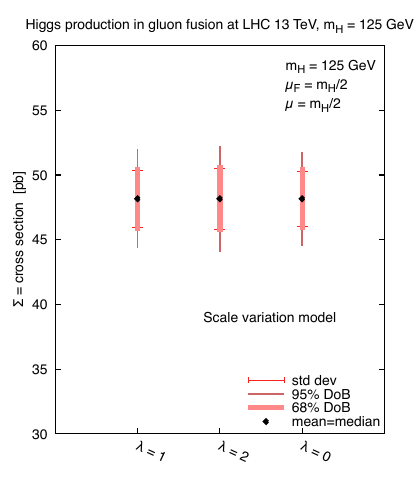}
  \caption{Dependence of the probability distribution of the observable at various orders on the parameter of the prior (left plot),
    and statistical estimators for the distribution at N$^3$LO for the same choices of parameter.}
  \label{fig:SCpriorscan}
\end{figure}

We finally consider variations of the prior parameter $\gamma$.
On top of our default choice ($\gamma=1$), we take a smaller value $\gamma=0$ (more agnostic choice)
and a larger value $\gamma=2$ (assuming a larger prior value for $\lambda$ of order 2, see discussion in Sect.~\ref{sec:SCprior}).
The results of these variations are shown in Fig.~\ref{fig:SCpriorscan},
for the distributions of the cross section (left plot) and for the statistical estimators at N$^3$LO
(right plot).

The most striking feature of the first plot is that the width of the plateau is independent of the prior.
This is a consequence of the fact that the plateau is determined by the theta function in the likelihood,
and is thus a feature that does not depend on the prior.
What changes when varying the prior is the functional behaviour of the tails of the distributions,
and due to the normalization constraint this also modifies the ``hight'' of the plateau.
The difference are however very mild. The only distribution that changes significantly is
the one given the knowledge of the LO only, that we recall is fully determined from the prior and it
thus does not contain any relevant information.

The stability of the result is also manifest when looking at the summary plot (on the right in the figure).
All bands are basically identical.
Note that one could obtain a larger difference by changing the functional form of the prior,
e.g.\ replacing the exponential form with a power law. However, even in this case only
the tails of the distributions will be affected, thus still leading to small (though perhaps more marked) differences.
We conclude that this method, despite its limitations, has the advantage
of being essentially independent of the choice of the prior.

\section{Dealing with scale dependence}
\label{sec:scaleindep}

So far we have presented two models that allow to construct a probability distribution for the missing higher orders
of a given perturbative expansion \emph{at a fixed scale}.
We have already seen in the examples that the same method on the same observable at different scales
produces different results.
Hopefully with sufficiently many orders the dependence on the scale will be mild, but nevertheless this dependence is undesirable.
In this section we propose an approach to produce a prediction for the observable and its uncertainty that is scale independent.

\subsection{The scale as a model parameter}
\label{sec:scalepar}

\begin{figure}[t]
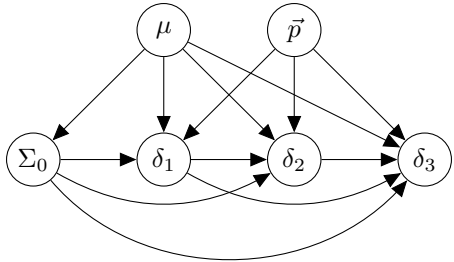

\centering
  \tikz[baseline=(current bounding box.north)]{ %
    \node[latent] (d0) {$\Sigma_0$} ; %
    \node[latent, right=of d0] (d1) {$\delta_1$} ; %
    \node[latent, right=of d1] (d2) {$\delta_2$} ; %
    \node[latent, right=of d2] (d3) {$\delta_3$} ; %
    \node[latent, above=of d2] (pars) {$\vec p$} ; %
    \node[latent, above=of d1] (mu) {$\mu$} ; %
    \edge {mu} {d0,d1,d2,d3} ; %
    \edge {pars} {d1,d2,d3} ; %
    \edge {d0} {d1} ; %
    \edge {d1} {d2} ; %
    \edge {d2} {d3} ; %
    \path (d0) edge [->, >={triangle 45}, bend right=30] (d2) ;
    \path (d0) edge [->, >={triangle 45}, bend right=50] (d3) ;
    \path (d1) edge [->, >={triangle 45}, bend right=30] (d3) ;
  }
  \caption{The most general Bayesian network of models of inference of missing higher orders,
    including explicitly the scale $\mu$ as a parameter over which one can marginalize.}
  \label{fig:netScaleIndep}
\end{figure}

The idea to remove the scale dependence from the probability distribution is very simple:
promoting the scale $\mu$ to be a parameter of the model.
The corresponding Bayesian network is depicted in Fig.~\ref{fig:netScaleIndep}.
In this way, the scale dependence is easily removed by simply \emph{marginalizing} over $\mu$:
\begin{align}\label{eq:PSigmanomu}
P(\Sigma|\delta_n,...,\delta_1,\Sigma_0)
  &= \frac{P(\Sigma,\delta_n,...,\delta_1,\Sigma_0)}{P(\delta_n,...,\delta_1,\Sigma_0)} \nonumber\\
  &= \frac{\int d\mu\, P(\Sigma,\delta_n,...,\delta_1,\Sigma_0,\mu)}{P(\delta_n,...,\delta_1,\Sigma_0)} \nonumber\\
  &= \frac{\int d\mu\, P(\Sigma|\delta_n,...,\delta_1,\Sigma_0,\mu) P(\delta_n,...,\delta_1,\Sigma_0,\mu)}{P(\delta_n,...,\delta_1,\Sigma_0)} \nonumber\\
  &= \int d\mu\, P(\Sigma|\delta_n,...,\delta_1,\Sigma_0,\mu)\, P(\mu|\delta_n,...,\delta_1,\Sigma_0).
\end{align}
The equation above gives the probability of the observable $\Sigma$ given the knowledge of the first orders up to N$^n$LO.
The latter are seen not as the values of the perturbative contribution at a given scale,
but in a more abstract sense.
This probability is expressed as the integral over $\mu$ of the same probability but at fixed $\mu$,
which was given in Eq.~\eqref{eq:PSigmaj} and was the object discussed so far in this paper,
and the \emph{posterior} distribution for $\mu$ given the first $n+1$ orders,
which is in turn given by
\begin{equation}\label{eq:Posteriormu}
P(\mu|\delta_n,...,\delta_1,\Sigma_0) = \frac{P(\delta_n,...,\delta_1,\Sigma_0,\mu)}{P(\delta_n,...,\delta_1,\Sigma_0)}
= \frac{P(\delta_n,...,\delta_1,\Sigma_0|\mu)P_0(\mu)}{\int d\mu\,P(\delta_n,...,\delta_1,\Sigma_0|\mu)P_0(\mu)},
\end{equation}
where $P_0(\mu)$ is the prior distribution for the scale $\mu$.
A number of comments are in order.

The first and most important comment is related to the interpretation of this procedure.
Indeed, the fact that $\mu$ is a parameter of the model implies that it has some sort of ``physical'' meaning,
since it is possible to make inference on it and construct a posterior distribution, Eq.~\eqref{eq:Posteriormu}.
That is, inference selects values of the scale that are ``better'' than others.
This seems to be in direct contradiction with the fact that the renormalization scale is unphysical,
and thus any value is equally good.
In other words, there is no physical motivation for preferring some values of the scale
and disfavouring others.\footnote
{This is not entirely true. If the scale $\mu$ is chosen to be too different from the hard scale of the process $Q$,
then large logarithms of the ratio $\mu/Q$ invalidate the perturbative expansion.
This observation however allows only to disfavour some extreme regions of the scale,
while leaving a large region of allowed values.}

This apparent contradiction is solved by noting that the probabilistic inference on $\mu$
selects values that are ``better'' \emph{according to the model}.
Indeed the posterior distribution for $\mu$ is proportional to $P(\delta_n,...,\delta_1,\Sigma_0|\mu)$,
that is the model-dependent probability of the sequence of the first $n$ orders at the scale $\mu$.
Inference selects scales for which this probability is higher, namely scales
for which the perturbative expansion fits well with the assumptions of the model.
Therefore, not only this procedure is perfectly acceptable,
but it also does what we would have desired, namely selecting the scales
with the best perturbative expansion (according to the assumptions of the model),
within a well defined probabilistic framework.

As a second comment, we observe that since the scale has become a parameter of the model, a prior distribution $P_0(\mu)$ has to be specified.
This prior distribution will contain our prejudices on what are the most appropriate scales,
and it thus represents in some sense a ``residual scale dependence'' of the result.
However, we will see that if the prior distribution is sufficiently broad and non-informative,
the results will depend very little on its precise form and size.

Another comment is related to the fact that the only new ingredient needed to compute
the scale-independent distribution is just the prior $P_0(\mu)$.
Indeed, the distribution $P(\Sigma|\delta_n,...,\delta_1,\Sigma_0,\mu)$ in Eq.~\eqref{eq:PSigmanomu}
was the subject of the previous sections, and the posterior distribution for $\mu$
depends on the prior and on $P(\delta_n,...,\delta_1,\Sigma_0|\mu)$,
which is the basic model-dependent object discussed in Sect.~\ref{sec:GB} and \ref{sec:ModelScaleVar},
which is also needed in the computation of $P(\Sigma|\delta_n,...,\delta_1,\Sigma_0,\mu)$,
see Sect.~\ref{sec:generalfeatures}.
Therefore, no additional computations are needed to ``remove'' the scale dependence:
scale-independent results can be obtained with just a simple (numerical) integration.

We can use Eq.~\eqref{eq:PSigmaderiv} to write explicitly the scale-independent distribution Eq.~\eqref{eq:PSigmanomu}
in terms of fundamental objects.
We find, approximating the observable with the order $n+j$,
\begin{align}\label{eq:PSigmanomu2}
P(\Sigma|\delta_n,...,\delta_1,\Sigma_0)
  &= \int d\mu\, P(\Sigma|\delta_n,...,\delta_1,\Sigma_0,\mu)\, P(\mu|\delta_n,...,\delta_1,\Sigma_0)  \\
  &\simeq \int d\mu \int d\delta_{n+j}\cdots d\delta_{n+1}\,
 \delta\(\Sigma-\Sigma_{n+j}(\mu)\)
    \frac{P(\delta_{n+j},...,\delta_{n+1},\delta_n,...,\delta_1,\Sigma_0|\mu)P_0(\mu)}{P(\delta_n,...,\delta_1,\Sigma_0)}, \nonumber
\end{align}
having used Eq.~\eqref{eq:Pdeltadecomp} and Eq.~\eqref{eq:Posteriormu}.
As we did in Sect.~\ref{sec:generalfeatures}, we can use the delta function to compute one
of the $\delta_k$ integrals. Alternatively, we can consider moments of the distribution,
and use the delta function to integrate over $\Sigma$. We have for the generic $p$ moment
\begin{align}\label{eq:SigmaMoments}
\av{\Sigma^p}_{\delta_n,...,\delta_1,\Sigma_0}
  &=\int d\Sigma\; \Sigma^p\, P(\Sigma|\delta_n,...,\delta_1,\Sigma_0) \nonumber\\
  &\simeq \int d\mu \int d\delta_{n+j}\cdots d\delta_{n+1}\,
    \Sigma_{n+j}^p(\mu)
    \frac{P(\delta_{n+j},...,\delta_{n+1},\delta_n,...,\delta_1,\Sigma_0|\mu)P_0(\mu)}{P(\delta_n,...,\delta_1,\Sigma_0)},
\end{align}
from which we can compute for instance the mean of the distribution and its variance.
It is interesting to notice that in the cases in which the probability
$P(\delta_m,...,\delta_1,\Sigma_0|\mu)$ is symmetric for $\delta_k\to-\delta_k$,
as it is the case of the geometric behaviour model,
the computation of the mean of the distribution is simplified as the contributions from unknown higher orders integrate to zero.
Therefore, in such a symmetric case, the mean reduces to
\begin{align}\label{eq:SigmaMean}
\av{\Sigma}_{\delta_n,...,\delta_1,\Sigma_0}
  &= \int d\mu \; \Sigma_n(\mu) \frac{P(\delta_n,...,\delta_1,\Sigma_0|\mu)P_0(\mu)}{P(\delta_n,...,\delta_1,\Sigma_0)},
\end{align}
which depends on the \emph{known} N$^n$LO result $\Sigma_n(\mu)$.
This is the same result that we would obtain when approximating the observable using only the known orders ($j=0$).
If we were working at fixed scale, this would just be the number that comes out of the computation.
But since this number is scale dependent, it generates a distribution,
Eq.~\eqref{eq:PSigmanomu2} with $j=0$, whose mean is given by Eq.~\eqref{eq:SigmaMean}.
This is a very interesting consequence of the procedure: even a fixed-order computation at order $n$
cannot be regarded as an ``exact'' prediction at that order, because it depends on the unphysical scale.
The distribution Eq.~\eqref{eq:PSigmanomu2} with $j=0$ represents the best we can say about the N$^n$LO result,
ignoring the missing higher orders.
Because the N$^n$LO is a distribution, one can also compute its standard deviation.
This standard deviation would represent a true ``scale uncertainty'' on the \emph{known} finite order,
but it would not contain any uncertainty from missing higher orders.
This is very different from the usual approach, where the ``canonical scale uncertainty''
is used as an estimate of the missing higher order uncertainty.

We conclude by discussing the form of the prior distribution for $\mu$.
In principle, any value of $\mu$ is allowed, and each value is equally valid as they all lead to the same exact result.
However, in practice some values have to be avoided.
For instance, in theories like QCD at small scales the coupling grows and invalidates
the perturbative hypothesis. Therefore, the scale (in QCD) cannot become too low.
Similarly, very large scales (in QCD) are not advisable,
as they lead to small values of the coupling, slowing down the convergence of the expansion.
At the same time, physical processes are characterized by one (or more) physical scale(s),
and the various orders $\delta_k$ will contain logarithms of the ratio of such scale(s)
with the unphysical scale $\mu$.
If $\mu$ is very different from the physical scale(s) the logarithms became large and invalidate the perturbative expansion.
Therefore, the most convenient and meaningful approach is to use the value of the physical scale(s) to determine
a ``central value'' $\mu_0$ of the scale $\mu$, and then vary it in a range
that satisfy the previous (or any other) constraints.
Considering that the scale dependence is logarithmic,
the easiest option is a flat distribution in the logarithm of the scale, namely
\begin{equation}\label{eq:muPrior}
  P_0(\mu) = \frac1{2\log F}\frac1{\mu} \theta\(\log F - \abs{\log\frac{\mu}{\mu_0}}\), \qquad F>1,
\end{equation}
where $F$ is factor that sets the size of the interval of allowed scales,
assumed to be symmetric about $\mu_0$ (which is not a necessary condition).
Alternatively, one could consider a distribution peaked in $\mu_0$,
for instance a gaussian, perhaps with hard limits to avoid entering the low- and large-scale regions.
However, in the spirit of letting the model select the scale that leads to the best convergence properties,
a flat distribution seems more natural.
The factor $F$ should be sufficiently large to contain a region where good convergence is achieved.
This may be either selected by eye looking at the behaviour of the expansion as a function of the scale,
or with a step-by-step approach where $F$ is enlarged if the posterior for $\mu$ turns out to be peaked
close or at the boundary of the allowed region.

We stress that, despite the simplicity of this approach to obtain scale-independent results,
this method is very innovative and provides for the first time a consistent way to deal with scale dependence
which is compatible with physical requirements.

\subsection{A pre-example: application to the Cacciari-Houdeau model}

Let us now start to investigate the implication of this method to obtain scale-independent results.
To begin with, we consider the original Cacciari-Houdeau model, described in Sect.~\ref{sec:CH}.
We are not really interested in obtaining results within the CH model,
since we have proposed a new model that can be seen as an improved version of it.
Rather, we want to use it to point out the importance of using the normalized variables $\delta_k$ for defining the model.

To see this, let us start by computing the posterior distribution for $\mu$
given the knowledge of the LO. In the CH model, indeed, the LO is treated as a source of information.
The posterior is given by (using the notation of Sect.~\ref{sec:CH})
\begin{align}
  P(\mu|c_0)
  &= \frac{P(c_0|\mu)P_0(\mu)}{P(c_0)} \nonumber\\
  &= \frac{P_0(\mu)}{P(c_0)}\int d\bar c\, P(c_0|\bar c,\mu)P_0(\bar c) \nonumber\\
  &\propto P_0(\mu)\int_{\abs{c_0(\mu)}}^\infty \frac{d\bar c}{\bar c^2}
= \frac{P_0(\mu)}{\abs{c_0(\mu)}},
\end{align}
where we have used Eq.~\eqref{eq:CHlik} and Eq.~\eqref{eq:CHprior}, and assumed that the hidden parameter $\bar c$
and the scale $\mu$ are uncorrelated a priori.
This computation shows that the posterior distribution for $\mu$ given the LO
is larger for values of $\mu$ for which the LO $c_0(\mu)$ is smaller.\footnote
{In this notation we are assuming that if the LO starts at $\Ord(\as^{k_0})$ the coefficient $c_0$
incorporates this factor $\as^{k_0}$, so that the expansion Eq.~\eqref{eq:SigmaPert} holds.}
This is not particularly clever, as very often NLO and higher corrections are positive,
so a larger value of the LO should be favoured instead, to reduce the impact of higher orders
and lead to a more well-behaved expansion.
This is for instance what happens in the Higgs production process that we have used as an example so far.

The underlying reason for the failure of this procedure is that from the LO only
it is impossible to decide which scale should be favoured.
However, in this model the situation does not improve when adding the NLO.
Indeed the posterior becomes
\begin{equation}
P(\mu|c_0,c_1) \propto \frac{P_0(\mu)}{\max\(\abs{c_0(\mu)}, \abs{c_1(\mu)}\)^2},
\end{equation}
which is likely still mostly determined by $c_0$, as $\abs{c_1}$ will typically be smaller than $\abs{c_0}$
being a perturbative correction, so that it will never win in the max function.
Therefore, the problem with using the information from the LO to make inference in the model
is not solved by adding orders.
In order to avoid this issue one should really treat the LO as just a prefactor,
and start the inference procedure from the NLO, as we do in this work.
This is another important motivation for using the normalized quantities $\delta_k$ in the definition of the model.

\subsection{Scale-independent geometric behaviour model}
\label{sec:GBscaleindep}

We now move to the geometric behaviour model discussed in Sect.~\ref{sec:GB}.
This model is more interesting and we will use the whole machinery introduced in Sect.~\ref{sec:scalepar}
to obtain scale-independent distributions and uncertainties.

To begin with, we want to stress that in this case, making inference on the normalized perturbative expansion,
gives the desired posterior distribution for the scale.
We start noticing that the knowlege of the LO $\Sigma_0$ does not change the distribution of $\mu$,
\begin{equation}
P(\mu|\Sigma_0)
= \frac{P(\Sigma_0|\mu)P_0(\mu)}{P(\Sigma_0)}
= P_0(\mu),
\end{equation}
which comes from the fact that $\Sigma_0$ does not appear in the likelihood of the model, Eq.~\eqref{eq:GBlik},
and therefore $P(\Sigma_0|\mu)=P(\Sigma_0)$.
Equivalently, if one considers instead the normalized LO $\delta_0$ for which a likelihood exists,
it cannot change the distribution for $\mu$ because $\delta_0=1$ is scale independent.
This is due to the fact, stressed already several times, that the LO does not carry any information
on the behaviour of the expansion, not even when scale dependence is taken into account.
The first non-trivial information comes from the NLO coefficient $\delta_1(\mu)$,
that changes the distribution for $\mu$. The general computation is somewhat more complicated than in the CH case,
so we stick to the simple case in which the prior for $a$ is flat ($\omega=0$), so that we get
\begin{equation}
P(\mu|\delta_1) \propto P_0(\mu) \[\frac1{1+\epsilon} + \log\frac1{\abs{\delta_1(\mu)}}\].
\end{equation}
This probability is clearly larger for scales such that $\abs{\delta_1(\mu)}$ is smaller
(at this order the dependence is only logarithmic, but at higher orders power behaving contributions appear).
This is the expected behaviour, namely inference selects scales for which perturbative corrections are smaller.
Note that for such scales the LO $\Sigma_0(\mu)$ may be larger to ``compensate''\footnote
{Clearly, it's the other way round: when the LO is bigger, the higher orders should be consequently smaller.}
for the next orders (this is what happens in the Higgs case).

\begin{figure}[t]
  \centering
  \includegraphics[width=0.7\textwidth,page=1]{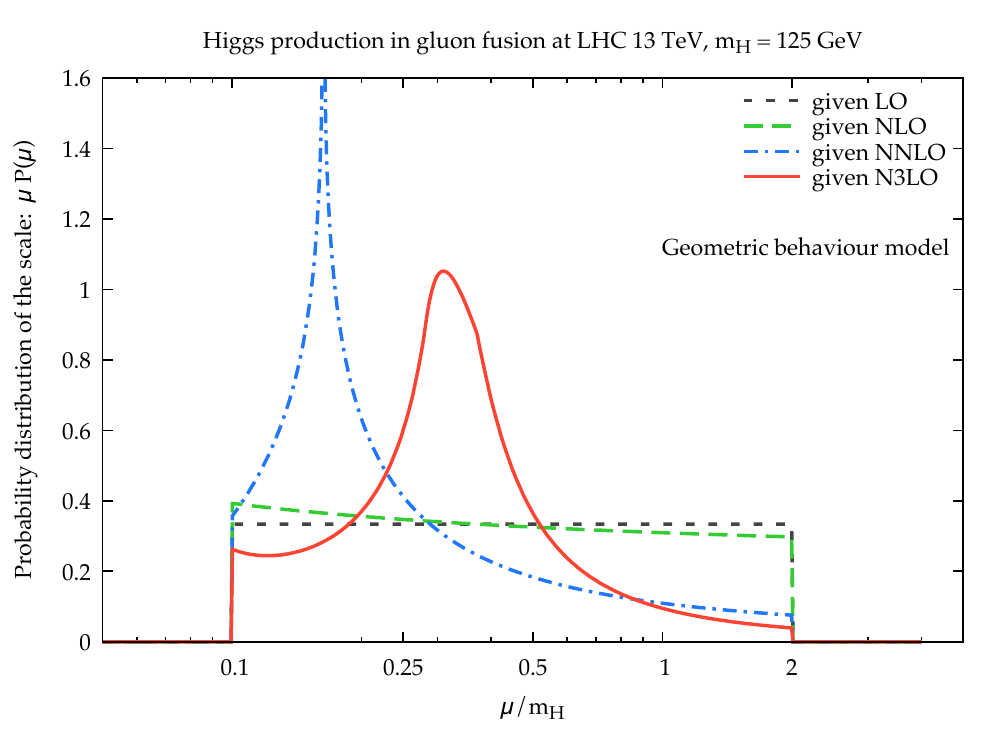}
  \caption{Posterior distribution for the scale $\mu$ within the geometric behaviour model with different states of knowledge,
    for the Higgs production process.}
  \label{fig:GBmuPost}
\end{figure}

Understanding the behaviour of the posterior for $\mu$ beyond NLO is difficult analytically.
Therefore we now switch to numerical results.
In Fig.~\ref{fig:GBmuPost} we show the posterior for $\mu$, Eq.~\eqref{eq:Posteriormu},
for our example of Higgs production, for four different states of knowledge:
LO (black), NLO (green), NNLO (blue) and N$^3$LO (red).
The first curve corresponds to our prior, that has been taken to be flat in $\log\mu$,
Eq.~\eqref{eq:muPrior}, ranging from $\mh/10$ to $2\mh$.
This choice of range is motivated by the fact that scales between $\mh/4$ and $\mh/2$
lead to better convergence properties of the expansion (see e.g.\ Ref.~\cite{Bonvini:2016frm}, and also Fig.~\ref{fig:ggHscale})
and are thus used to define the center of the distribution,
while the width is set by requiring that the lowest scale ($\mh/10=12.5$~GeV) is still characterised
by a sufficiently small strong coupling.
Note that the whole range of scales allowed by this prior spans a factor of 20,
which is way larger than the ususal range of scales considered in standard computations
(typically, once the hard scale of the process is identified, the central scale is chosen within a factor two or four).
This means that in our computation we consider also scales
that are usually considered too far from the hard scale of the process.

We observe that, as commented before through the analytic computation, the knowledge of the NLO
leads to a preference towards smaller scales, even though the preference is mild because the dependence
on the NLO coefficient $\delta_1$ is logarithmic.
The knowledge of the NNLO changes the situation more dramatically, giving a net preference
for scales around $\mu\sim20$~GeV, and suppressing strongly high scales.
The peak corresponds to the scale at which the NNLO correction is zero,
which incidentally corresponds to the region where the NNLO scale dependence has a plateau, Fig.~\ref{fig:ggHscale}.
At N$^3$LO, the distribution is still peaked, but with a smoother bump towards somewhat larger scales, approximately $\mh/3$.
The reason is that the N$^3$LO correction is zero at a higher scale, close to $\mh/2$,
and so there is a trade off between the preference of the NNLO correction and that of the N$^3$LO correction.
The high scale region is further suppressed.

\begin{figure}[t]
  \centering
  \includegraphics[width=0.9\textwidth,page=1]{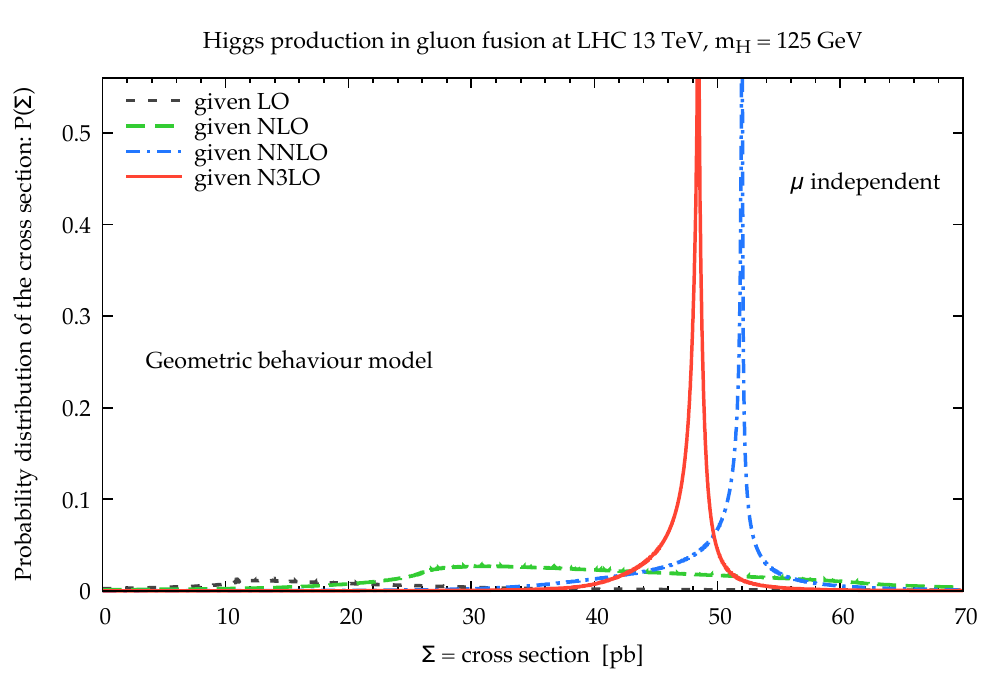}
  \caption{Similar to Fig.~\ref{fig:GBdistr}, after removing scale dependence.}
  \label{fig:GBdistrScaleindep}
\end{figure}

We now move to the distribution for the observable $\Sigma$.
We use the approximation based on the first unknown higher order, as we have seen that it captures well the full uncertainty.
The results are shown in Fig.~\ref{fig:GBdistrScaleindep}.
We immediately observe that these distributions are asymmetric,
which is a consequence of the fact that scale dependence is non-trivial.
At the first two orders, the distributions are very broad and not predictive.
However, we notice that the procedure to remove the scale dependence has the effect
of favouring larger values with respect to the fixed-scale results, in particular
covering (at NLO) the region favoured by the next orders with a high probability.
The distribution at NNLO features a high peak towards large values of the cross section, at $\Sigma\simeq51$~pb.
This peak is a direct consequence of the peak in the posterior of $\mu$, Fig.~\ref{fig:GBmuPost},
and indeed it corresponds to the value of the cross section at the plateau of the NNLO, Fig.~\ref{fig:ggHscale}.
Note however that the distribution is very asymmetric, and the mean of the distribution is to the left of the peak,
more in line with the next order.
Finally, the distribution at N$^3$LO is the narrowest, with a marked peak
and a slight asymmetry.

\begin{figure}[t]
  \centering
  \includegraphics[width=0.495\textwidth,page=5]{./images/ggH_uncertainty_summary_paper.pdf}
  \includegraphics[width=0.495\textwidth,page=1]{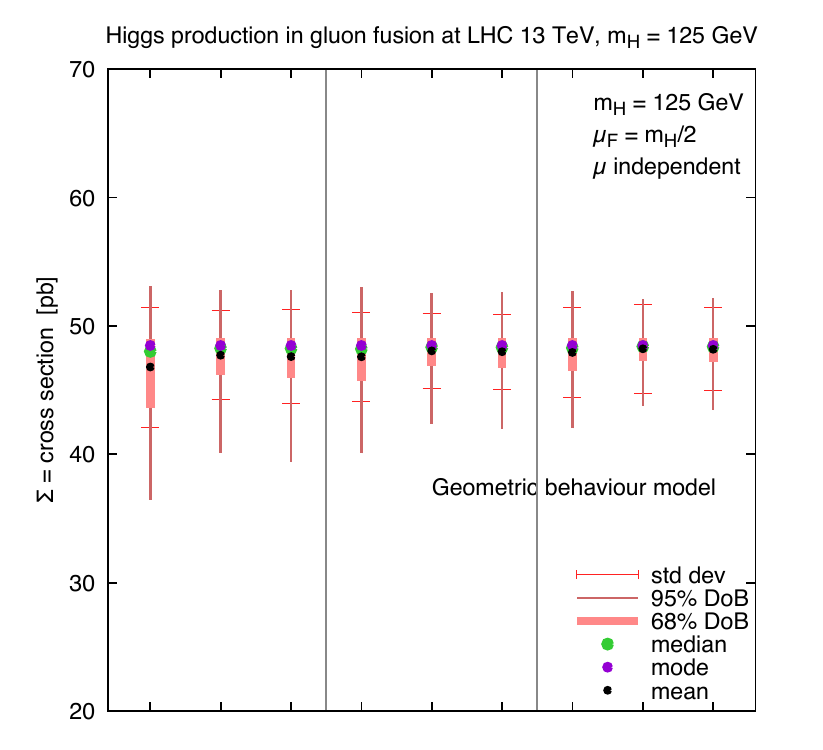}
  \caption{Left plot: similar Fig.~\ref{fig:GBsummary}, after removing scale dependence.
    As a reference the conventional scale variation ``error'' is also shown for $\mu=\mh/2$.
    Right plot: dependence on the prior for $\mu$ of the most precise result (given knowledge of N$^3$LO).
    The first three results correspond to a larger support $\mh/16<\mu<4\mh$,
    the next three to the default support $\mh/10<\mu<2\mh$,
    and the last three to a smaller support $\mh/8<\mu<\mh$.
    In each block, the first result correspond to a flat prior, the second to a triangular prior,
    and the third to a truncated gaussian prior.}
  \label{fig:GBsummaryScaleindep}
\end{figure}

To better understand these results, we show at each order the mean, mode, median, standard deviation and degree of belief intervals
in Fig.~\ref{fig:GBsummaryScaleindep} (left plot).
The pattern found is similar to that of the fixed-scale result, Fig.~\ref{fig:GBsummary}.
One difference is that the uncertainties here are somewhat larger, especially at low orders.
This however implies that the convergence pattern is improved, as for instance this time the 68\%
DoB band is always contained in that of the previous order
(which is mostly due to the fact that the NLO band is increased).
The mean and median are rather similar in all cases, while the mode is
not always close, due to the asymmetry of the distributions.
At N$^3$LO the prediction is rather precise, certainly much more precise than at previous orders.
The mean at N$^3$LO is close to the canonical result at $\mu=\mh/2$.

These results clearly depend on the choice of the prior used for the scale $\mu$.
Therefore, we now consider variations of the prior, in order to understand to what extend the results depend on it.
We consider two types of variations: the support (range of allowed values), and the shape.
Previously we used a support $\mh/10<\mu<2\mh$, and we now consider a larger support $\mh/16<\mu<4\mh$
and a smaller one $\mh/8<\mu<\mh$
Secondly, on top of the uniform prior used before, we consider a symmetric triangular prior (with the same support),
and a gaussian like prior, obtained cutting the tails after two standard deviations,
and adjusting the parameters such that the support is again the same.
The results are shown in the right plot of Fig.~\ref{fig:GBsummaryScaleindep} for the most precise result based on the knowledge of the N$^3$LO.
We notice that enlarging the support has the effect of enlarging the uncertainty,
and the converse is also true.
Also, the flat prior gives obviously the larger uncertainty, while the triangular and gaussian priors
give similar results.
Overall, we see that the dependence on the prior is rather mild, certainly milder than the dependence on the result
on the scale when it is kept fixed. For instance, in this case the mean, mode and median are more or less independent
of the prior.
We conclude, as anticipated, that the ``residual scale dependence'' induced by the freedom in choosing the prior for $\mu$
is very much reduced with respect to the canonical scale dependence of the result.
We can thus consider these results as almost scale independent.

\subsection{Scale-independent scale variation model}
\label{sec:SCscaleindep}

We now apply the approach to remove scale dependence applied to the scale variation model.
Note that it may seem somewhat contradictory to use the information on the scale dependence twice,
once for estimating the higher orders and once for removing the scale dependence.
However, there is nothing problematic in practice, because as in the previous case the approach
to remove the scale dependence just selects, through inference, scales at which the model
provides the best performance, independently of how the model itself works.

To begin with, we show the posterior for $\mu$ in this model.
According to the results of Sect.~\ref{sec:SCinf}, from Eq.~\eqref{eq:Posteriormu} we get
\begin{equation}
P(\mu|\delta_n,....,\delta_1,\Sigma_0) \propto
\frac{P_0(\mu)}{r_0(\mu)\cdots r_{n-1}(\mu)} \int_{\lambda_n(\mu)}^\infty d\lambda\,\lambda^{\gamma-n}e^{-\lambda},
\end{equation}
where $\lambda_n(\mu)$ is given in Eq.~\eqref{eq:lambdamdef}.
The integral can be computed analytically, but written in this way it's clear that
the integral is larger the smaller $\lambda_n(\mu)$.
If the numbers $r_k$ were independent of $\mu$, the posterior would thus be larger
for scales for which the largest $\delta_k(\mu)/r_{k-1}(\mu)$ in $k\in[1,n]$ is small.
Namely, it would select the scale for which the perturbative corrections were small,
which is indeed what we expect this approach to do.
Of course, the numbers $r_k$ do depend on $\mu$, so this picture is modified by their presence in the prefactor.
However, their dependence is mild (see Fig.~\ref{fig:ggHrk}),
and typically milder than the dependence of the $\delta_k$'s,
so the argument above gives a sufficiently good description of how the marginalization over $\mu$ works in this model.

\begin{figure}[t]
  \centering
  \includegraphics[width=0.7\textwidth,page=2]{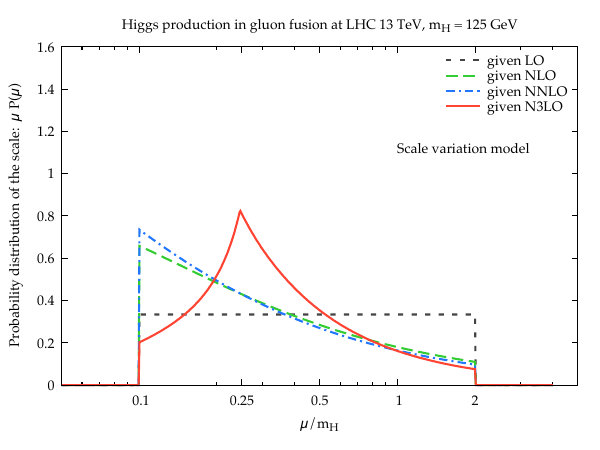}
  \caption{Same as Fig.~\ref{fig:GBmuPost}, but for the scale variation model.}
  \label{fig:SCmuPost}
\end{figure}

In order to understand exactly what happens, we now move to the numerical analysis.
In Fig.~\ref{fig:SCmuPost} we show the posterior of $\mu$ for this model,
in analogy with the discussion of the previous section.
We see that this time, in the range under consideration, both NLO and NNLO
tend to favour small scales.
At N$^3$LO, too small scales are disfavoured, and the distribution becomes peaked at $\mu\sim\mh/4$.
As in the previous case, high scales are strongly disfavoured.

\begin{figure}[t]
  \centering
  \includegraphics[width=0.9\textwidth,page=2]{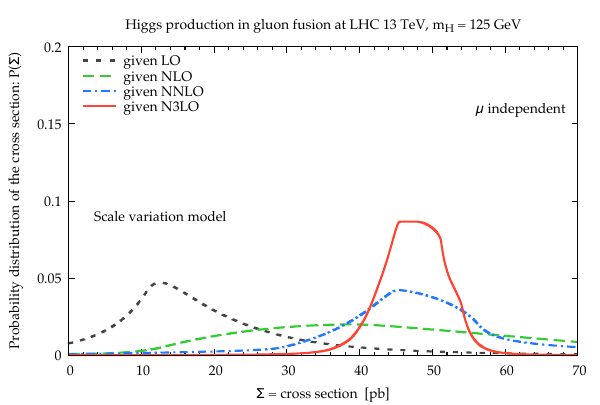}
  \caption{Same as Fig.~\ref{fig:GBdistrScaleindep}, but for the scale variation model.}
  \label{fig:SCdistrScaleindep}
\end{figure}

The probability distribution for the cross section is shown in Fig.~\ref{fig:SCdistrScaleindep}.
Again, after removing the scale dependence the distribution becomes asymmetric.
Again, since at low orders smaller scales are favoured, larger values of the cross section are more probable,
in line with the results of the higher orders.
The distribution at N$^3$LO has a sort of plateau, and strongly decreasing tails,
giving a well localised result.
However, the uncertainty seems rather large, and specifically larger than the result at fixed scale, Fig.~\ref{fig:SCdistr}.

\begin{figure}[t]
  \centering
  \includegraphics[width=0.495\textwidth,page=6]{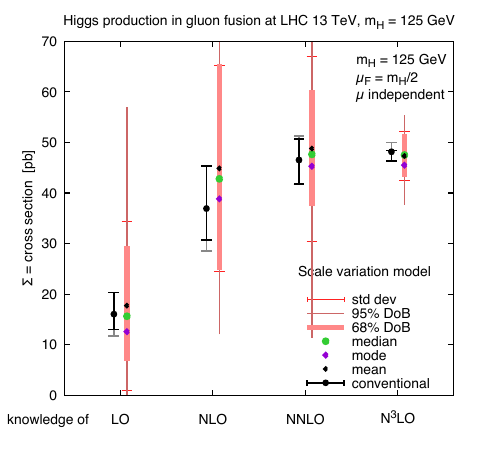}
  \includegraphics[width=0.495\textwidth,page=2]{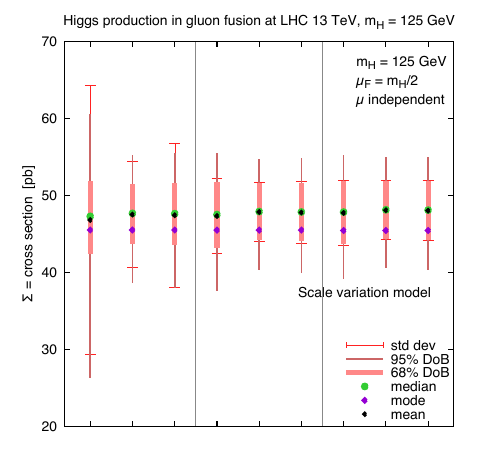}
  \caption{Same as Fig.~\ref{fig:GBsummaryScaleindep}, but for the scale variation model.}
  \label{fig:SCsummaryScaleindep}
\end{figure}

Finally, in Fig.~\ref{fig:SCsummaryScaleindep} (left), we show the summary of the results using the
usual quantifiers: mean, mode, median, standard deviation, degree of belief intervals.
With respect to the fixed-scale results, the uncertainties are larger, and the compatibility
between different orders consequently improved.
The mean of the distribution is rather stable from NLO onwards, suggesting that perhaps these uncertainties are overestimated
(as we pointed out also when discussing the model at fixed scale, Sect.~\ref{sec:ModelScaleVar}).
At N$^3$LO, because the tails die rapidly, the 95\% DoB interval is only slightly larger
than the 68\% DoB interval, suggesting that also after computing the scale independent result
the interpretation of this model as a provider of some sort of ``absolute error'' is still valid.

In the same figure the right plot shows the dependence of the most precise result
on the choice of prior.
The variations are the same introduced in Sect.~\ref{sec:GBscaleindep}.
We see that for a wider support of the prior (first three results)
the uncertainty increases quite significantly, with the exception of the triangular prior (second result).
The explanation for this is that when pushing the smallest scale to be $\mh/16$,
the calculation of the $r_k$, that probes scales up to a factor of 4 smaller than the current $\mu$,
gets contributions from scales as small as $\mh/64\sim2$~GeV, where $\as$ becomes large.
Indeed, from Fig.~\ref{fig:ggHrk} we see that $r_2$ and $r_3$ explode below $\sim\mh/10$,
implying that for those scales the uncertainty in this model becomes huge.
Therefore, when this region is included in the support for $\mu$,
it contaminates the whole distribution with a very broad contribution, thereby leading
to the large uncertainties seen in the figure.
The triangular distribution is an exception because it goes to zero at the endpoint,
thereby suppressing part of the problematic region.
Reducing the range or changing the shape, instead, has a much milder, almost negligible effect.
We conclude that in this model it is crucial to keep the region of allowed scales
within a sensible range, in order to avoid artificially large contributions.
In particular low scales may be problematic because of the large strong coupling.

\section{Validation and applications}
\label{sec:applications}

In this section, we will consider a number of examples of applications of our methods.
First, we focus on examples with a known sum, that will serve as a validation of the procedure.
Some of them are just mathematical series without a physical meaning, others are physics examples.
We then move to some applications where the true sum is not known.
We focus on observables that are known to a high order, so that the results are more significant.

\subsection{Convergent series}
\label{sec:convser}

We start by considering the simplest case, namely a convergent series.
This example may not be too significant as we know that perturbative expansions are divergent,
but it represents a good check of the machinery.
We consider a quasi-geometric series, namely
\begin{equation}\label{eq:convser}
\Sigma = \sum_{k=0}^\infty \as^k A^k \cos(B k)
= \frac{1-A\as\cos(B)}{1+A^2\as^2-2A\as\cos(B)}.
\end{equation}
The presence of the parameter $B$ induces some ``oscillations'' in the perturbative expansion,
and in the limit $B=0$ (or a multiple of $\pi$) we recover a geometric series.
This series is convergent for $\abs{A\as}<1$, and it is bounded by a geometric series $\sum_k\abs{A\as}^k$.
In the following, we fix $A=4$ and $B=2$,
to have sufficiently large perturbative corrections and a sufficiently non-trivial pattern.
We take $\as$ to be $\as(\mz)=0.118$, so that the sum is $\Sigma=0.740531$.
Note that the effective expansion parameter of this series is $A\as$,
but we made a distinction because we want to also induce an artificial scale dependence,
to test the scale variation model and the procedure to remove scale dependence.
This is obtained by changing the scale of $\as$ from $\mz$ to a generic $\mu$,
compensating for this by adding the proper scale-dependent contributions to the
expansion coefficients, according to the procedure described in Sect.~\ref{sec:appScale}.

\begin{figure}[t]
  \centering
  \includegraphics[width=0.545\textwidth,page=1]{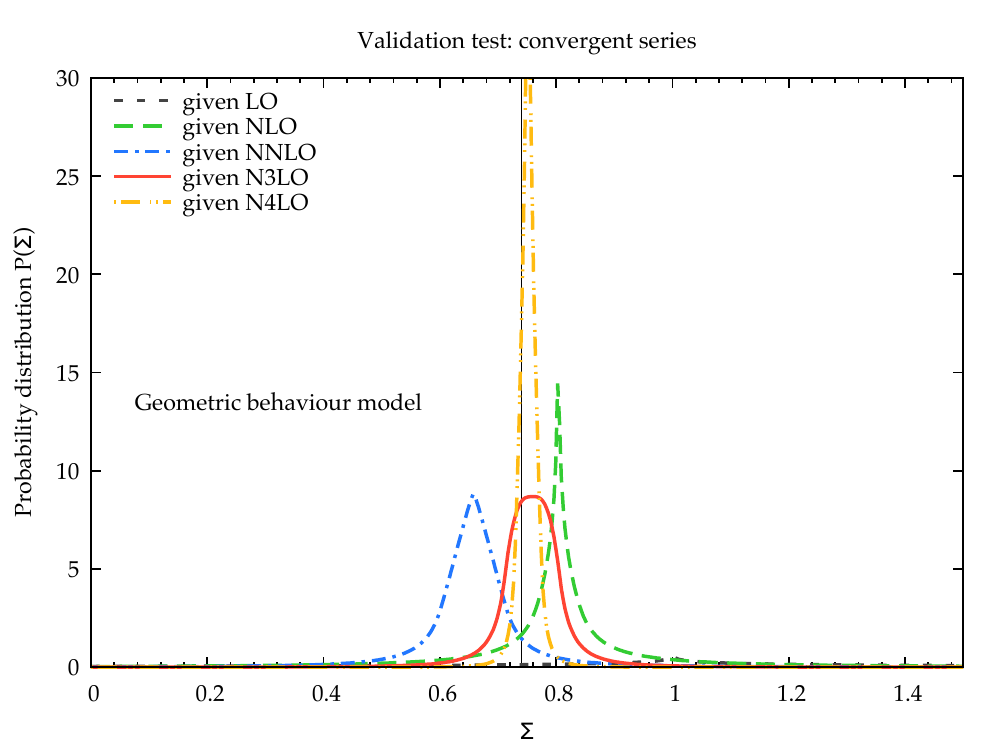}
  \includegraphics[width=0.445\textwidth,page=1]{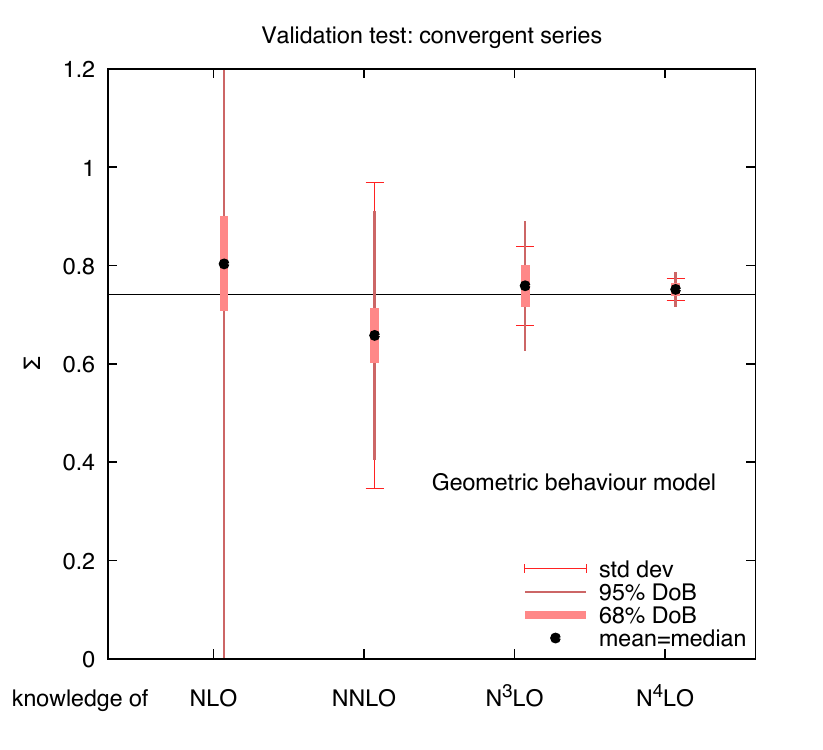}
  \caption{Geometric behaviour model applied to the toy convergent series Eq.~\eqref{eq:convser}.
    The left plot shows the distributions for the observable, and the right plot shows a summary of the quantifiers of these distributions.
    The thin solid black lines represent the exact result.}
  \label{fig:Conv1}
\end{figure}

Let us start by ignoring the scale dependence, and consider thus the expansion Eq.~\eqref{eq:convser} with $\as=\as(\mz)$.
In this case, we can only use the geometric behaviour model of Sect.~\ref{sec:GB}.
Of course, the series Eq.~\eqref{eq:convser} satisfies the assumption of the model Eq.~\eqref{eq:GBhyp},
and so the model is expected to work well.
We assume to know the first five orders, up to N$^4$LO.
In Fig.~\ref{fig:Conv1} (left plot) we show the probability distribution for the next order
given the knowledge of the first orders up to N$^4$LO.
We see that by adding information the distributions become narrower,
and they move closer to the exact result, represented by the vertical black thin line.
In particular, the last two distributions (given the knowledge of the N$^3$LO and N$^4$LO)
are in excellent agreement with the exact result.
In the same figure (right plot) we show the quantifiers (mean, standard deviation, degree of belief intervals)
of these distributions (ignoring the one using only the knowledge of the LO, that is purely determined by priors and is thus not significant).
We see that all result are well compatible with the exact result, which is always within
one standard deviation, and almost always also within the $68\%$ DoB interval.
Increasing the knowledge, the uncertainty clearly shrinks, reaching a very high precision at N$^4$LO
(and also fairly good at N$^3$LO).
These results show that, at least with this somewhat trivial example,
the geometric behaviour model works very well, as expected.

\begin{figure}[t]
  \centering
  \includegraphics[width=0.485\textwidth,page=2]{./images/plot_Pdistr_convser_paper.pdf}
  \includegraphics[width=0.505\textwidth,page=1]{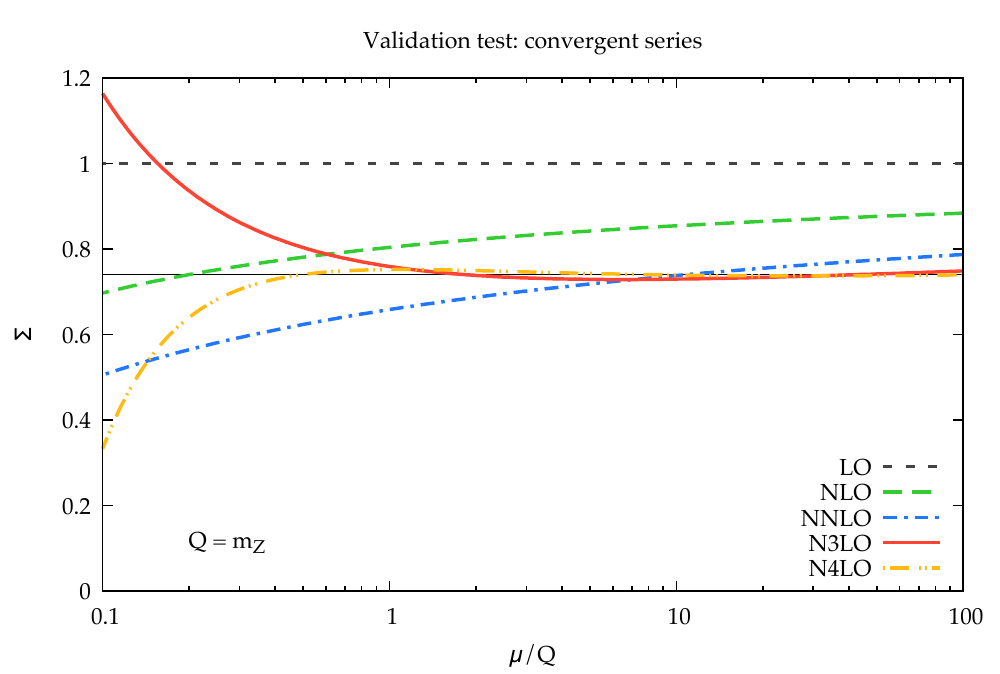}
  \caption{Left plot: the distributions obtained applying the scale variation model
    to the toy series Eq.~\eqref{eq:convser} assuming $Q=\mz$ and $\mu=Q$.
    Right plot: the scale dependence of the observable. The exact result is shown as a thin solid black line.}
  \label{fig:Conv2}
\end{figure}

Let us now assume that the series Eq.~\eqref{eq:convser} is a QCD observable, so that we can consider its scale dependence.
We can thus use the scale variation model of Sect.~\ref{sec:ModelScaleVar}.
The probability distributions obtained with this model are shown in Fig.~\ref{fig:Conv2} (left plot).
Here we see something strange and undesired, namely the distributions become broader
when adding the knowledge of the NNLO and of the N$^3$LO.
Note that all of them are well compatible with the exact result, so the method is accurate,
but the uncertainties are large, even at N$^4$LO, so the method is not precise.
This peculiar pattern can be understood by looking at the scale dependence of the (fake) observable,
Fig.~\ref{fig:Conv2} (right plot).
We observe that the convergence pattern of the expansion deteriorates quickly for scales $\mu<Q\equiv\mz$,
and it improves visibly at higher scales. Therefore, for this observable high scales have to be favoured.

\begin{figure}[t]
  \centering
  \includegraphics[width=0.545\textwidth,page=3]{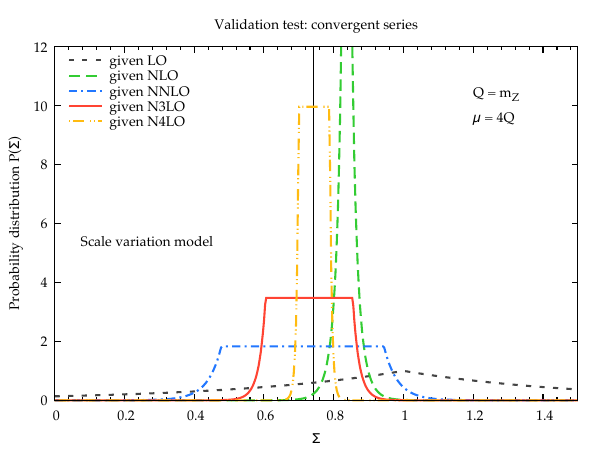}
  \includegraphics[width=0.445\textwidth,page=2]{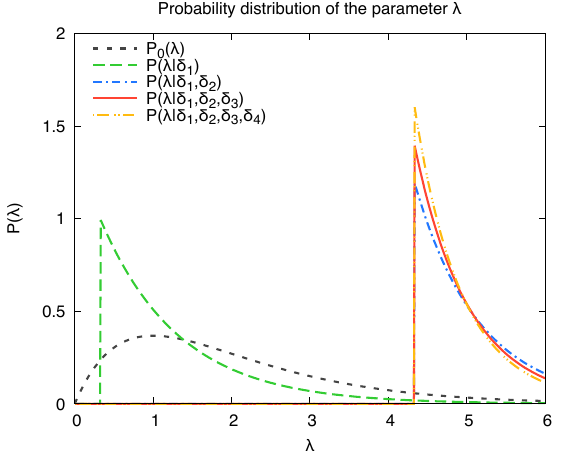}
  \caption{Left plot: same as Fig.~\ref{fig:Conv2} but for $\mu=4Q$.
    Right plot: the posterior distribution for $\lambda$ at $\mu=4Q$.}
  \label{fig:Conv3}
\end{figure}

Let us take for example $\mu=4Q$ and look again at the distributions of the scale variation model.
In Fig.~\ref{fig:Conv3} (left plot) we see that indeed the distributions have a better pattern,
and the distribution with the largest amount of information is rather precise
(even though not at the level of the geometric behaviour model).
However, the distribution at NLO is still narrower than that of the next order.
This undesired behaviour is an artefact due to the fact that the LO is scale independent.
Indeed, had we used literally the scale independence of the LO, the number $r_0$ would be zero,
forcing $\lambda$ to be infinite, making the model not predictive.
In Sect.~\ref{sec:notations_rk} and Sect.~\ref{sec:SChyp} we have suggested to arbitrarily set $r_0=1/2$
when the LO is scale independent. This value is sufficiently large to give unimportant restrictions to $\lambda$,
with the drawback that it allows small values of $\lambda$ leading to an artificially narrow distribution at NLO.
This can be appreciated by looking at Fig.~\ref{fig:Conv3} (right plot),
where we show the posteriors for $\lambda$. We see that the knowledge of the NLO
gives a very mild restriction on $\lambda$, namely $\lambda\geq\abs{\delta_1}/r_0$, which is small because $r_0$ is large.
Conversely, at the next order, when the first real information on the scale dependence ($r_1$) is used,
the smallest allowed value of $\lambda$ becomes much bigger.
This implies that the posterior $P(\lambda|\delta_1,\Sigma_0,\mu)$,
used in the construction of the green curve of Fig.~\ref{fig:Conv3} (left),
is inaccurate and thus leads to an inaccurate prediction.
We conclude that, when the LO is scale independent, the knowledge of the NNLO is needed in order to obtain
the first meaningful prediction through this model.
Note also that the subsequent orders do not push the lowest value of $\lambda$ forward,
similarly to what happens in the Higgs case.
This suggest again that this model is not well suited for precise predictions,
as the assumption itself is not optimal. For a possibly better model, see Sect.~\ref{sec:SCviol}.

\begin{figure}[t]
  \centering
  \includegraphics[width=0.545\textwidth,page=4]{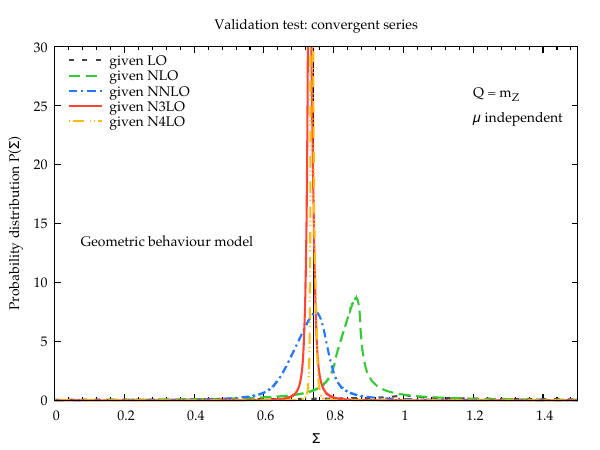}
  \includegraphics[width=0.445\textwidth,page=3]{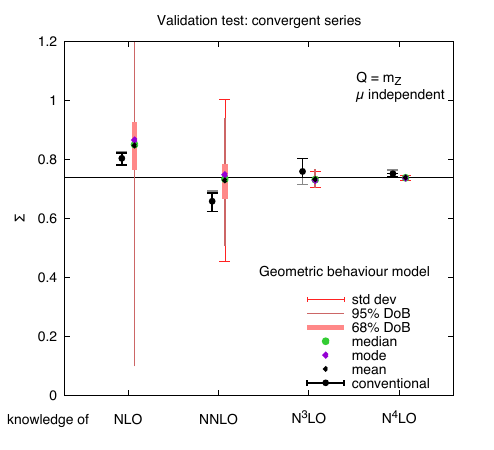}
  \includegraphics[width=0.545\textwidth,page=5]{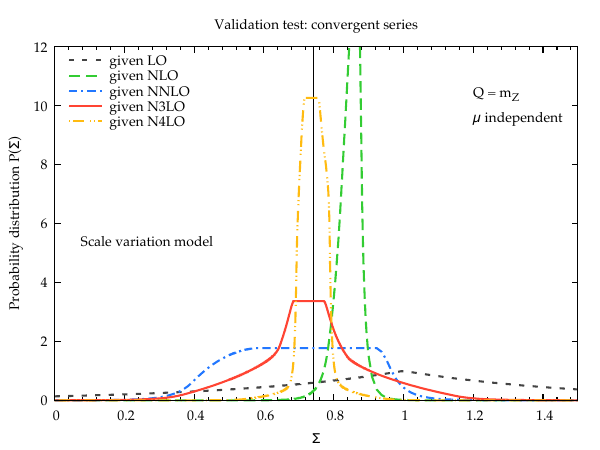}
  \includegraphics[width=0.445\textwidth,page=4]{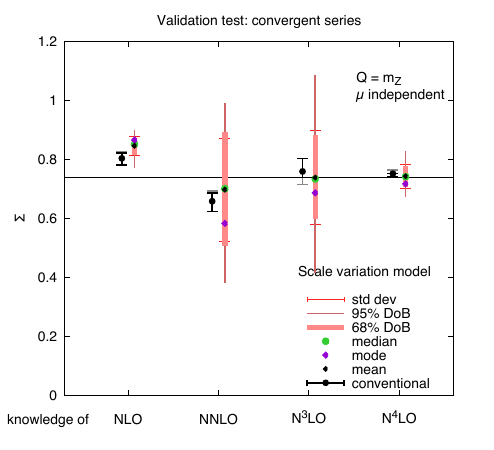}
  \caption{Results after applying the procedure of Sect.~\ref{sec:scaleindep} to remove the scale dependence,
    using a flat prior in the range $Q<\mu<50Q$. The upper plots show the distributions (left) and the summary (right)
    for the geometric behaviour model, while the lower plots are the same for the scale variation model.
    The thin solid black lines represent the exact result.}
  \label{fig:Conv4}
\end{figure}

We finally use the procedure of Sect.~\ref{sec:scaleindep} to obtain scale independent results.
According to Fig.~\ref{fig:Conv2} (right), it is clear that scales larger than $Q$ have to be favoured.
We consider a large range, namely $Q<\mu<50Q$, to be used as the support of our (flat) prior for $\mu$.
The distributions obtained after marginalizing over $\mu$ are shown in Fig.~\ref{fig:Conv4} (left)
for both the geometric behaviour model (above) and the scale variation model (below),
and the summary of the distributions through the usual quantifiers are also shown (right).
In these last plots we also show the conventional scale ``error'' assuming a central scale $\mu=Q$.
For the geometric behaviour model, we see that the convergence pattern is still excellent,
and the uncertainty shrinks more rapidly than in the result at fixed scale, providing very precise predictions
both at N$^3$LO and at N$^4$LO, perfectly compatible with the exact result.
For the scale variation model the situation is slightly improved with respect to the fixed scale result,
but the uncertainties are still rather large, and of course well compatible with the exact result
(with the exception of the NLO, for the reasons mentioned above).
We conclude that all the methods proposed so far work as expected, and in particular the geometric behaviour model
performs extremely well for a series that is compatible with its assumptions.

\subsection{Divergent series}
\label{sec:Div}

We now move to a somewhat opposite case, namely a factorially divergent series.
Provided the expansion parameter is sufficiently small, the expansion has the behaviour of an asymptotic series
(see Fig.~\ref{fig:asympt}). Therefore, it should be a decent representative of actual perturbative expansions,
and thus a good validation test for our models.
We consider the expansion
\begin{equation}\label{eq:divser}
\Sigma = \as \sum_{k=0}^\infty \as^k A^k k! = -\frac1A \exp\(-\frac1{A\as}\)\Gamma\(0, -\frac1{A\as}\),
\end{equation}
where the analytic sum (which is the Borel sum, or equivalently the function of which the series is the asymptotic expansion),
is strictly speaking valid only for $A<0$ (or, if $A$ is complex, for values of $A$ that are not real and positive).
However, for $A>0$ the Borel sum can be still computed up to an ambiguous contribution,
which is exponentially suppressed in the coupling and we thus ignore it.
Note that we have included a factor of $\as$ in front, so that the expansion starts at $\Ord(\as)$.
This choice allows to have a LO that is scale dependent, to avoid the issues discussed
in the previous case when using the scale variation model.
The series Eq.~\eqref{eq:divser} is assumed to be written at $\mu=Q\equiv\mz$, with $\as=\as(\mz)=0.118$,
and the expansion at generic $\mu$ can be obtained with the procedure described in Sect.~\ref{sec:appScale}.
In the following, we will consider both cases $A=-1$ and $A=1$,
dubbed respectively as factorially divergent series with alternating signs and with same signs.

\begin{figure}[t]
  \centering
  \includegraphics[width=0.545\textwidth,page=1]{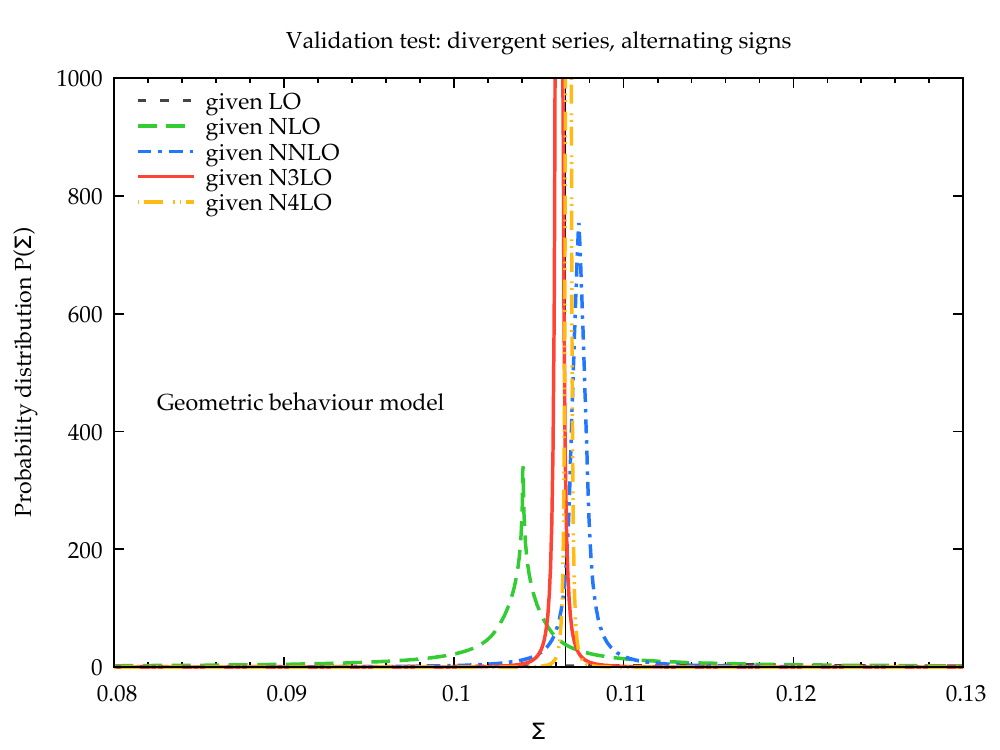}
  \includegraphics[width=0.445\textwidth,page=1]{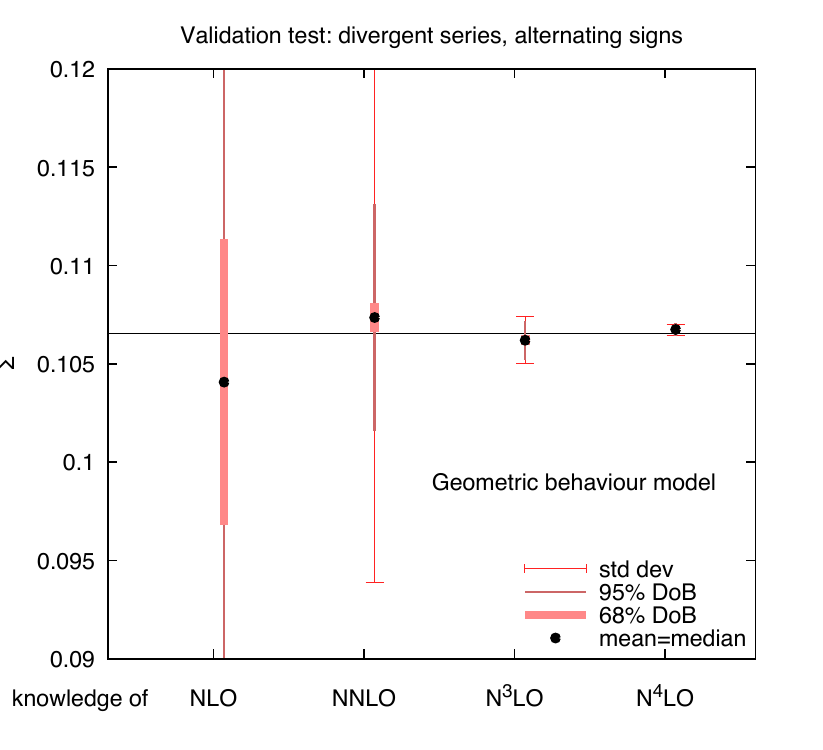}
  \includegraphics[width=0.545\textwidth,page=1]{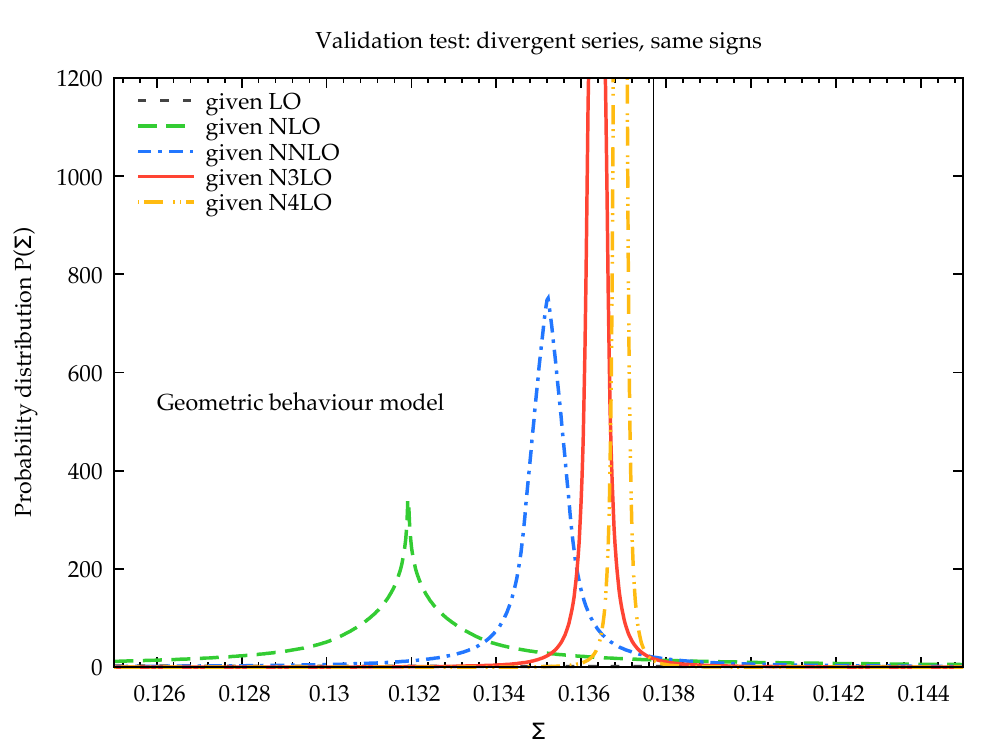}
  \includegraphics[width=0.445\textwidth,page=1]{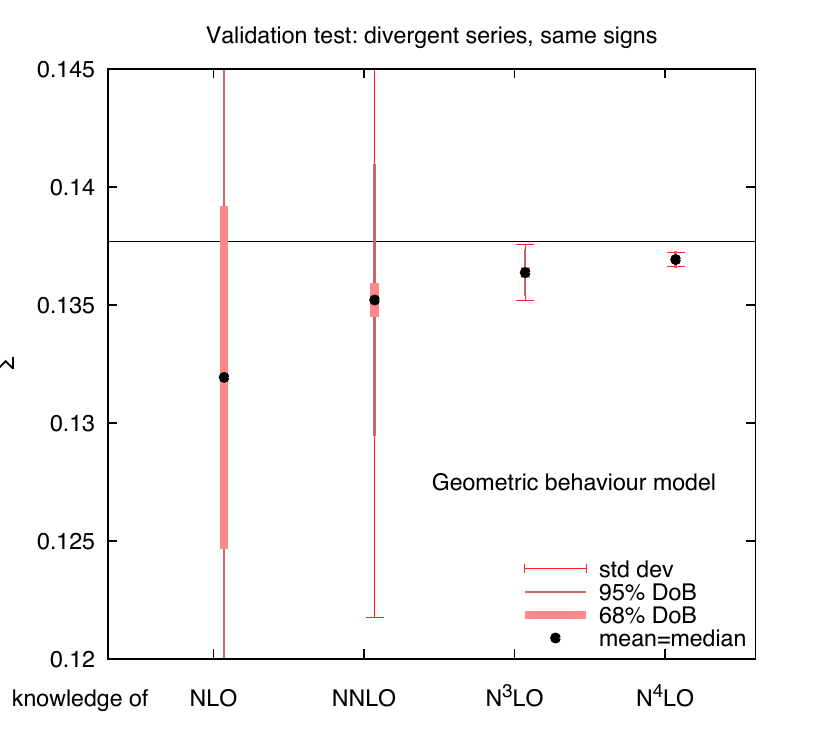}
  \caption{Geometric behaviour model applied to the toy divergent series Eq.~\eqref{eq:divser}, for $A=-1$ (upper plots) and $A=1$ (lower plots).
    The left plot shows the distributions for the observable, and the right plot shows a summary of the quantifiers of these distributions.
    The thin solid black lines represent the exact result.}
  \label{fig:Div1}
\end{figure}

Let us start again by ignoring the scale dependence and thus using only the geometric behaviour model.
The results (distributions and summary in terms of quantifiers) are shown in Fig.~\ref{fig:Div1}.
We observe that in the case of alternating signs ($A=-1$) the method works very well,
with the distributions shrinking and converging towards the exact result.
This is a consequence of the expansion parameter being sufficiently small
to give an approximately convergent pattern at low orders,
as we hope is the case for physical observables.
The prediction is thus accurate, and increasing the order becomes also very precise.
In the case of all same sings ($A=1$) the situation is different.
The distributions shrink and move towards the exact result, without ``reaching'' it.
In this case, it is clear that the shape of the distribution is inadequate,
as the exact result always lies in the tail of the distributions.
This behaviour is not surprising, as this series violates the assumptions
of the model in a ``maximal'' way (the series can be considered as the bound of a physical expansion).

\begin{figure}[t]
  \centering
  \includegraphics[width=0.485\textwidth,page=2]{./images/plot_Pdistr_factdivaltr_paper.pdf}
  \includegraphics[width=0.505\textwidth,page=1]{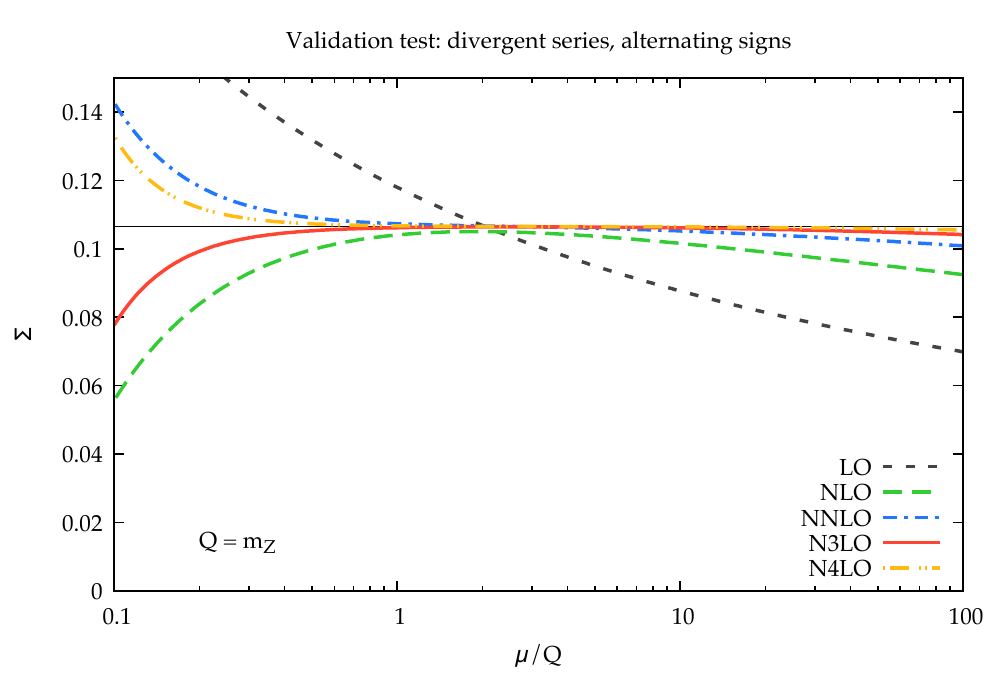}
  \includegraphics[width=0.485\textwidth,page=2]{./images/plot_Pdistr_factdivsame_paper.pdf}
  \includegraphics[width=0.505\textwidth,page=1]{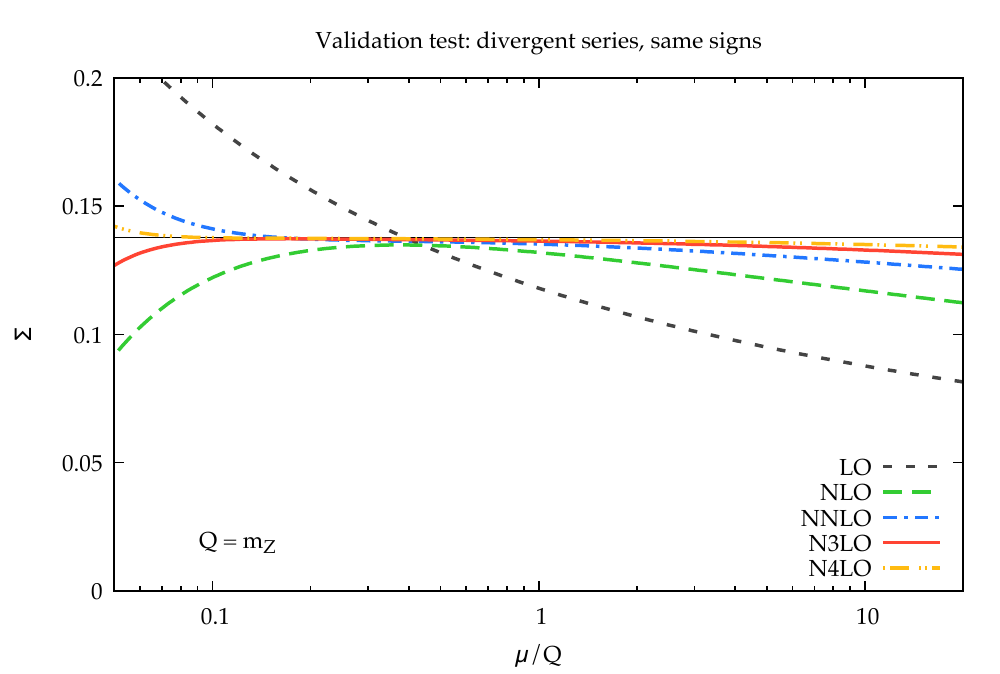}
  \caption{Left plots: the distributions obtained applying the scale variation model
    to the toy series Eq.~\eqref{eq:divser} assuming $Q=\mz$ and $\mu=Q$, for alternating signs (above) and same signs (below).
    Right plots: the scale dependence of the observable in either case. The exact result is shown as a thin solid black line.}
  \label{fig:Div2}
\end{figure}

Let us now include the scale dependence, and consider the scale variation model.
In Fig.~\ref{fig:Div2} (left plots) we show the distributions obtained with this model in the two cases, for $\mu=Q\equiv\mz$.
For alternating signs, the accuracy of the model is manifest, with the distributions
always covering the exact result (within the plateau region when present),
and getting narrower increasing the order. However, the precision is not at the level of the
geometric behaviour model, as usual.
When the coefficients of the series have all the same sign, a similar pattern
to that of the geometric behaviour model is observed.
In the same figure, the right plots show the (artificial) scale dependence of our observable,
as we constructed it. Note that the range is different, as in one case (alternating signs)
larger scales lead to a better convergence,
while in the other case (same signs) lower scales give a better expansion
(this peculiar behaviour is just a consequence of the artificial way in which scale dependence is introduced).

\begin{figure}[p]
  \centering
  \includegraphics[width=0.48\textwidth,page=3]{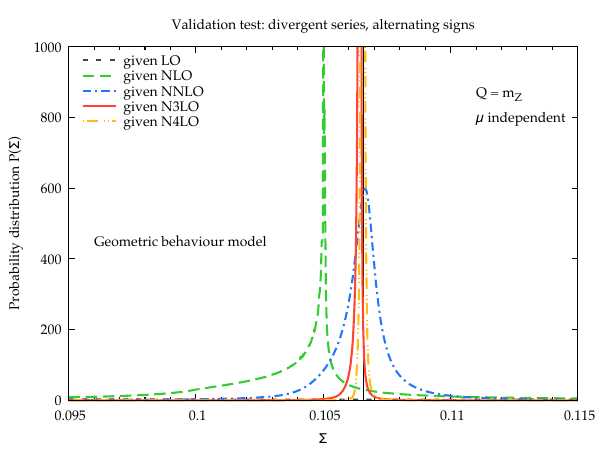}
  \includegraphics[width=0.39\textwidth,page=3]{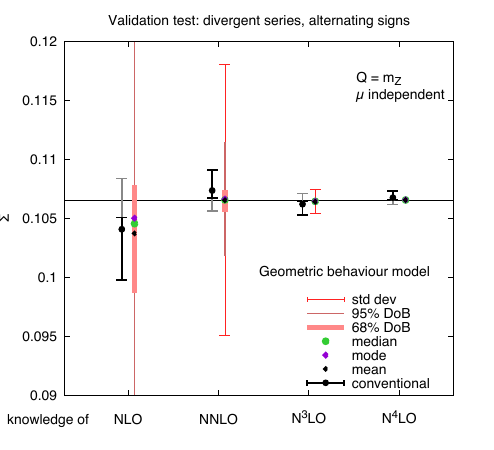}\\
  \includegraphics[width=0.48\textwidth,page=4]{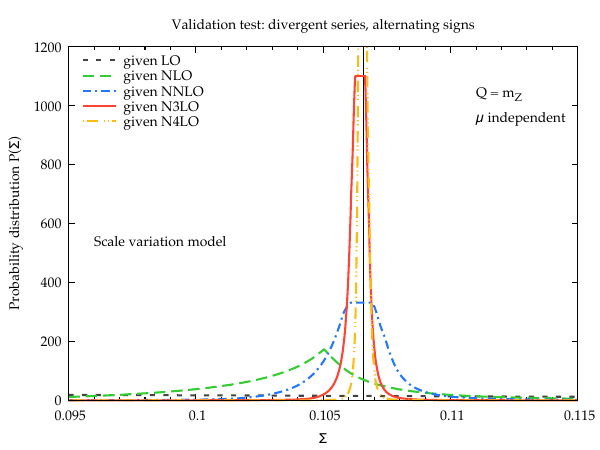}
  \includegraphics[width=0.39\textwidth,page=4]{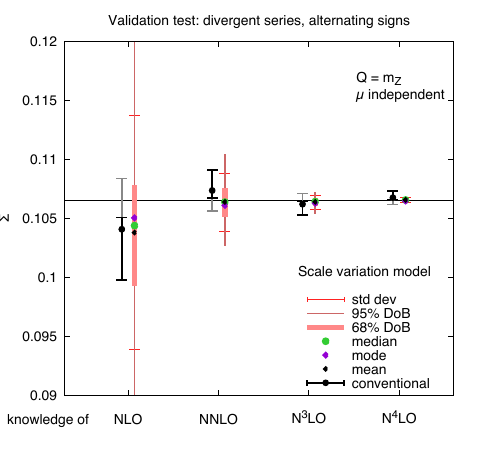}\\
  \includegraphics[width=0.48\textwidth,page=3]{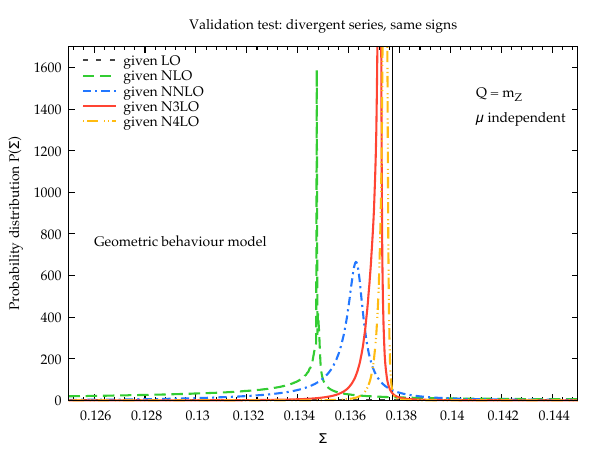}
  \includegraphics[width=0.39\textwidth,page=3]{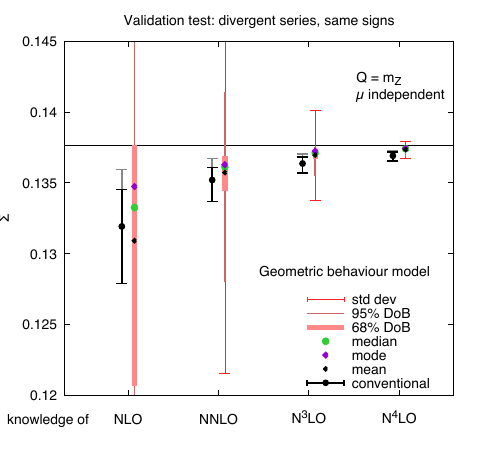}\\
  \includegraphics[width=0.48\textwidth,page=4]{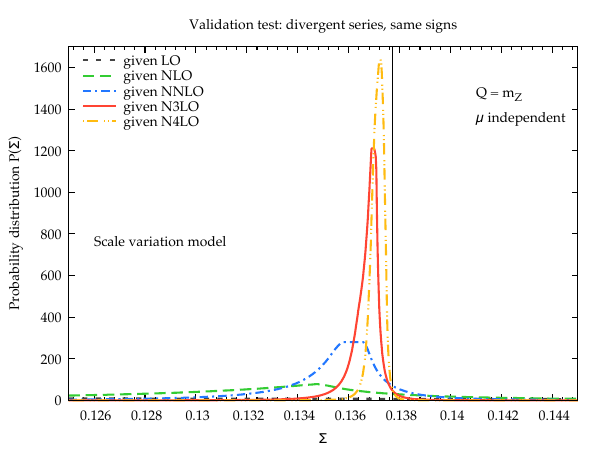}
  \includegraphics[width=0.39\textwidth,page=4]{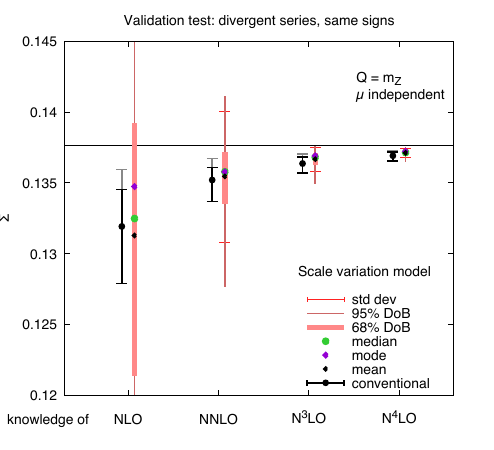}
  \caption{Scale independent results for alternating signs (upper plots) and same signs (lower plots).
    The thin solid black lines represent the exact result.}
  \label{fig:Div3}
\end{figure}
\afterpage{\FloatBarrier}

According to this observation, we now apply the procedure of Sect.~\ref{sec:scaleindep}
to remove the scale dependence, using flat priors for $\mu$ in the range $Q/2<\mu<16Q$ for alternating signs
and $Q/8<\mu<8Q$ for same signs.
The results are shown in Fig.~\ref{fig:Div3}.
For the alternating signs series (upper plots) removing the scale dependence improves the precision
of both the geometric behaviour model and of the scale variation model, while keeping excellent accuracy.
This provides an important validation of all our methods.
For the same sign series marginalizing over the scale improves a bit the accuracy,
especially for the geometric behaviour model, where the exact result becomes compatible at least within
95\% DoB also at high orders.
While not perfect in this case, our methods are certainly superior to the conventional scale ``error'', also shown in the plots.

Note that the methods can improve if the sign pattern is taken into account.
The two expansions considered are characterized by either all positive corrections or corrections with alternating signs.
If the patter is recognised within the model, then the prediction of the next orders can be performed accordingly.
We will investigate this possibility in Sect.~\ref{sec:sign}.

\subsection{Anharmonic oscillator in Quantum Mechanics}
\label{sec:ana}

We now consider a physics example, namely a quartic anharmonic oscillator in quantum mechanics.
Using the notation of Ref.~\cite{Graffi:1990pe}, its hamiltonian is given by
\begin{equation}
H = p^2 +x^2 + \alpha x^4,
\end{equation}
where $\alpha$ is the ``coupling'', namely the perturbative expansion parameter.
The energy levels of this anharmonic oscillator as a perturbative expansion in $\alpha$ are known to diverge~\cite{Bender:1969si}.
The ground-state energy is given by
\begin{equation}
E_0(\alpha) = 1+2\sum_{k=1}^\infty \frac{A_k}{2^k} \alpha^k,
\end{equation}
where the first 75 $A_k$ coefficients can be found in Ref.~\cite{Bender:1969si}.
We recall that these coefficients have alternating signs, namely $A_k = (-1)^{k-1}\abs{A_k}$.
They grow faster than a power, but slower than a factorial.
The exact result for this ground state energy has been computed in Ref.~\cite{Graffi:1990pe} through a Borel-Pad\'e method.
We consider the value $\alpha=0.1$, which is the smallest value reported in that paper, corresponding to the sum $E_0(0.1)=1.0653$.
For this value the series already starts diverging quite soon:
the optimal truncation value of the asymptotic expansion is $\kas=6$.

\begin{figure}[b]
  \centering
  \includegraphics[width=0.545\textwidth,page=1]{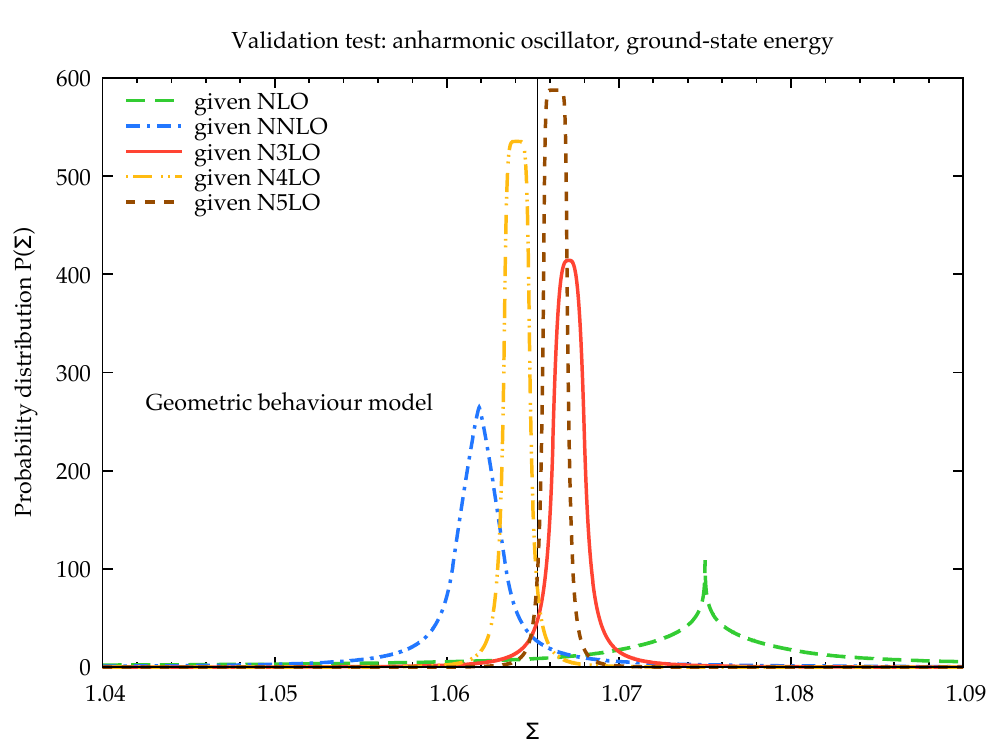}
  \includegraphics[width=0.445\textwidth,page=1]{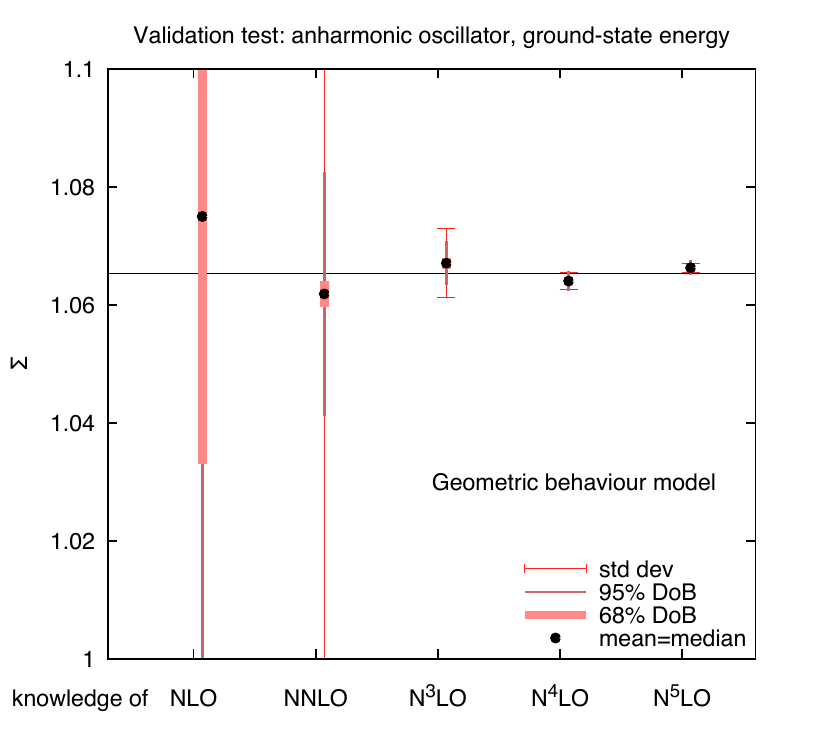}
  \caption{Geometric behaviour model applied to the ground-state energy of the quartic anharmonic oscillator in quantum mechanics.
    The left plot shows the distributions for the observable, and the right plot shows a summary of the quantifiers of these distributions.
    The thin solid black lines represent the exact result.}
  \label{fig:Anharmonic}
\end{figure}

Since this is a quantum mechanical system, there is no unphysical scale dependence,
and therefore we can only use the geometric behaviour model of Sect.~\ref{sec:GB}.
The distributions for the ground-state energy given different status of knowledge are shown
in Fig.~\ref{fig:Anharmonic} (left).
We observe that starting from N$^3$LO the distributions are quite narrow, and oscillate
around the exact value, because of the alternating signs of the $A_k$ coefficients.
In fact, the prediction does not become very accurate and precise, essentially because 
we are very close to the asymptotic point $\kas=6$ that sets the limit of validity
of the assumptions of the model.

This can be appreciated also by looking at the summary of the distributions through the usual quantifiers
in the right plot of the same figure.
It seems surprising that the quality of the uncertainty, though perfectly acceptable
(the exact result is always within the 95\% DoB interval), is not as good as in the similar
case of the factorially divergent series discussed in Sect.~\ref{sec:Div}.
The reason is that the effective coupling, namely the parameter determining the power growth of the expansion,
is larger in this case than it was in the factorially divergent case.
In fact, we are now in a condition for which the expansion is barely perturbative,
and therefore all our assumptions, including the basic ones of Sect.~\ref{sec:basics},
are barely satisfied.
Nevertheless, with all these limitations in mind, the method works rather well.
Note that taking into account the sign pattern (alternating signs) in the model,
as described in Sect.~\ref{sec:sign}, can help improving the accuracy of the results.

\subsection{Higgs production in the threshold limit}
\label{sec:ggHn3ll}

We conclude the examples used for the validation of the models with a QCD process.
Since, to our knowledge, there are no QCD observables in the perturbative regime
for which the exact result is known, we use a trick.
Namely, we consider a purely all-order resummed result at a finite logarithmic accuracy,
and consider it as if it were the full (exact) result.
We expand it in powers of $\as$ to obtain the first few perturbative orders, to which our methods are applied.

This procedure obviously misses contributions that are not logarithmically enhanced,
which may constitue an important part of the observable.
Therefore, it cannot be regarded as a validation of a complete QCD perturbative expansion,
but only of a part of it (the logarithmically enhanced contributions) which may in general behave differently
from the full result.
This limitation can be minimized by considering threshold resummation of an inclusive observable.
Indeed, threshold logarithms reproduce a number of structures that are present in the full
result (e.g., plus distributions), and indeed they often offer a good approximation
to the full result~\cite{Catani:2001ic}.
Additionally, to all orders threshold contributions diverge factorially~\cite{Catani:1996yz}, as the full expansion does.
Therefore, a threshold approximation can be regarded as a good representative of a full QCD perturbative expansion.

The process we consider is once again Higgs production in gluon fusion. Its threshold resummation
is known at a high logarithmic accuracy, N$^3$LL (see e.g.\ Ref.~\cite{Bonvini:2014joa}),
and it is well known to provide a good approximation of the full result order by order (see e.g.\ Ref.~\cite{Bonvini:2016frm}).
More precisely, we take the so-called $\psi$-soft$_2$ resummation~\cite{Bonvini:2014joa},
with the so-called default choice for the constant terms~\cite{Bonvini:2016frm}.
Using the same setup of Sect.~\ref{sec:ggH}, at the scale $\mu=\mh$ (and $\muf=\mh/2$)
the partial sums of the perturbative expansion of the cross section are given by\footnote
{These numbers have been computed using the \href{https://www.roma1.infn.it/~bonvini/troll/}{\texttt{TROLL}} code~\cite{Bonvini:2014joa,Bonvini:2016frm}.}
\begin{equation}\label{eq:XSggHn3ll}
\Sp(\mh) = \left\{ 13.0, 28.0, 37.8, 42.4 \right\}\; \text{pb}.
\end{equation}
These numbers are very similar to the exact ones, Eq.~\eqref{eq:XSggH}:
they are slightly smaller, but given the large size of the perturbative corrections,
this approximation captures well the behaviour of the expansion.
The ``exact'' result, namely the all-order resummed one, is instead given by
\begin{equation}\label{eq:XSggHn3llexact}
\Sigma = 45.0\; \text{pb}.
\end{equation}
Note that this result is in fact scale dependent, as it is not really exact.
The scale dependence is induced by terms beyond N$^3$LL and beyond the threshold approximation.
In practice, this dependence is rather mild, and certainly much milder than the dependence of
the fixed-order result. Therefore, to a first approximation, we simply ignore it and use the value
of Eq.~\eqref{eq:XSggHn3llexact}, which is computed at $\mu=\mh/2$ where the convergence is faster,
as if it were the exact one.

\begin{figure}[p]
  \centering
  \includegraphics[width=0.66\textwidth,page=1]{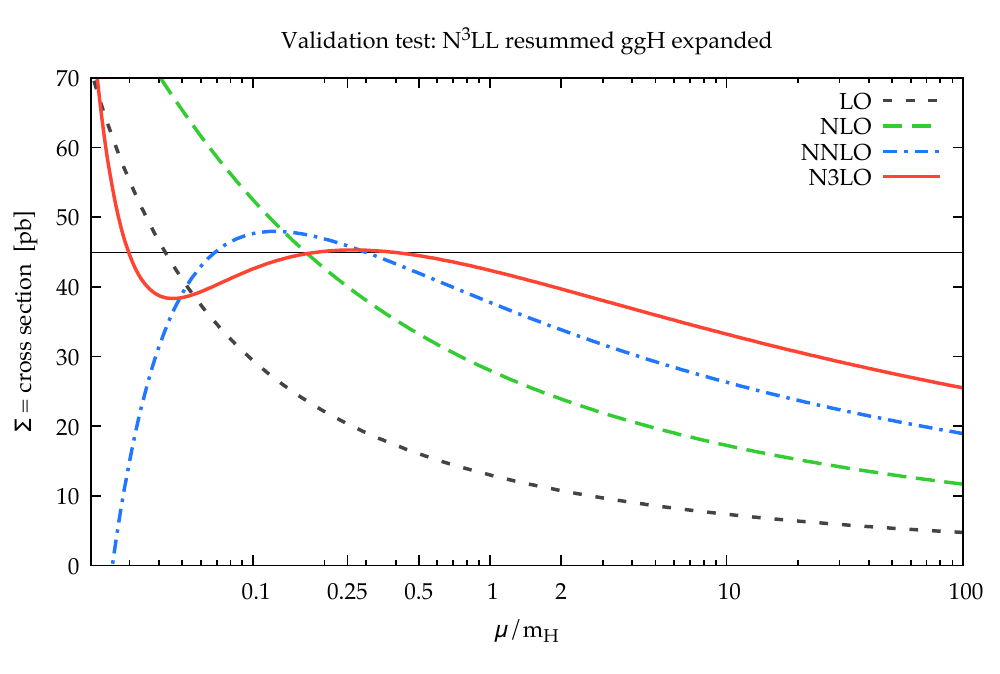}
  \caption{Scale dependence of the purely resummed expanded Higgs cross section (settings as in Fig.~\ref{fig:ggHscale}).
    The thin solid black line represents the ``exact'' result.}
  \label{fig:ggHn3llscale}
\end{figure}

\begin{figure}[p]
  \centering
  \includegraphics[width=0.445\textwidth,page=1]{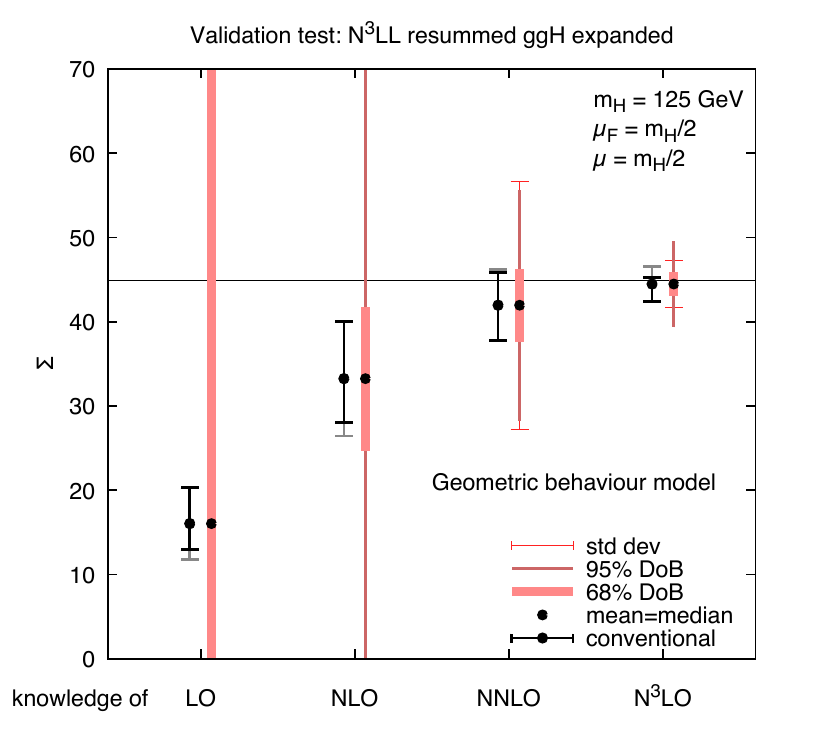}
  \includegraphics[width=0.445\textwidth,page=3]{./images/ggHresN3LL_uncertainty_summary_paper.pdf}\\
  \includegraphics[width=0.445\textwidth,page=2]{./images/ggHresN3LL_uncertainty_summary_paper.pdf}
  \includegraphics[width=0.445\textwidth,page=4]{./images/ggHresN3LL_uncertainty_summary_paper.pdf}
  \caption{The geometric behaviour model (upper plots) and scale variation model (lower plots)
    at fixed scale $\mu=\mh/2$ (left plots) and after marginalizing over the scale (right plots)
    for the resummed Higgs production cross section at N$^3$LL.
    The thin solid black lines represent the all-order resummed result, used as ``exact'' result in this example.}
  \label{fig:ggHn3ll}
\end{figure}
\afterpage{\FloatBarrier}

We start by showing the scale dependence of the various orders in this example in Fig.~\ref{fig:ggHn3llscale}.
We observe that this plot looks very similar to that of Fig.~\ref{fig:ggHscale},
as expected from the goodness of the threshold approximation.
Therefore, the probability distributions that we get from the models are also very similar,
and we thus do not report them here.
We also observe that the ``exact'' result is rather close to the plateau of the N$^3$LO result.

In Fig.~\ref{fig:ggHn3ll} we show the summaries of the uncertainties using the usual quantifiers.
We see that all results (with the exception of the scale variation model at LO)
are well compatible with the ``exact'' result,
especially after marginalizing over the scale.
These plots are very similar to those of Figs.~\ref{fig:GBsummary}, \ref{fig:SCsummary},
\ref{fig:GBsummaryScaleindep} and \ref{fig:SCsummaryScaleindep},
and thus provide a strong validation of those results.
We can also see that the canonical scale variation ``error'' is acceptable at NNLO and N$^3$LO,
but it underestimates the uncertainty at NLO and strongly at LO,
and this even at the ``optimal'' scale $\mu=\mh/2$ used in the plots:
the pattern may become much worse at different (e.g.\ higher) scales.

\subsection{$e^+e^-$ into hadrons at N$^4$LO}
\label{sec:epem}

We now consider some applications of our methods to real physical processes of interest for which the exact result is not known.
We start from the classical process of $e^+e^-\to\text{hadrons}$, which is known in QCD up to N$^4$LO~\cite{Baikov:2012zn}.
We define the observable $\Sigma$ as the ratio of the function $R(s)$ describing the process\footnote
{In fact, $R(s)$ is the ratio of the cross section for $e^+e^-\to\text{hadrons}$ to that of $e^+e^-\to\mu^+\mu^-$,
but since we normalize to the LO, the denominator is immaterial.}
to its LO $R_0(s)$,
\begin{equation}
\Sigma = \frac{R(s)}{R_0(s)} = 1+\frac{\as(s)}{\pi}+1.40923\(\frac{\as(s)}{\pi}\)^2-12.805\(\frac{\as(s)}{\pi}\)^3-80.434\(\frac{\as(s)}{\pi}\)^4+\Ord(\as^5)
\end{equation}
where $s$ is the center of mass energy squared of the collision, the renormalizaion scale is $\mu=\sqrt{s}$,
and we have assumed $n_f=5$ active flavours for the computation of the numerical coefficients of the expansion~\cite{Baikov:2012zn}.
We choose the collider energy to be $\sqrt{s}=10$~GeV, so that the bottom quark is active ($n_f=5$)
and the scale is sufficiently large to be in the perturbative regime, and at the same time sufficiently small
to neglect the contribution from a virtual $Z$ boson.

\begin{figure}[t]
  \centering
  \includegraphics[width=0.53\textwidth,page=1]{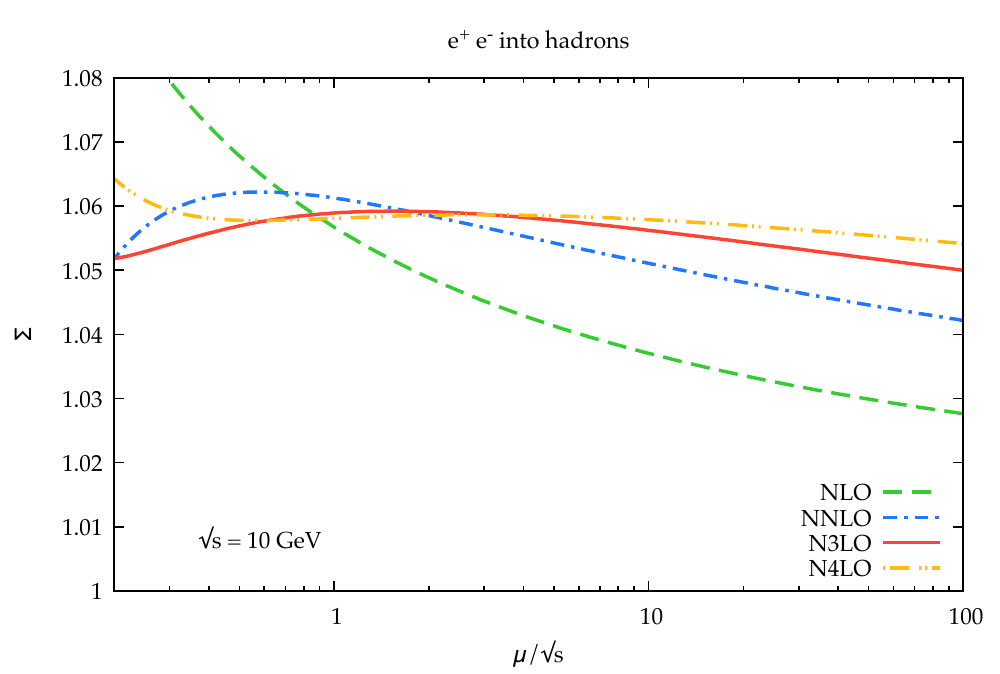}
  \includegraphics[width=0.46\textwidth,page=1]{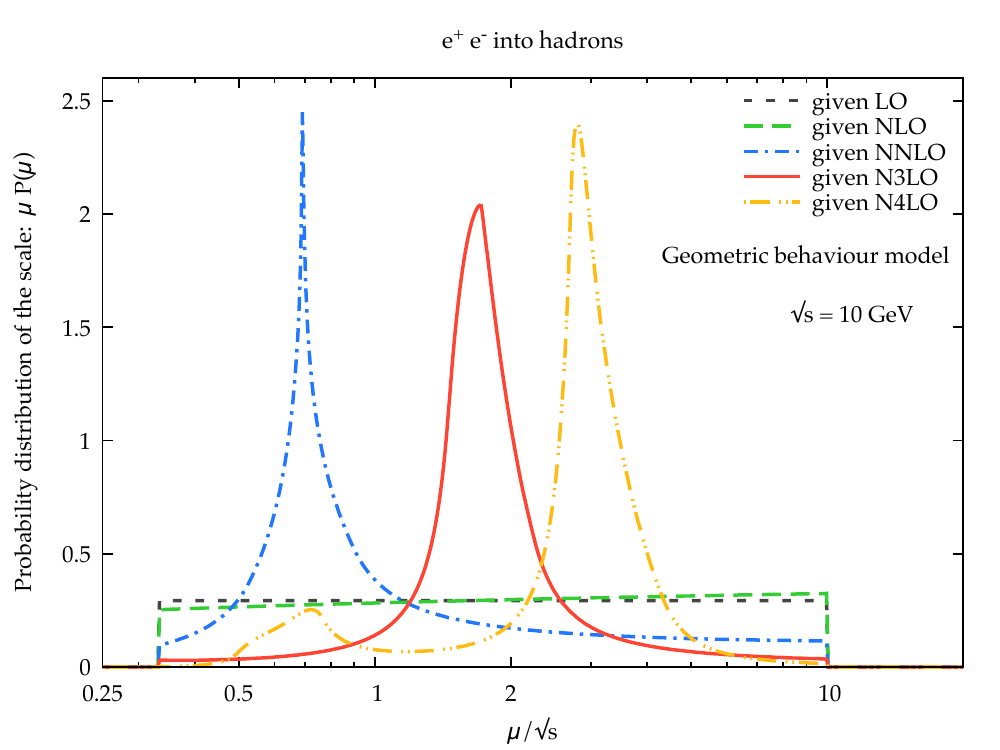}
  \caption{Left: scale dependence of the $e^+e^-\to\text{hadrons}$ process at $\sqrt{s}=10$~GeV.
    Right: posterior distribution for the scale for the geometric behaviour model.}
  \label{fig:epemscale}
\end{figure}

The ``raw'' result of this process, shown as a function of the renormalization scale, is depicted in Fig.~\ref{fig:epemscale} (left).
With the exception of the LO, which is scale independent (it corresponds to the lower edge in the plot),
we observe that increasing the order the scale dependence flattens out.
At N$^4$LO, the result is very stable upon scale variation.
We also observe an apparently convergent pattern at all scales except the smallest ones,
where of course the larger value of $\as$ makes the perturbative expansion badly behaved.

\begin{figure}[p]
  \centering
  \includegraphics[width=0.48\textwidth,page=1]{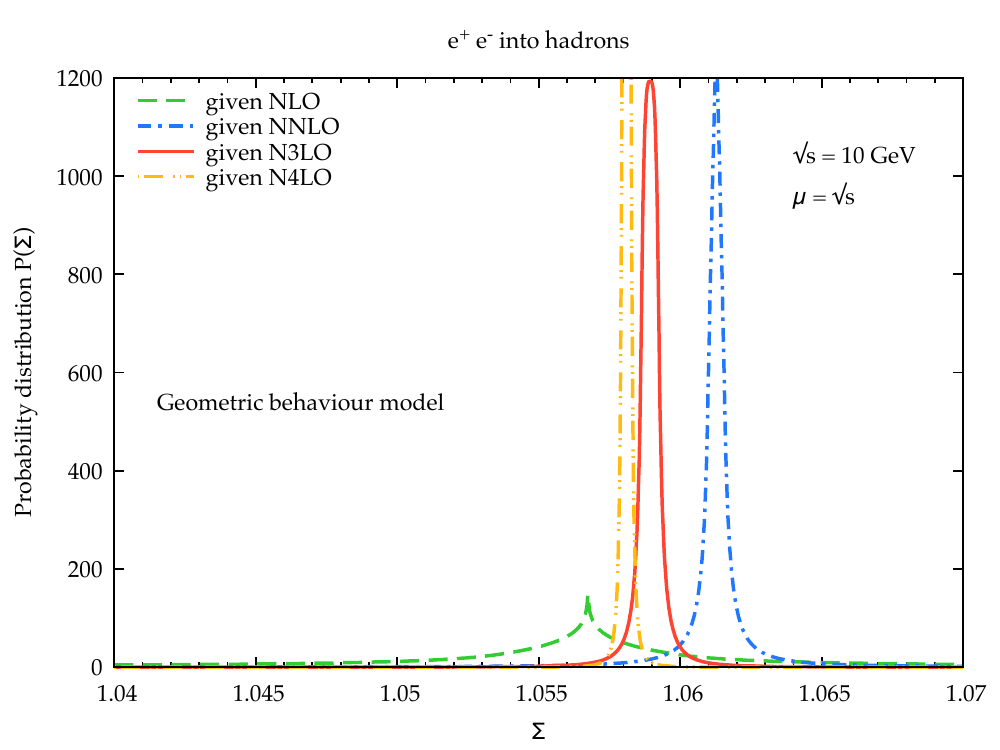}
  \includegraphics[width=0.39\textwidth,page=1]{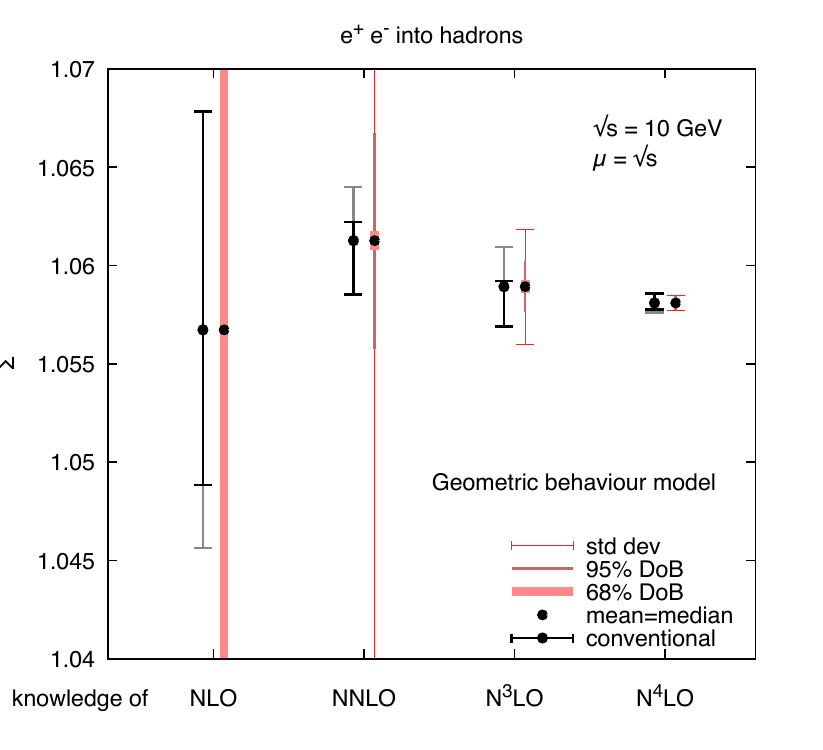}\\
  \includegraphics[width=0.48\textwidth,page=3]{./images/plot_Pdistr_epemhadrons_paper.pdf}
  \includegraphics[width=0.39\textwidth,page=3]{./images/epemhadrons_uncertainty_summary_paper.pdf}\\
  \includegraphics[width=0.48\textwidth,page=2]{./images/plot_Pdistr_epemhadrons_paper.pdf}
  \includegraphics[width=0.39\textwidth,page=2]{./images/epemhadrons_uncertainty_summary_paper.pdf}\\
  \includegraphics[width=0.48\textwidth,page=4]{./images/plot_Pdistr_epemhadrons_paper.pdf}
  \includegraphics[width=0.39\textwidth,page=4]{./images/epemhadrons_uncertainty_summary_paper.pdf}
  \caption{Fixed scale and scale independent results for the $e^+e^-\to\text{hadrons}$ process,
    for both the geometric behaviour model (upper plots) and the scale variation model (lower plots).}
  \label{fig:epemplots}
\end{figure}
\afterpage{\FloatBarrier}

We now apply our probabilistic methods to this expansion.
We consider both the geometric behaviour model and the scale variation model,
both at fixed scale $\mu=\sqrt{s}$ and after marginalizing over the scale,
using a flat prior in the range $\sqrt{s}/3<\mu<10\sqrt{s}$.
The distributions and their summary in terms of the usual quantifiers are shown in Fig.~\ref{fig:epemplots}.
Let us start from the geometric behaviour model.
At fixed scale, the distributions from NNLO onwards are quite narrow and localized,
giving rise to precise predictions.
However, at least at NNLO, the prediction is not very accurate,
as the next orders tend to favour values towards the tail of the NNLO distribution.
Indeed, the 68\% DoB region is very small, and the next orders are compatible only within the 95\% DoB interval.
The fact that the precision of the NNLO distribution is not faithful is also
seen from the large standard deviation of the distribution, due to the slowly decreasing tails.
We observe that after marginalizing over the scale the pattern improves significantly.
The NNLO distribution becomes very asymmetric, covering with higher probability
a region of smaller cross section. Indeed, in this case the 68\% DoB interval is larger and covers the next orders.
We see once again that marginalizing over the scale, on top of removing a bias, leads to more
accurate results.
A peculiar feature of the scale independent result is the bimodal distribution at N$^4$LO.
This is a consequence of the scale dependence, shown in Fig.~\ref{fig:epemscale} (left),
where we see that there are two scales for which the N$^4$LO correction vanishes.
One value, approximately $\mu\sim3\sqrt{s}$, leads to a larger cross section which
is also close to the NNLO result, which makes this scale more favourable,
thus producing the higher peak of the distribution.
The other value, approximately $\mu\sim0.6\sqrt{s}$, leads to a smaller cross section
and produces the secondary peak. This value is less favoured because the
N$^3$LO is larger here, but at a slightly higher scale all corrections are rather small,
and in particular the NNLO correction also vanishes.
This implies that the posterior distribution for $\mu$, shown in Fig.~\ref{fig:epemscale} (right),
is bimodal itself, with the secondary peak in correspondence of the peak of the posterior at NNLO.
Note that in the end the fact that the distribution for $\Sigma$ is bimodal is not a problem,
and indeed the uncertainty shown through the quantifiers is rather small (though slightly asymmetric).

The scale variation model gives, as usual, larger distributions and larger uncertainties,
though the tails are exponentially suppressed and therefore the ``support'' of the distribution is more localized.
The accuracy of the results is good, with a mild exception for the NNLO result at fixed scale,
that covers the next orders only within the 95\% interval.
As expected, this is improved after marginalizing over the scale, which in turn also gives
more peaked distributions at higher orders.
The uncertainty estimate of the scale independent results is certainly more conservative
than that obtained with the geometric behaviour model, and consequently very reliable but less precise.

\subsection{Higgs decay to $gg$ at N$^4$LO}
\label{sec:Hgg}

Other QCD observables known at N$^4$LO are the decay width of the Higgs boson into
a $b\bar b$ pair~\cite{Baikov:2005rw,Herzog:2017dtz} and into gluons\footnote{In the heavy top
effective theory where the interaction vertex between the Higgs and the gluons is pontlike.}~\cite{Herzog:2017dtz}.
The former shares various similarities with the $e^+e^-\to\text{hadrons}$ process discussed in Sect.~\ref{sec:epem},
and the application of our methods leads to very similar results.
Therefore, we do not consider it and just focus on the decay into gluons.
We define the observable $\Sigma$ as the decay width 
\begin{align}\label{eq:Hgg0}
\Sigma &= \Gamma_{H\to gg}(\mu),
\end{align}
which, for $\mu=\mh$, leads to the expansion~\cite{Herzog:2017dtz}
\begin{align}\label{eq:Hgg}
\Sigma &= \Gamma_0(\mh)\[1+5.70305\as(\mh) +15.5120\as^2(\mh) +12.666\as^3(\mh) +69.329\as^4(\mh)+\Ord(\as^5)\],
\end{align}
where the numerical coefficients have been computed with $n_f=5$, as appropriate,
and the LO at the generic scale $\mu$ is given by
\beq
\Gamma_0(\mu) = 
\as^2(\mu) \frac{G_F \mh^3}{36\sqrt{2}\pi^3}
=
\as^2(\mu) \times 14.43\; \text{MeV}.
\eeq
Differently from the previous example (and from the $H\to b\bar b$ decay) the LO now is scale dependent.
When computing this result at another scale, we must then account for the change at LO,
using the procedure described in Sect.~\ref{sec:appScale}.

\begin{figure}[!tb]
  \centering
  \includegraphics[width=0.53\textwidth,page=1]{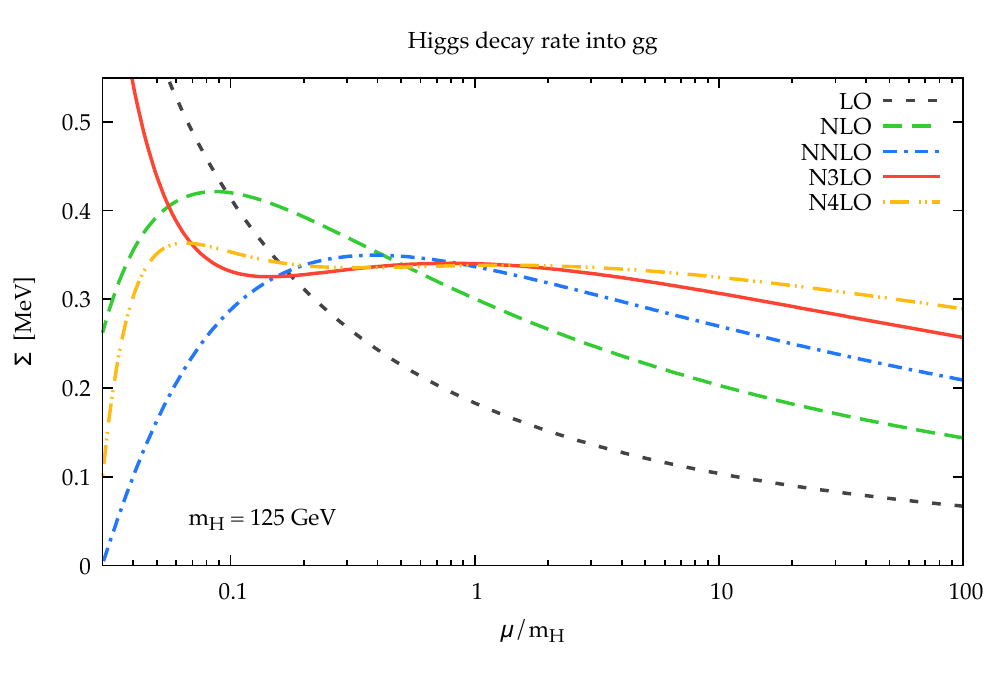}
  \includegraphics[width=0.46\textwidth,page=1]{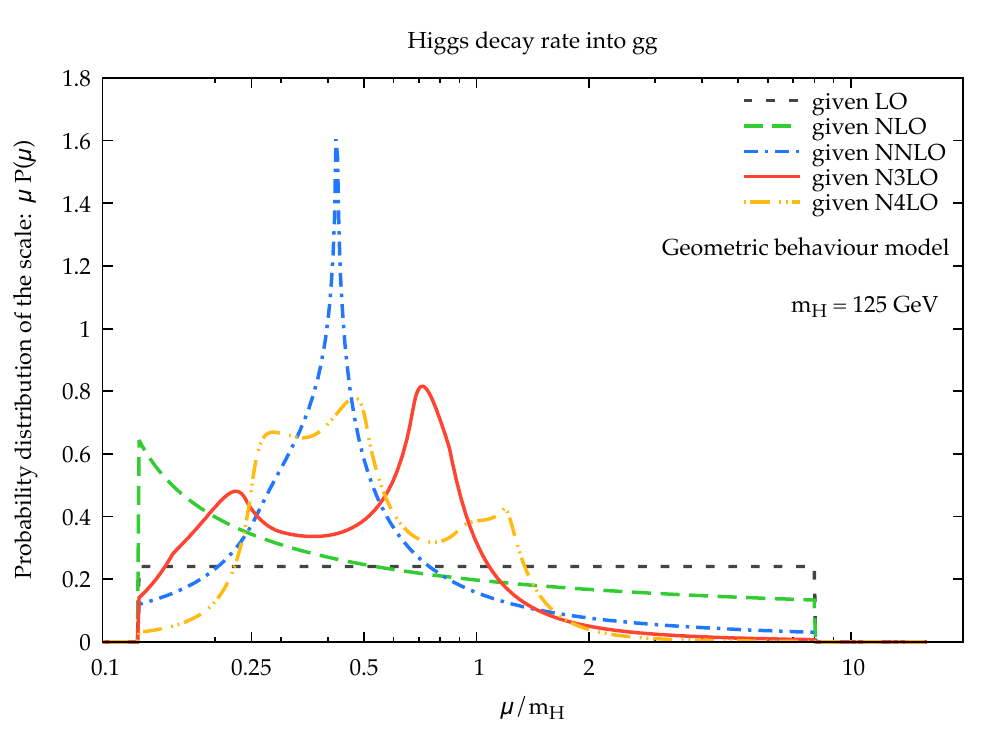}
  \caption{Left: scale dependence of the $H\to gg$ decay width.
    Right: posterior distribution for the scale for the geometric behaviour model.}
  \label{fig:Hggscale}
\end{figure}

In Fig.~\ref{fig:Hggscale} (left) we show the ``raw'' results as a function of $\mu$.
We note that the behaviour of this plot is similar to that of Higgs production in gluon fusion, Fig.~\ref{fig:ggHscale},
due to the fact that the two processes are obviously related, and in particular they both start at $\Ord(\as^2)$.
In this case, one extra order is known, thus providing an interesting and useful validation.
Note however that the perturbative correction are somewhat smaller in this case,
leading to a better convergence pattern.

\begin{figure}[p]
  \centering
  \includegraphics[width=0.48\textwidth,page=1]{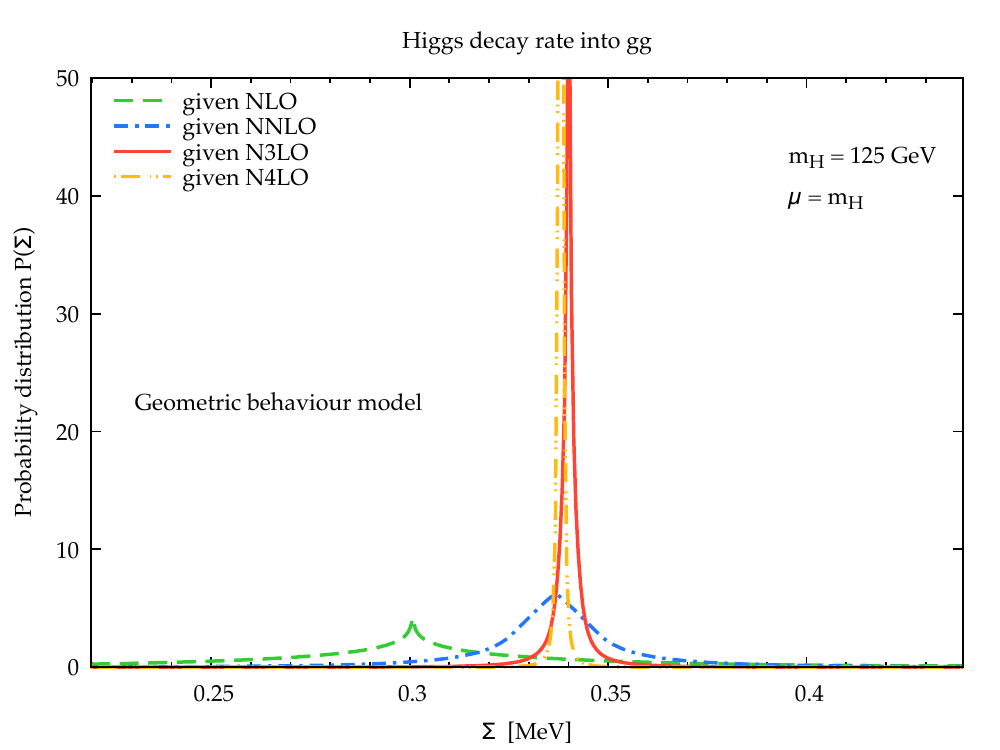}
  \includegraphics[width=0.39\textwidth,page=1]{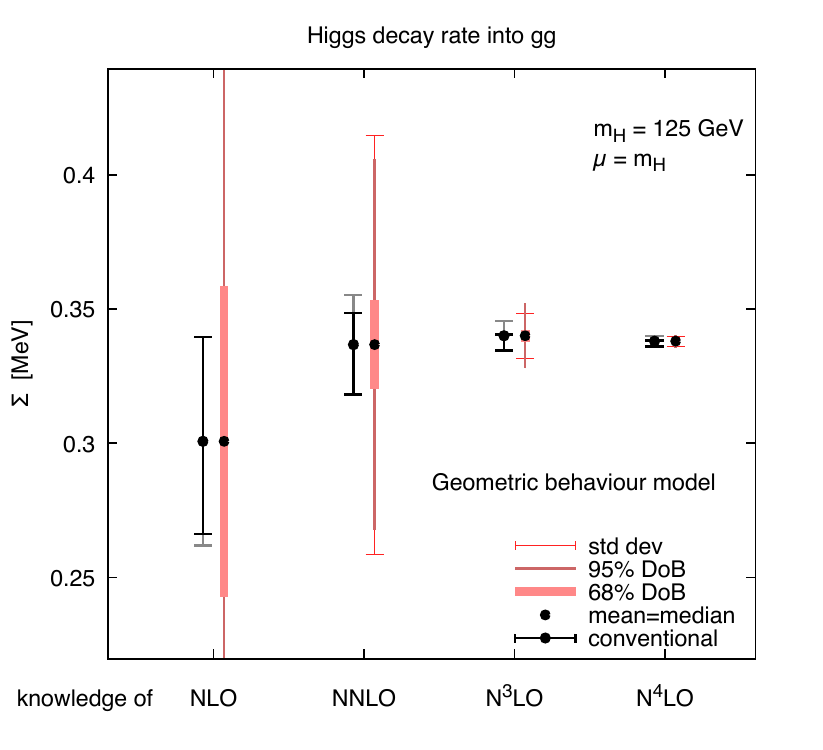}\\
  \includegraphics[width=0.48\textwidth,page=3]{./images/plot_Pdistr_Hgg_paper.pdf}
  \includegraphics[width=0.39\textwidth,page=3]{./images/Hgg_uncertainty_summary_paper.pdf}\\
  \includegraphics[width=0.48\textwidth,page=2]{./images/plot_Pdistr_Hgg_paper.pdf}
  \includegraphics[width=0.39\textwidth,page=2]{./images/Hgg_uncertainty_summary_paper.pdf}\\
  \includegraphics[width=0.48\textwidth,page=4]{./images/plot_Pdistr_Hgg_paper.pdf}
  \includegraphics[width=0.39\textwidth,page=4]{./images/Hgg_uncertainty_summary_paper.pdf}
  \caption{Fixed scale and scale independent results for the $H\to gg$ decay width,
    for both the geometric behaviour model (upper plots) and the scale variation model (lower plots).}
  \label{fig:Hggplots}
\end{figure}

We move immediately to the results of our methods, shown in Fig.~\ref{fig:Hggplots},
in the same format as Fig.~\ref{fig:epemplots}.
The scale $\mu=\mh$ used in the plots is a good one, because the perturbative expansion at this scale is very benign.
Therefore, the probability distributions at fixed scale are
rather good and converge well, with the next order always included within the 68\% DoB interval
with no exceptions.
Moreover, the precision at N$^4$LO of both models is comparable.

\afterpage{\FloatBarrier}

When we marginalize over the scale, using a flat prior for $\mu$ in the range $\mh/8<\mu<8\mh$,
we obtain slightly larger uncertainties, with improved convergence
(in particular the mean of the distributions at lower orders are more in line with the higher order results).
We observe that the geometric behaviour model generates again multimodal distributions,
this time both at N$^3$LO and N$^4$LO.
The explanation is still the same, namely that there is more than one region of scales for which the
observable converges well, thus producing multimodal posterior distributions for the scale,
as shown in Fig.~\ref{fig:Hggscale} (right).
The marginalization over the scale produces a prediction that is slightly smaller (though compatible)
with the one at $\mu=\mh$, which is an obvious consequence of the fact that for $\mu\sim\mh$ both the N$^3$LO and the N$^4$LO
have a maximum (see Fig.~\ref{fig:Hggscale}, left).

These results allow us to conclude that the estimate of the uncertainties using our methods
are robust and reliable. Given the similarity of this process to the case of Higgs production,
we can extrapolate that the methods are reliable also in that case.

\subsection{A resummed logarithmic-ordered expansion}
\label{sec:ggHres}

So far we have applied our methods to fixed-order perturbative expansions.
However, we have stressed several times that our methods can work also for
more general expansions, provided they behave in a perturbative way.
In particular, one can consider the results obtained using all-order resummations,
and consider the sequence of increasing logarithmic accuracy, namely LL, NLL, NNLL and so on.
Such an expansion is supposed to behave perturbatively, with a structure given by
(we consider for simplicity the case of a single-logarithmic enhancement)
\begin{equation}
\Sp = \sum_{k=0}^\infty g_k(\as L) \as^k,
\end{equation}
where the $g_k$ coefficients are all-order functions of their argument, $L$ being the resummed logarithm.
The resummation assumes $\as L\sim 1$, so the $g_k$ coefficients are formally coefficients of $\Ord(1)$,
and the explicit power of $\as$ determines a fully fledged perturbative expansion.

In this section, we consider
threshold resummation applied to the Higgs production process, which is known to a high order, N$^3$LL~\cite{Bonvini:2014joa}.
We have already used this result in Sect.~\ref{sec:ggHn3ll} to construct a fixed-order perturbative expansion
by expanding the resummed result. Here, instead, we take the full resummed result at various logarithmic orders,
matched (as appropriate for phenomenology) to the corresponding fixed-order result.
Namely, the perturbative expansion corresponds to the sequence LO+LL, NLO+NLL, NNLO+NNLL, N$^3$LO+N$^3$LL,
whose values at $\mu=\mh$ (and $\muf=\mh/2$) are given by
\begin{equation}\label{eq:XSggHres}
\Sp(\mh) = \left\{ 15.9, 38.7, 47.1, 48.5 \right\}\; \text{pb}
\end{equation}
(we used the same setting as Sect.~\ref{sec:ggHn3ll}, namely the default of Ref.~\cite{Bonvini:2016frm}).
Since the coefficients of the perturbative expansion are all-order functions of $\as$,
their scale dependence cannot be reconstructed using the procedure of Sect.~\ref{sec:appScale}.
Therefore, to simplify the treatment of this process, we just consider the geometric behaviour model at fixed scale.\footnote
{Of course, it is possible to retrieve the correct scale dependence of the resummed result numerically,
  e.g.\ from the \href{https://www.roma1.infn.it/~bonvini/troll/}{\texttt{TROLL}} code~\cite{Bonvini:2014joa,Bonvini:2016frm}.
  This would however slow down the computation of the uncertainty in \texttt{THunc}.}
To investigate the dependence of the results on the scale, we consider also the expansion at another scale,
$\mu=\mh/2$, leading to the sequence
\begin{equation}\label{eq:XSggHres2}
\Sp(\mh/2) = \left\{ 20.1, 46.2, 50.1, 48.6\right\}\; \text{pb}.
\end{equation}
Note that at this scale the value of the coupling is different, $\as(\mh/2)=0.1252$.

\begin{figure}[t]
  \centering
  \includegraphics[width=0.48\textwidth,page=1]{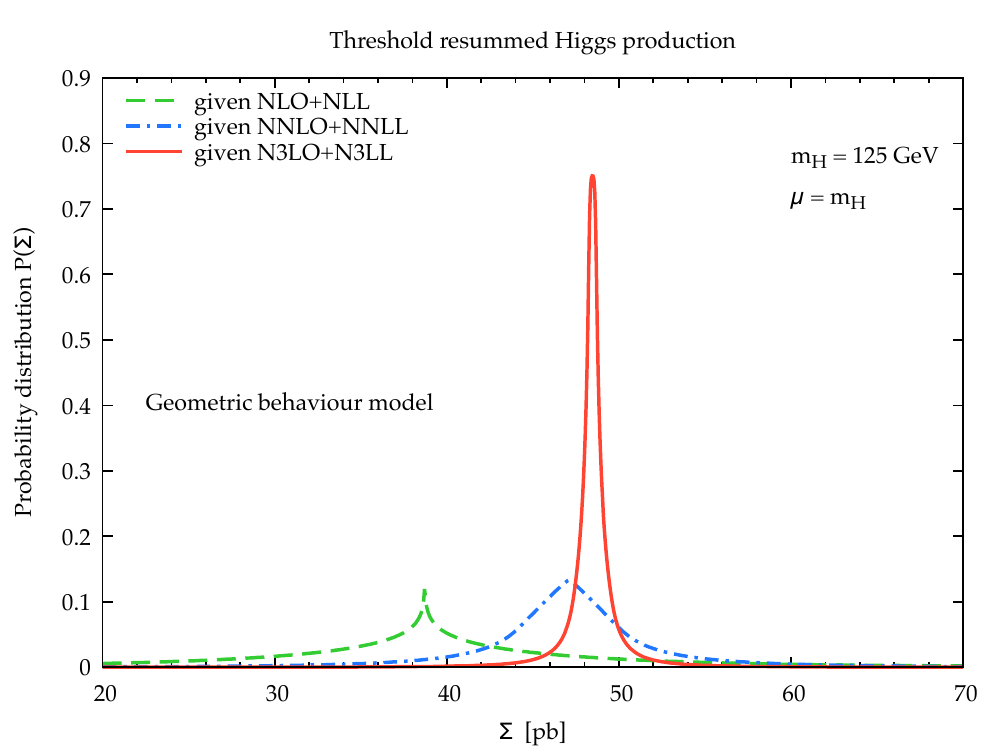}
  \includegraphics[width=0.39\textwidth,page=1]{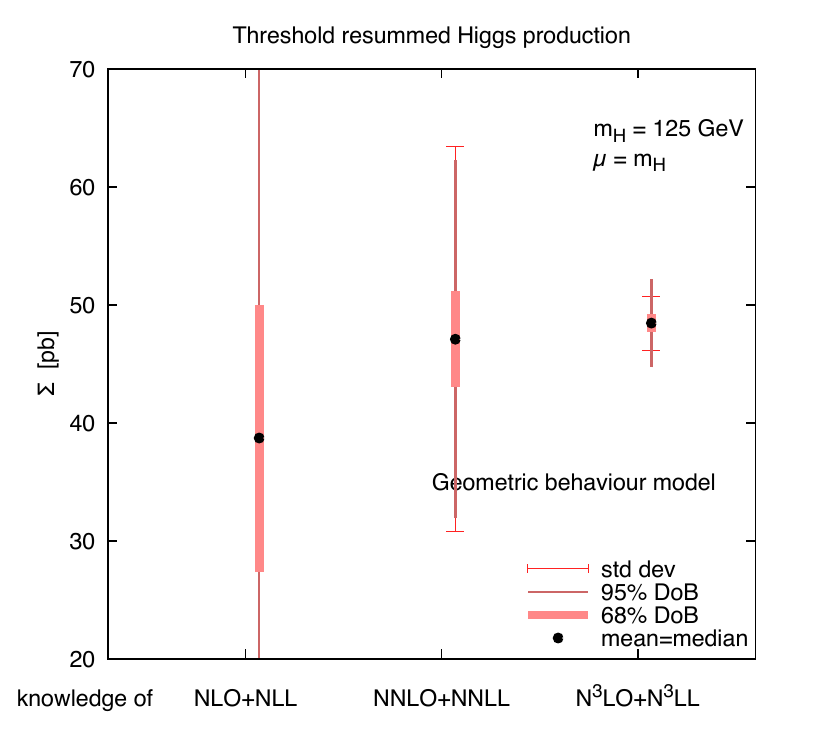}\\
  \includegraphics[width=0.48\textwidth,page=2]{./images/plot_Pdistr_ggHres_paper.pdf}
  \includegraphics[width=0.39\textwidth,page=2]{./images/ggHres_uncertainty_summary_paper.pdf}
  \caption{Probability distributions and their summaries for the Higgs production process at resummed level matched to fixed order,
    for fixed scale $\mu=\mh$ (upper plots) and $\mu=\mh/2$ (lower plots) using the geometric behaviour model.}
  \label{fig:ggHresplots}
\end{figure}

We show in Fig.~\ref{fig:ggHresplots} the results of the geometric behaviour model applied at each of the scales considered.
Because the resummation speeds up the (apparent) perturbative convergence~\cite{Bonvini:2016frm},
these results are somewhat more precise (smaller uncertainty) than those
coming from the fixed-order expansion.
Nonetheless, they remain reliable, as one can appreciate from the fact that the 68\% DoB
interval always covers the next order.
These results are also fully compatible with those coming from the fixed-order perturbative expansion.
A thorough study of the uncertainties of this (or any other) process is beyond the scope of this work
and will be done elsewhere.

\section{Treating correlations between different theory predictions}
\label{sec:corr}

The approach and models presented so far can be used to compute the uncertainty
on a given perturbative expansion, namely for the prediction of a given observable.
In many cases, one needs to consider more than one observable at a time in a physics analysis.
In these cases, correlations between the uncertainties of different observables are fundamental.

We stress that when we say different observables we actually also mean the same physical observable at different kinematic points.
Consider for example a differential observable, like a transverse momentum distribution.
The methods presented so far would work independently for any value of the transverse momentum $\pt$,
as each value of $\pt$ leads to a different perturbative expansion,
in principle with very different perturbative behaviours.\footnote
{Consider for example the region of small $\pt$ where the presence of logarithms of $\pt$
invalidate the perturbative expansion and requires the all-order resummation of these logarithms.}
It's clear that in such cases there are strong correlations between different values of $\pt$,
approaching 100\% for close values of $\pt$.
But there may be also important correlations for $\pt$ values very far from each other,
induced by the knowledge of the full cross section (integral of the distribution).
Similarly, when one consider different distributions for the same physical process,
there likely are correlations on the uncertainty from missing higher orders, e.g.\ due to
the fact that all distributions must integrate to the same total cross section.

The situation is somewhat different when considering different processes.
Here it is less clear whether the uncertainties should be correlated or not.
Note that there is a class of uncertainties, due to the uncertain values of parameters
such as couplings or masses, that are clearly (and easily) correlated between different observables of different processes.
However, here we are dealing with the uncertainty due to missing higher orders.
The presence of correlations in this case implies that the perturbative expansions of the different observables
of different processes are somehow related.
This is not to be excluded a priori, as indeed perturbative expansions often show some similarities.\footnote
{This is for instance the case when there is a common dynamical mechanism giving the bulk of the higher order corrections,
as it happens in kinematic regions with logarithmic enhancements.}
However, quantifying such similarities is hard in general and subject to a large degree of arbitrariness.

Finding a general and satisfactory solution to all these cases is complicated (if not impossible)
and it will not be subject of this work.
In the following, we limit ourselves to discuss some simple cases and propose some strategies.
This section has just to be considered as a starting point for future work in this direction.

\subsection{Correlations for the same observable at different kinematic points}
\label{sec:corrSameObs}

Let us start our discussion from the case where there certainly are correlations,
namely the case of the same observable at different kinematic points.
Let us call by $\vec v$ and $\vec w$ generic configurations of kinematic variables,
and by $\Sigma(\vec v)$ and $\Sigma(\vec w)$ the observable that depends on those variables.
The information on the correlation between $\Sigma(\vec v)$ and $\Sigma(\vec w)$ is contained in the joint distribution
\begin{equation}
P(\Sigma(\vec v), \Sigma(\vec w) | \delta_n(\vec v),...,\delta_1(\vec v),\Sigma_0(\vec v), \delta_n(\vec w),...,\delta_1(\vec w),\Sigma_0(\vec w)),
\end{equation}
where each term of the perturbative expansion is computer either at $\vec v$ or at $\vec w$.
We recall that, in the limit $\vec w\to\vec v$, the distribution must lead to a positive 100\% correlation.

Constructing such joint distribution satisfying the aforementioned requirement is far from trivial.
The most promising way that we can imagine consists in considering not the values of the observable
at given kinematic points, but the observable as a \emph{function} of the kinematics.
The probability to be computed is that of the function,
and the requirement that the correlation of adjacent points tends to one is automatic
provided obvious conditions of continuity and smoothness are included.
Also, the condition that the integral of the distribution is the inclusive cross section is easy to implement.
Note that, in general, with this approach the probability distribution of a single point would not necessarily correspond
with that obtained from the application of the methods introduced in the previous sections.
In particular, one could expect to obtain a smaller uncertainty with this approach,
given the additional constraints included.

How to realize in practice this approach is again subject of arbitrary choices.
A promising option consists in using the proposal of Ref.~\cite{Tackmann:2020xxx}.
The idea is simple.
There are classes of higher order contributions whose functional structure is known to all orders from resummation techniques.
These do not capture all the possible missing higher orders,
but in some kinematic regions they are quite accurate, and they can be constructed
such that in other regions they anyway represent to some accuracy the full result.
In this way, the only missing information for constructing the next orders is
the value of some numerical coefficients which the functional structure of the resummation
depends upon.
In other words, rather than using a generic set of functions to model the observable
one restricts the attention to a physically motivated subset parametrized by a small number of real parameters.
In this way the inference is to be done on the parameters themselves.
Moreover, these parameters are typically power series in the coupling,
and therefore one can use straight away the probabilistic methods introduced in this work.\footnote
{More precisely, since these parameters are not physical observables,
they are not scale independent to all orders and therefore the scale variation model
of Sect.~\ref{sec:ModelScaleVar} cannot be used.
One can instead use the geometric behaviour model of Sect.~\ref{sec:GB}.
The removal of scale dependence of Sect.~\ref{sec:scaleindep} cannot be applied to the parameters themselves,
again because they are not physical and they do legitimately depend on unphysical scales,
but it can be applied at the end to the physical observable, which has to be scale independent.}
The actual implementation of this approach will be studied elsewhere.

\subsection{Correlations among observables of different processes}
\label{sec:corrDiffObs}

We now move to discussing the correlations among observables belonging to different physical processes.
In this case, it is not clear at all that the uncertainty due to missing higher orders should be correlated:
the computations are different, the number of known orders is in general different,
the perturbative series describing the various observables are different.
Therefore, at first sight, one is tempted to say that the uncertainty from missing higher orders
between observables of different process are uncorrelated.

This conclusion is likely correct for very different processes.
However, some processes may be characterized by perturbative expansions with common features,
for instance when the underlying physical mechanism at the origin of
some kind of corrections is the same.
This is for instance the case when the bulk of perturbative corrections is given by
logarithmically enhanced contributions, which in turn originate from a specific
structure of particle emissions.
Similarly, processes like $Z$ and $W$ production at proton-proton colliders
(or equivalently neutral-current and charged-current structure functions in DIS experiments) are very similar,
as they share the same structure of the interaction and thus the form of the perturbative corrections.
In these cases it makes sense to assume that correlations between the uncertainties from missing higher orders
of the given processes are present.

The tough question is how to quantify such correlations.
To keep the discussion very general, we may assume that \emph{some} of the parameters
characterizing the probabilistic model are common to two given observables $\Sigma$ and $\Sigma'$
of different processes.
If this is the case, the joint distribution can be written as
\begin{align}\label{eq:joint2obs}
  P(\Sigma &,\Sigma' | \delta_n,...,\delta_1,\Sigma_0, \delta_n',...,\delta_1',\Sigma_0') \\
  &\propto P(\Sigma, \Sigma', \delta_n,...,\delta_1,\Sigma_0, \delta_n',...,\delta_1',\Sigma_0') \nonumber\\
  &= \int d\vec p\, P(\Sigma, \Sigma', \delta_n,...,\delta_1,\Sigma_0, \delta_n',...,\delta_1',\Sigma_0',\vec p) \nonumber\\
  &= \int d\vec p\, P(\Sigma | \delta_n,...,\delta_1,\Sigma_0, \vec p)
P(\Sigma'|\delta_n',...,\delta_1',\Sigma_0',\vec p)
P(\delta_n,...,\delta_1,\Sigma_0|\vec p)
P(\delta_n',...,\delta_1',\Sigma_0'|\vec p)
P_0(\vec p) \nonumber
\end{align}
where $\vec p$ represents the vector of parameters of the model that are shared between the observables.
Of course the model may depend on additional parameters, but these are assumed to be independent for the two observables,
and therefore do not enter our decomposition.
Eq.~\eqref{eq:joint2obs} depends on the probability of each observable given its respective perturbative expansion
\emph{and} the common parameters $\vec p$, and similarly on the probability of the perturbative expansions
themselves given $\vec p$.
The actual form of these probability distributions depend on what is $\vec p$, but once the model is specified
and the set of common parameters identified, then Eq.~\eqref{eq:joint2obs} can be used directly,
as it depends on ingredients introduced in the previous sections.

The choice of the common parameters $\vec p$ depends on the processes under consideration and it is the trickiest part of the job.
For instance, one model-independent parameter that could be used to construct such correlations is the unphysical scale $\mu$,
or better its ratio to the hard scale of the process.
In principle, the scale dependence of each process is independent, however it is induced in all cases by the
dependence of the coupling on the scale.
For this reason, one can assume that the variation of the observable induced by a variation of the scale
is somehow related, and use it to construct the sought correlation.
This reasoning may however lead to some problems, as for instance obtaining such correlations
requires using the same prior $P_0(\mu/Q)$, with $Q$ the hard scale of each process,
which makes impossible to adjust this prior to process-dependent conditions.\footnote
{For instance, if the hard scale $Q$ of one of the (QCD) processes under consideration is small,
one does not want to let the prior probe values much smaller than $Q$, otherwise
the observable would get contribution with too large strong coupling.}
In agreement with this conclusion, in recent work~\cite{AbdulKhalek:2019bux,AbdulKhalek:2019ihb}
on the inclusion of theory uncertainties in PDF fits, the dependence on the renormalization scale is not considered
as a source of correlation among different \emph{classes} of processes,
while it is considered for similar processes like $Z$ and $W$ production belonging to the same class (Drell-Yan).

An alternative way to construct correlations between different but somehow related processes
consists in using again the approach of Ref.~\cite{Tackmann:2020xxx}, as we described in Sect.~\ref{sec:corrSameObs}.
Indeed, some of the parameters used in the construction of the resummed results at the basis of that approach
are process dependent, but some others are more universal, and common among different processes.
For instance, the cusp anomalous dimension
is one of the key ingredients of soft-gluon resummations for any process.
Therefore, in Eq.~\eqref{eq:joint2obs} one can take $\vec p$ to be the model parameters
describing the perturbative expansion of the common coefficients (such as the cusp anomalous dimension).
The probability distribution would be constructed exactly as one would do for Sect.~\ref{sec:corrSameObs}.
We conclude that the approach of Ref.~\cite{Tackmann:2020xxx} used in conjunction with our methods
is very promising for estimating theory uncertainties from missing higher orders in a way that produces
meaningful correlations between different observables, and deserves further studies.

We conclude by mentioning that the situation becomes more complicated for hadron-initiated processes,
due to the presence of parton distribution functions (PDFs).
Indeed, PDFs induce two additional problems.
One is that they depend on a different unphysical scale, the factorization scale $\muf$.
If the PDFs were known exactly, this would not be too problematic,
as one could extend the approach of Sect.~\ref{sec:scaleindep} to $\muf$ as well.
However, PDFs are non-perturbative objects and cannot be computed from the theory.
Therefore, they are determined by a fit to data, using a theoretical computation for the perturbative coefficients
describing an observable as input.
This means that PDFs indirectly depend on the accuracy of the computation of the perturbative coefficients,
which is the second problem.
Ideally, one would like to include theory uncertainties with proper correlations in the determination of the PDFs themselves,
as studied recently in Refs.~\cite{AbdulKhalek:2019bux,AbdulKhalek:2019ihb} (using canonical scale variation).
The trouble is that the $\muf$ dependence is thus entangled with the theory uncertainty,
and it is not obvious how to properly deal with it (see e.g.\ Ref.~\cite{Harland-Lang:2018bxd}).
This important topic deserves a dedicated study.

\section{Conclusions}
\label{sec:conclusions}

Having a reliable estimate of the uncertainty on theoretical predictions is of fundamental importance for (particle) physics.
A particularly important source of uncertainty is the one coming from the finite perturbative accuracy
of theoretical computations.
Quantifying the uncertainty due to unknown (or missing) higher orders is crucial,
especially for those theories like QCD that are characterized by large perturbative corrections.
In these cases the missing higher orders may be sizeable, and underestimating them may lead to wrong conclusions
when comparing theoretical predictions with precise data.

The standard practice to estimate the size of the missing higher orders is based
on the ``canonical scale variation'' method, where the unphysical scale(s) is varied
by a factor of 2 about an arbitrary central scale. Since scale dependence is formally subleading,
the effect of the variation is formally higher order. The problem of this approach
appears when one tries to promote this information to an uncertainty.
Not only the procedure is totally arbitrary and has no probabilistic foundation;
it also often underestimates the true uncertainty, and it is thus not reliable.

A groundbreaking approach
to define and compute the uncertainty from missing higher orders
has been introduced by Cacciari and Houdeau in 2011~\cite{Cacciari:2011ze}.
This method uses a Bayesian approach to infer the probability distribution for an observable
from its perturbative expansion, under some conditions that define the Cacciari-Houdeau (CH) model.
Later, the CH model has been modified to account for the divergence of the perturbative expansion~\cite{Bagnaschi:2014wea}.
While this approach is formally superior to canonical scale variation,
it also has limitation. Indeed, in some cases the CH model is not very robust
and it does not predict reliably the uncertainty from the missing higher orders.
This behaviour is due to the assumptions of the model itself, and not to the approach per se.

In this work we built upon the CH model to construct more general, flexible and reliable models,
using the same Bayesian approach. Moreover, we have introduced an innovative way to
eliminate an ambiguity of theoretical computations in quantum field theory,
namely the dependence on unphysical scale(s).
We have validated the models using perturbative expansions with a known sum,
and tested their quality on some observables that are known to a high perturbative order.

The models proposed in this work are essentially two, with a number of possible variants discussed in the appendices.
One model is a generalization of the CH model, and assumes that the various orders $k$
of the perturbative expansion are bounded by a geometric behaviour $ca^k$,
where both $c$ and $a$ are (hidden) model parameters.
The main difference with respect to the CH model is that $a$,
that plays the role of the expansion parameter of the geometric bound,
in the CH approach is fixed to be the coupling constant, possibly up to an externally supplied fixed factor.
Promoting $a$ to a model parameter allows to select its most appropriate values
based only on information from the expansion itself, through a probabilistically valid inference procedure.
This improvement alone already makes the model much more reliable and robust.
In practical applications, it performs very well, providing precise and accurate uncertainties.
These are also pretty stable upon variations of the assumptions, as the bulk of the distributions
is rather insensitive to the precise form of the prior, which mostly affects the tails.

The other model uses the information from the scale variation,
similarly to the canonical scale variation method, but in a probabilistically well defined way.
In particular, it assumes that the next order is bounded by a factor $\lambda$
times a properly defined scale variation coefficient of the current order.
The factor $\lambda$ is a hidden parameter of the model and it is inferred from the known orders.
Oversimplifying, this model can be interpreted as the canonical scale variation method where the
size of the variation, rather than being fixed to a factor of 2, is variable and inferred from the expansion itself.
We have noticed that this model predicts somewhat larger uncertainties than the geometric behaviour model,
thus providing a more conservative estimate of the theory uncertainty.
This uncertainty is very stable upon variation of the prior distribution.

There are a number of improvements that can be applied to the models,
discussed in the appendices.
Two of them are particularly interesting.
One suggests to impose an additional condition on the scale dependence of the inferred higher orders:
indeed, not only the various perturbative contributions should get smaller and smaller
by increasing the order, but also their scale dependence should, as it is always formally of the next order
and must thus behave perturbatively.
Requesting a reduction of the scale dependence provides strong constraints on the missing higher orders,
resulting in smaller uncertainties (i.e., higher precision).
Another improvement consists in looking for a sign pattern in the various perturbative orders,
to infer the next orders according to the same pattern.
Also this procedure allows to reduce the uncertainty, producing more precise results.
Moreover, different conditions can be combined together to provide
more stringent models that can further improve the precision.

Which model to choose is a duty of the user.
A ``perfect'' or ``correct'' model does not exist, as the only way to be sure about the size of the missing
higher orders is computing them, and any estimate is subject to arbitrary assumptions and biases.\footnote
{Conversely, ``bad'' or ``wrong'' models do exist, and can be discarded based e.g.\ on an unsuccessful validation.}
The Bayesian approach highlights the presence of such subjective assumptions,
making them very explicit (the model, the priors).
Users, based on their beliefs about the expansion, can freely choose the model and priors, declaring them. 
This said, we now give a \emph{personal} preference.
From the studies presented in this work, it is clear that the geometric behaviour model
provides precise results without sacrificing accuracy. Given the simplicity of the model
and the fast implementation provided, we suggest its use as default.
Moreover, this is the only model
that is applicable beyond a quantum field theory, and it is thus more general and universal.
Since the plain scale variation model is simple and fast, for QCD observables we also recommend
its application as a ``double check'', consistently with its conservative interpretation.
Should one be interested in providing more aggressive results,
the best performance is obtained by combining all the proposed assumptions,
Eqs.~\eqref{eq:GBhyp}, \eqref{eq:SChyp} and \eqref{eq:SChyp2},
and keeping track of the sign pattern as discussed in App.~\ref{sec:sign}.
Such an approach makes the most of the known orders, and it is thus
particularly useful when only a low number of orders is available.
However, we suggest its use only for very specific and dedicated studies and not as a general-purpose approach,
also due to its slow numerical implementation.
In all cases, variations of the priors should be considered to assess the stability of the results upon
changes of the subjective arbitrary input.

Independently of the model used, we have proposed a general way to remove
the dependence on the renormalization scale from the probability distribution of the observable
(the method can be extended also to other unphysical scales).
Indeed, each model uses as basic elements the coefficients of the perturbative expansion,
which however depend on the unphysical scale and would thus lead to infinitely many
different results depending on the choice of the scale.
The solution is actually very simple: promoting the unphysical scale to be a parameter of the model.
The inference works exactly as in the general case, Eqs.~\eqref{eq:inference1} and \eqref{eq:inference2},
with the difference that among the hidden parameters there is also the scale, with its own prior.
In this way, after marginalizing over the scale, the dependence disappears.
The original arbitrariness in the choice of scale is almost completely removed
--- only a mild arbitrariness remains in the choice of the prior.
It has to be stressed that the inference on the scale acts differently depending on the model,
selecting those values of the scale for which the given model performs better.
In this way, the most accurate and precise results are obtained for free,
without the need of choosing wisely the scale corresponding to a ``good'' perturbative expansion.
We also stress that with this procedure the ``best'' prediction for an observable is
given by the mean of its probability distribution,
in contrast with standard methods, where it
corresponds to the value of the perturbative expansion computed at the highest known order at an arbitrary ``central'' scale.
Therefore not only the method provides a reliable uncertainty, but also a better central value for the observable.
For scale dependent observables, we always recommend the use of this procedure.

We have finally considered the problem of quantifying the correlations between various theoretical predictions
and their uncertainties. We believe that this argument cannot be addressed in full generality,
but specific solutions need to be put in place depending on the observables and the processes under considerations.
Some proposals have been suggested, but a thorough study is delayed to future work.

The results of this paper can be used to compute the uncertainty on any physical observable
through a publicly released code named \texttt{THunc}, available at
\begin{center}
  \href{https://www.roma1.infn.it/~bonvini/THunc}{\tt www.roma1.infn.it/$\sim$bonvini/THunc}
\end{center}
The two main models proposed in this work are implemented analytically and lead to a very fast evaluation of the uncertainties.
Other models (including those proposed in the appendices or alternatives invented by the user)
can be implemented straight away through a custom model feature, at the price of a slower numerical evaluation.

\acknowledgments
{
I am grateful to Enrico Franco, Giulio D'Agostini and Sara Borroni for very useful discussions on probability theory related to this work.
I also thank Luca Rottoli and Simone Marzani for useful comments to the manuscript,
and Franz Herzog, Juan Rojo and Stefano Forte for discussions.
This work was supported by the Marie Sk\l{}odowska Curie grant HiPPiE@LHC under the agreement n.~746159.
}

\appendix
\section{Algorithms}
\label{sec:appAlgos}

\subsection{Computing renormalization scale dependence at fixed order}
\label{sec:appScale}

In this appendix we show how to reconstruct the renormalization scale dependence of physical observables
starting from the perturbative expansion at a given scale $\mu_0$. We consider the case of QCD for definiteness,
but the results can be easily extended to other quantum field theories.

Let us assume we know the expansion of an observable as
\begin{equation}\label{eq:Snmu0}
\Sigma_n(\mu_0) = \sum_{k=0}^n \bar c_k \az^{k+k_0},
\end{equation}
where $\bar c_k\equiv c_k(\mu_0)$ are the (known) coefficients of the expansion computed at the scale $\mu=\mu_0$
up to fixed order $k=n$,
and $\az=\as(\mu_0)$ is the value of the strong coupling at this scale, assumed to be known as well.
We also included in the expansion an offset $k_0$ to account for expansions that start at $\Ord(\as^{k_0})$.
We now wish to rewrite the perturbative expansion at a different scale $\mu$
\begin{equation}\label{eq:Snmu}
\Sigma_n(\mu) = \sum_{k=0}^n c_k(\mu) \as^{k+k_0}(\mu),
\end{equation}
which differs from $\Sigma_n(\mu_0)$, Eq.~\eqref{eq:Snmu0}, by subleading terms, because $\Sigma$ is scale independent
(note that this is true only if $\Sigma$ is a physical observable).

Because the scale dependence of the coupling is known, one can reconstruct the scale dependence of the expansion coefficients
by imposing the scale independence of $\Sigma$.
Using the definition of the $\beta$-function from the Callan-Symanzik equation
\begin{align}
\mu^2\frac{d}{d\mu^2} \as(\mu)
 &= \beta(\as(\mu)) \nonumber\\
 &= -\as^2(\mu)\Big[\beta_0+\beta_1 \as(\mu) + \beta_2 \as^2(\mu) +\beta_3 \as^3(\mu) +\beta_4 \as^4(\mu) +\ldots\Big]
\end{align}
one can obtain a system of differential equations for the coefficients by imposing
\begin{equation}
0 = \mu^2\frac{d}{d\mu^2} \Sigma = 
\sum_{k=0}^n \[\mu^2\frac{d}{d\mu^2} c_k(\mu) \as^{k+k_0}(\mu) + c_k(\mu)(k+k_0) \as^{k+k_0-1}(\mu) \beta(\as(\mu))\] +\Ord(\as^{n+1+k_0}).
\end{equation}
Expanding the term in square brackets in powers of $\as$, the system can be solved iteratively order by order,
using as initial condition the values $\bar c_k$ of the coefficients at $\mu_0$.

This way to proceed, though perfectly valid, is not very convenient in practice.
It is indeed more efficient to consider directly the solution of the Callan-Symanzik equation
to express $\az\equiv \as(\mu_0)$ in terms of $\as(\mu)$, plugging this into Eq.~\eqref{eq:Snmu0},
and finally expanding in powers of $\as(\mu)$ and read off the resulting coefficients $c_k(\mu)$
by comparison with Eq.~\eqref{eq:Snmu}.
The solution of the running coupling equation up to five loops~\cite{Baikov:2016tgj}
and expanded in powers of $\as$ as appropriate is given by
\begin{align}
\az &= \as(\mu)\Bigg[1+\as(\mu)\beta_0\lr + \as^2(\mu)\(\beta_0^2\lr^2+\beta_1\lr\)
+ \as^3(\mu)\(\beta_0^3\lr^3+\frac52\beta_1\beta_0\lr^2+\beta_2\lr\) \nonumber\\
&\qquad\qquad +\as^4(\mu) \(\beta_0^4 \lr^4  + \frac{13}{3} \beta_0^2 \beta_1 \lr^3  + \(3 \beta_0\beta_2 +\frac32 \beta_1^2\)\lr^2
   +\beta_3 \lr\) \nonumber\\
&\qquad\qquad +\as^5(\mu) \(\beta_0^5 \lr^5  + \frac{77}{12} \beta_0^3 \beta_1 \lr^4
  + \(6\beta_0^2\beta_2 +\frac{35}6 \beta_0\beta_1^2\)\lr^3
  + \frac72 \(\beta_0\beta_3 +\beta_1\beta_2\)\lr^2
   +\beta_4 \lr\) \nonumber\\
&\qquad\qquad + \Ord(\as^6) \Bigg],
\end{align}
with
\begin{equation}
\lr = \log\frac{\mu^2}{\mu_0^2}.
\end{equation}
Obtaining the sought result is now a mere exercise of replacement and expansion.
For completeness, we report here the results for the first few coefficients $c_k(\mu)$
for three relevant values of $k_0$, namely $k_0=0,1,2$.
For $k_0=0$ (processes starting at $\Ord(\as^0)$ at LO) we have
\begin{align}
c_1(\mu) &= \bar c_1 \nonumber\\
c_2(\mu) &= \bar c_2 + \beta_0 \bar c_1 \lr \nonumber\\
c_3(\mu) &= \bar c_3 + \(\beta_1 \bar c_1 + 2 \beta_0 \bar c_2\) \lr + \beta_0^2 \bar c_1 \lr^2 \nonumber\\
c_4(\mu) &= \bar c_4 + \(\beta_2 \bar c_1 + 2 \beta_1 \bar c_2 + 3\beta_0 \bar c_3\) \lr + \(\frac52\beta_0\beta_1\bar c_1+3\beta_0^2\bar c_2\)\lr^2 + \beta_0^3 \bar c_1 \lr^3 \nonumber\\
c_5(\mu) &= \bar c_5 + \(\beta_3 \bar c_1 + 2 \beta_2 \bar c_2 + 3\beta_1 \bar c_3 + 4\beta_0\bar c_4\) \lr
           + \[\(\frac32\beta_1^2+3\beta_0\beta_2\)\bar c_1 + 7\beta_0\beta_1\bar c_2 +6\beta_0^2\bar c_3\]\lr^2 \nonumber\\
         & + \(\frac{13}3\beta_0^2\beta_1\bar c_1+4\beta_0^3\bar c_2\)\lr^3
           + \beta_0^4 \bar c_1 \lr^4 \nonumber\\
c_6(\mu) &= \bar c_6 + \(\beta_4 \bar c_1 + 2 \beta_3 \bar c_2 + 3\beta_2 \bar c_3 + 4\beta_1\bar c_4 + 5\beta_0\bar c_5\) \lr \nonumber\\
         &+ \[\frac72\(\beta_1\beta_2+\beta_0\beta_3\)\bar c_1 + \(4\beta_1^2+8\beta_0\beta_2\)\bar c_2
           +\frac{27}2\beta_0\beta_1\bar c_3 +10\beta_0^2\bar c_4\]\lr^2 \\
         & + \[\(\frac{35}6\beta_0\beta_1^2+6\beta_0^2\beta_2\)\bar c_1 + \frac{47}3\beta_0^2\beta_1\bar c_2 + 10\beta_0^3\bar c_3\]\lr^3
           + \(\frac{77}{12}\beta_0^3\beta_1\bar c_1 + 5\beta_0^4\bar c_2\)\lr^4
           + \beta_0^5 \bar c_1 \lr^5. \nonumber
\end{align}
For $k_0=1$ (processes starting at $\Ord(\as)$ at LO) we have
\begin{align}
c_1(\mu) &= \bar c_1 + \beta_0 \bar c_0 \lr \nonumber\\
c_2(\mu) &= \bar c_2 + \(\beta_1 \bar c_0 + 2 \beta_0 \bar c_1\) \lr + \beta_0^2 \bar c_0 \lr^2 \nonumber\\
c_3(\mu) &= \bar c_3 + \(\beta_2 \bar c_0 + 2 \beta_1 \bar c_1 + 3\beta_0 \bar c_2\) \lr
           + \(\frac52\beta_0\beta_1\bar c_0 +3\beta_0^2\bar c_1\)\lr^2 + \beta_0^3 \bar c_0 \lr^3 \nonumber\\
c_4(\mu) &= \bar c_4 + \(\beta_3 \bar c_0 + 2 \beta_2 \bar c_1 + 3\beta_1 \bar c_2 + 4\beta_0\bar c_3\) \lr
           + \[\(\frac32\beta_1^2+3\beta_0\beta_2\)\bar c_0 + 7\beta_0\beta_1\bar c_1 +6\beta_0^2\bar c_2\]\lr^2 \nonumber\\
         & + \(\frac{13}3\beta_0^2\beta_1\bar c_0+4\beta_0^3\bar c_1\)\lr^3
           + \beta_0^4 \bar c_0 \lr^4 \nonumber\\
c_5(\mu) &= \bar c_5 + \(\beta_4 \bar c_0 + 2 \beta_3 \bar c_1 + 3\beta_2 \bar c_2 + 4\beta_1\bar c_3 + 5\beta_0\bar c_4\) \lr \nonumber\\
         &+ \[\frac72\(\beta_1\beta_2+\beta_0\beta_3\)\bar c_0 + \(4\beta_1^2+8\beta_0\beta_2\)\bar c_1
           +\frac{27}2\beta_0\beta_1\bar c_2 +10\beta_0^2\bar c_3\]\lr^2 \\
         & + \[\(\frac{35}6\beta_0\beta_1^2+6\beta_0^2\beta_2\)\bar c_0 + \frac{47}3\beta_0^2\beta_1\bar c_1 + 10\beta_0^3\bar c_2\]\lr^3
           + \(\frac{77}{12}\beta_0^3\beta_1\bar c_0 + 5\beta_0^4\bar c_1\)\lr^4
           + \beta_0^5 \bar c_0 \lr^5. \nonumber
\end{align}
For $k_0=2$ (processes starting at $\Ord(\as^2)$ at LO) we have
\begin{align}
c_1(\mu) &= \bar c_1 + 2\beta_0 \bar c_0 \lr \nonumber\\
c_2(\mu) &= \bar c_2 + \(2\beta_1 \bar c_0 + 3 \beta_0 \bar c_1\) \lr + 3\beta_0^2 \bar c_0 \lr^2 \nonumber\\
c_3(\mu) &= \bar c_3 + \(2\beta_2 \bar c_0 + 3 \beta_1 \bar c_1 + 4\beta_0 \bar c_2\) \lr
           + \(7\beta_0\beta_1\bar c_0 +6\beta_0^2\bar c_1\)\lr^2 + 4\beta_0^3 \bar c_0 \lr^3 \nonumber\\
c_4(\mu) &= \bar c_4 + \(2\beta_3 \bar c_0 + 3 \beta_2 \bar c_1 + 4\beta_1 \bar c_2 + 5\beta_0\bar c_3\) \lr
           + \[\(4\beta_1^2+8\beta_0\beta_2\)\bar c_0 + \frac{27}2\beta_0\beta_1\bar c_1 +10\beta_0^2\bar c_2\]\lr^2 \nonumber\\
         & + \(\frac{47}3\beta_0^2\beta_1\bar c_0+10\beta_0^3\bar c_1\)\lr^3
           + 5\beta_0^4 \bar c_0 \lr^4 \nonumber\\
c_5(\mu) &= \bar c_5 + \(2\beta_4 \bar c_0 + 3 \beta_3 \bar c_1 + 4\beta_2 \bar c_2 + 5\beta_1\bar c_3 + 6\beta_0\bar c_4\) \lr \nonumber\\
         &+ \[9\(\beta_1\beta_2+\beta_0\beta_3\)\bar c_0 + \(\frac{15}2\beta_1^2+15\beta_0\beta_2\)\bar c_1
           +22\beta_0\beta_1\bar c_2 +15\beta_0^2\bar c_3\]\lr^2 \\
         & + \[\(\frac{59}3\beta_0\beta_1^2+20\beta_0^2\beta_2\)\bar c_0 + 37\beta_0^2\beta_1\bar c_1 + 20\beta_0^3\bar c_2\]\lr^3
           + \(\frac{57}2\beta_0^3\beta_1\bar c_0 + 15\beta_0^4\bar c_1\)\lr^4
           + 6\beta_0^5 \bar c_0 \lr^5. \nonumber
\end{align}
Note that the LO coefficient remains always the same, $c_0=\bar c_0$, as it is always scale independent.
In all cases we have written all the coefficients whose scale dependence can be computed
from the known terms of the $\beta$-function, namely up to N$^6$LO for $k_0=0$ and up to N$^5$LO for $k_0>0$.

\subsection{Computing the coefficients for the geometric behaviour model}
\label{sec:appAlgo}

In this appendix we describe an algorithm to compute the coefficients $a_k$
defined in Eq.~\eqref{eq:GBakdef}, needed for an efficient implementation of the geometric behaviour model.
We report here the definition Eq.~\eqref{eq:GBakdef},
\begin{equation}
a_{k+1}< a<a_k
\qquad\Leftrightarrow\qquad
\max\[1,\frac{\abs{\delta_1}}{a},...,\frac{\abs{\delta_m}}{a^m}\] = \frac{\abs{\delta_k}}{a^k}
,
\end{equation}
assuming $a_0\equiv\infty$ and $a_{m+1}\equiv0$.
This definition means that the range $0\leq a<\infty$ is partitioned in consecutive intervals
and, in each interval, the max function selects one of its arguments.
Since the arguments of the max function contain powers of $a$ that grow with $k$,
the $a_k$ coefficients are ordered according to
\begin{equation}\label{eq:akordering}
\infty\equiv a_0 > a_1\geq ...\geq a_k\geq a_{k+1}\geq ...\geq a_m\geq a_{m+1}\equiv0.
\end{equation}
In general, it is possible that two or more consecutive $a_k$'s are identical.
This happens when a given term is never the maximum of the list for any value of $a$.
This behaviour complicates the computation of these coefficients.

To construct an efficient algorithm, let us suppose that this does not happen,
namely that all $a_k$ are distinct.
If this was the case, then $a_k$ simply represents the value of $a$ for which
$\abs{\delta_k}/a^k=\abs{\delta_{k-1}}/a^{k-1}$.
Then, they would simply be given by ($\delta_0=1$)
\begin{equation}
a_k = \frac{\abs{\delta_k}}{\abs{\delta_{k-1}}} \qquad k=1,...,m.
\end{equation}
The first step of the algorithm consists in constructing these ``guesses'' for the $a_k$.
Then, the ordering condition Eq.~\eqref{eq:akordering} is checked.
If it is satisfied, it means that indeed all $a_k$ are distinct and thus the algorithm can stop and return the list.
Instead, if we find that for a given $k$
\begin{equation}
a_k<a_{k+1},
\end{equation}
it means that $\abs{\delta_k}/a^k$ can never be the output of the max function.
This implies that $\abs{\delta_k}/a^k$ can be removed from the argument of the max function,
and the check can be performed on what remains.
In practice, one can perform the comparison directly between the two adjacent terms, $\abs{\delta_{k+1}}/a^{k+1}$ and $\abs{\delta_{k-1}}/a^{k-1}$,
which leads to the replacement in the list of ``guesses''
\begin{equation}
a_k=a_{k+1} = \sqrt{\frac{\abs{\delta_{k+1}}}{\abs{\delta_{k-1}}}}.
\end{equation}
At this point the list of guesses of $a_k$ is modified, and one can perform again the ordering check.
If the ordering Eq.~\eqref{eq:akordering} is satisfied, then the algorithm stops, otherwise it keeps modifying the list
until the ordering is satisfied.
Note that, if there is a number of consecutive $\abs{\delta_i}/a^i$, $i=k,...,k+j-1$ that cannot be the output the max function,
then the corresponding coefficients are given by
\begin{equation}
a_k=a_{k+1}=...=a_{k+j} = \(\frac{\abs{\delta_{k+j}}}{\abs{\delta_{k-1}}}\)^{\frac1{1+j}}.
\end{equation}
Despite the iterative nature of this algorithm, given the small number of known orders in practical perturbative expansion,
it performs very well and it provides a fast evaluation of the inference in the geometric behavior model
(way faster than performing a numerical integration).

\section{Possible ways of improving the models}
\label{sec:models}

In the main body of the paper we focussed on two main models, with simple assumptions
that allowed to compute most of the integrals in the inference procedure analytically.
As a results the numerical implementation of these models is very fast.

In this appendix, we collect ideas to go beyond these simple models,
to improve their reliability and to provide alternatives that may lead to improved precision.
Most of them work at fixed scale, therefore the procedure of Sect.~\ref{sec:scaleindep}
can be applied afterwards without any modification.
In the \texttt{THunc} code, these or other models can be implemented through a ``custom model'' feature
(see Sect.~\ref{sec:custom}).

\subsection{Scale variation model with violation of the bound}
\label{sec:SCviol}

We have noticed in Sect.~\ref{sec:SCposterior} that the scale variation model tends to set the lower limit
of the parameter $\lambda$ from the first step of the inference, namely from the knowledge of the first non-trivial scale dependence.
We observed that the theta function imposed by the next orders is typically looser than that coming from the first order,
thereby possibly overestimating the uncertainty by requiring a too large value of $\lambda$.

A possible way to solve this issue is to relax the condition Eq.~\eqref{eq:SChyp} to a less stringent one,
namely allowing a violation of that bound.
This can be implemented in the model through a modified likelihood, Eq.~\eqref{eq:SClik}.
So, rather than having a strict theta function that forbids values of $\abs{\delta_k(\mu)}>\lambda r_{k-1}(\mu)$,
we allow them, but with an exponential suppression:
\begin{equation}\label{eq:SClikmod}
P(\delta_k|r_{k-1},\lambda,\mu) =
\frac{\rho_k}{2(1+\rho_k)\lambda r_{k-1}(\mu)} \exp\[\rho_k\(1 - \max\(1, \frac{\abs{\delta_k(\mu)}}{\lambda r_{k-1}(\mu)}\)\)\].
\end{equation}
This distribution is flat in the same region $\abs{\delta_k}\leq\lambda r_{k-1}$ that is allowed in the default model,
and it decreases exponentially in $\abs{\delta_k}/(\lambda r_{k-1})$ outside this region.
The strength of the exponential suppression is governed by the parameter $\rho_k$.
Since this generalization of the model is motivated by the behaviour of the first orders that may not be representative
of the expansion, it makes sense to allow more violation at the first orders and make the suppression stronger at higher orders.
This is achieved by using a different value of $\rho_k$ for each value of the order $k$.
A reasonable choice that makes the suppression strong enough not to bias too much the method is given by
\begin{equation}\label{eq:rhokdef}
\rho_k = 2k^2.
\end{equation}
The form Eq.~\eqref{eq:SClikmod} with $\rho_k$ given in Eq.~\eqref{eq:rhokdef} does not allow
to perform the inference in a simple analytic way.
Therefore, this variant of the model can be implemented numerically through the custom model feature of the \texttt{THunc} code,
see Sect.~\ref{sec:custom}.

\begin{figure}[t]
  \centering
  \includegraphics[width=0.435\textwidth,page=3]{./images/plot_Posterior_lambda_ggH_paper.pdf}
  \includegraphics[width=0.555\textwidth,page=1]{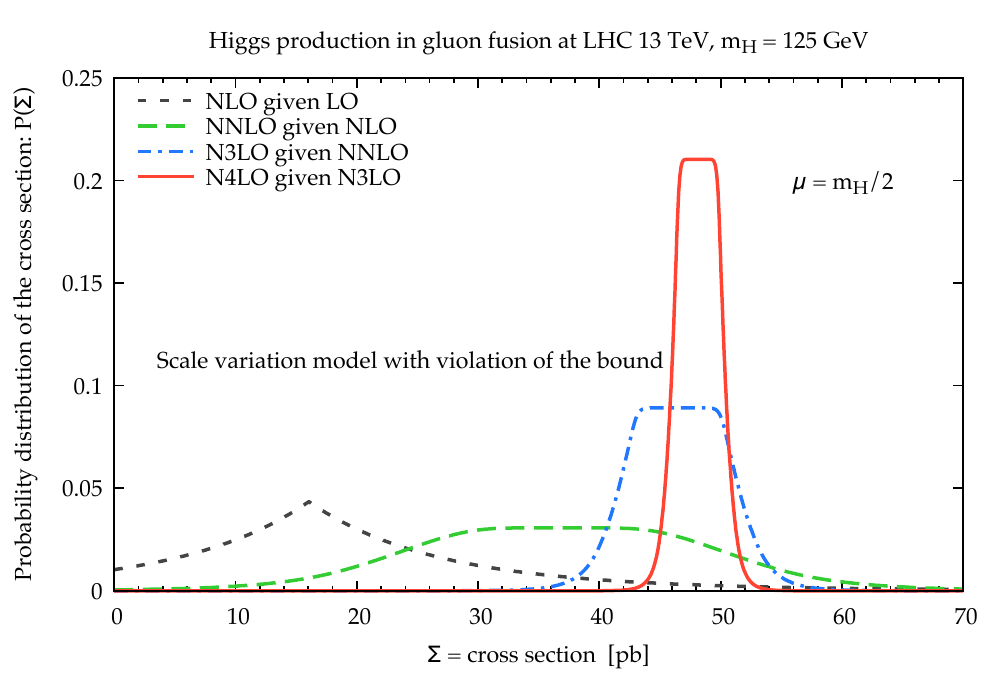}
  \caption{Posterior distribution for the parameter $\lambda$ (left)
    and probability distributions for the observable at the next order (right)
    for the Higgs production process within the modified scale variation model with violation of the bound.}
  \label{fig:SCmod}
\end{figure}

To see how this modification improves the behaviour of the model, we report in Fig.~\ref{fig:SCmod}
the posterior distribution for $\lambda$ (left plot) and the distributions for the next order (right plot)
for the example process of Higgs production in gluon fusion.
These results must be compared with Fig.~\ref{fig:SCposterior} (left plot) and
Fig.~\ref{fig:SCdistr} (solid lines), respectively.
We observe that the posterior for $\lambda$ is dramatically different, now allowing much smaller values of $\lambda$.
Note that the tail of this posterior at the various orders is the same, up to a factor due to the different normalization,
which is a consequence of the structure of the inference described in Sect.~\ref{sec:SCposterior}
that does not change at large $\lambda$.
Similarly, the tails of the distributions for the observable are unchanged,
but the central shape is different.
In particular the plateau has now smooth corners, and most importantly the distributions
at high orders are narrower, leading to a more precise (less conservative) prediction.
This modified scale variation method is thus better than the simpler one presented in Sect.~\ref{sec:ModelScaleVar},
but using it requires paying the price of a slower numerical implementation due to the
inability of solving the model analytically.

\subsection{Scale variation model with controlled scale dependence}
\label{sec:SCv2}

We have observed in Sect.~\ref{sec:SCresults} that making a prediction for the observable
using two unknown higher orders within the scale variation model leads to
very different results from those obtained using only the first missing higher order.
The reason for this comes from the fact that inference on $\delta_{n+2}$ requires the knowledge
of $r_{n+1}$, that is computed from the inferred $\delta_{n+1}$.
The problem is that there are values of $\delta_{n+1}$ allowed by the model condition $\abs{\delta_{n+1}}\leq \lambda r_n$
that generate too large values of $r_{n+1}$, which in turn allow very large values of $\delta_{n+2}$
from the next condition $\abs{\delta_{n+2}}\leq \lambda r_{n+1}$.
This behaviour however violates the spirit of the model, that 
assumes that the various $r_k$ are good estimators of the higher orders,
thus implying that they behave perturbatively as well.

The obvious solution to this problem is to require an additional constraint on the values of the $r_k$ coefficients.
Physically, this means that not only one requires that the higher order correction behaves perturbatively,
but also that its scale dependence is reduced (or at least not increased),
as one indeed expects.
Therefore, adding this condition to the model is very well justified physically, and would lead to a more stringent condition,
thus making such a variant of the scale variation model much more precise than its default version.
Additionally, the distribution obtained using two (or more) missing higher orders should be
way more stable and similar to that obtained using only the first missing higher order.

To implement it in practice, we shall assume that the $r_k$ satisfy the condition
\begin{equation}
r_k\lesssim r_{k-1}.
\end{equation}
This condition can be made more precise by introducing a new parameter $\eta$ and writing
\begin{equation}\label{eq:SChyp2}
r_k\leq \eta r_{k-1}.
\end{equation}
If one does not want to have an extra parameter in the model, it is possible to simply
fix $\eta$ to a given value (for instance 1) by using a prior for it given by a delta function.
Incorporating the condition in the model leads to a likelihood that contains the theta function
\begin{equation}
P(\delta_k|\delta_{k-1},...,\delta_1,\Sigma_0,\lambda,\eta,\mu)
\propto 
\theta(\eta r_{k-1}(\mu)-r_k(\mu)).
\end{equation}
Note that in principle this condition could be used alone, namely without also imposing the condition Eq.~\eqref{eq:SChyp},
thus defining a new model that should already lead to a decent constraint on the perturbative expansion.
However, a better performance can be obtained by requiring both Eq.~\eqref{eq:SChyp} and Eq.~\eqref{eq:SChyp2}.
In this case the likelihood becomes
\begin{equation}\label{eq:SClik2}
P(\delta_k|\delta_{k-1},...,\delta_1,\Sigma_0,\lambda,\eta,\mu) \propto
\theta\(\lambda r_{k-1}(\mu)-\abs{\delta_k(\mu)}\) \theta\(\eta r_{k-1}(\mu)-r_k(\mu)\).
\end{equation}
The trouble of this model is that computing the normalization factor
of this likelihood is very complicated because it depends on the scale dependence $r_k$ of the observable.
However, this normalization is very important, because its dependence on $\lambda$ and $\eta$
determines the structure of the inference.

\begin{figure}[t]
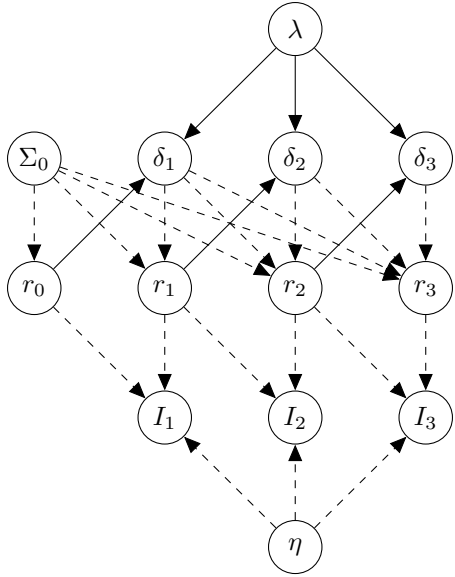

\centering
  \tikz[baseline=(current bounding box.north)]{ %
    \node[latent] (d0) {$\Sigma_0$} ; %
    \node[latent, right=of d0] (d1) {$\delta_1$} ; %
    \node[latent, right=of d1] (d2) {$\delta_2$} ; %
    \node[latent, right=of d2] (d3) {$\delta_3$} ; %
    \node[latent, above=of d2] (lam) {$\lambda$} ; %
    \node[latent, below=of d0] (r0) {$r_0$} ; %
    \node[latent, right=of r0] (r1) {$r_1$} ; %
    \node[latent, right=of r1] (r2) {$r_2$} ; %
    \node[latent, right=of r2] (r3) {$r_3$} ; %
    \node[latent, below=of r1] (I1) {$I_1$} ; %
    \node[latent, right=of I1] (I2) {$I_2$} ; %
    \node[latent, right=of I2] (I3) {$I_3$} ; %
    \node[latent, below=of I2] (eta) {$\eta$} ; %
    \edge[dashed] {d0} {r0} ; %
    \edge[dashed] {d0,d1} {r1} ; %
    \edge[dashed] {d0,d1,d2} {r2} ; %
    \edge[dashed] {d0,d1,d2,d3} {r3} ; %
    \edge[dashed] {r0,r1} {I1} ; %
    \edge[dashed] {r1,r2} {I2} ; %
    \edge[dashed] {r2,r3} {I3} ; %
    \edge[dashed] {eta} {I1,I2,I3} ; %
    \edge {lam} {d1,d2,d3} ; %
    \edge {r0} {d1} ; %
    \edge {r1} {d2} ; %
    \edge {r2} {d3} ; %
  }
  \caption{Bayesian network used to implement a constraint on the scale dependence of the missing higher orders.}
  \label{fig:SC2net}
\end{figure}

To circumvent this problem, we use a trick, namely we consider $r_k$ as explicit variables of the model,
and we introduce new binary variables $I_k$ defined by
\begin{equation}\label{eq:Ikdef}
I_k = \theta\(\eta r_{k-1}(\mu)-r_k(\mu)\).
\end{equation}
These new variables are considered as ``observables'', namely their values is assumed to be fixed
(to 1) by ``observation'' (in our case, this is an assumption and not an observation).
To clarify the structure of these variables and their relations, we show in Fig.~\ref{fig:SC2net}
the Bayesian network of this model.
The joint probability distribution for this model, up to order $m$, is given by
\begin{align}
  P(\delta_m,...,\delta_1&,\Sigma_0,r_m,...,r_0,I_m,...,I_1,\lambda,\eta,\mu)\nonumber\\
&= P(I_m|r_m,r_{m-1},\eta) P(r_m|\delta_m,...,\delta_1,\Sigma_0,\mu) P(\delta_m|r_{m-1},\lambda,\mu) \nonumber\\
&\times \ldots \nonumber\\
&\times P(I_1|r_1,r_0,\eta) P(r_1|\delta_1,\Sigma_0,\mu) P(\delta_1|r_0,\lambda,\mu) \nonumber\\
&\times P(r_0|\Sigma_0,\mu) P_0(\Sigma_0,\lambda,\eta,\mu) 
\end{align}
where
\begin{align}
  P(I_k|r_k,r_{k-1},\eta) &= \delta_{I_k, \theta(\eta r_{k-1}-r_k)} \label{eq:PIk}\\
  P(r_k|\delta_k,...,\delta_1,\Sigma_0,\mu) &= \delta(r_k-r_k(\mu)) \label{eq:Prk}\\
  P(\delta_k|r_{k-1},\lambda,\mu) &= \frac1{2\lambda r_{k-1}(\mu)} \theta\(\lambda r_{k-1}(\mu)-\abs{\delta_k(\mu)}\). \label{eq:Pdkxx}
\end{align}
Eqs.~\eqref{eq:PIk} and \eqref{eq:Prk} are delta functions because the respective variables $I_k$ and $r_k$ are deterministically defined
through Eq.~\eqref{eq:Ikdef} and Eq.~\eqref{eq:rkdef} respectively,
while Eq.~\eqref{eq:Pdkxx} is the same likelihood of the default model, Eq.~\eqref{eq:SClik}.
From the joint probability we may construct any conditional probability we desire.
Let us start from the distribution of the first missing higher order $\delta_{n+1}$
given the knowledge of the first orders up to $n$.
If we compute $P(\delta_{n+1}|\delta_n,...\delta_1,\Sigma_0,\mu)$ marginalizing over everything else including the $I_k$,
then we would get the same result of the default model, Sect.~\ref{sec:ModelScaleVar}.
Here instead we assume the various $I_k$ to be known and equal to 1, thus imposing the condition Eq.~\eqref{eq:SChyp2}.
In particular, $I_1=...=I_n=1$ gives constraints on the hidden parameter $\eta$,
which in turn puts the desired contraint on $\delta_{n+1}$ through the condition $I_{n+1}=1$.
So we compute
\begin{equation}
  P(\delta_{n+1}|\delta_n,...,\delta_1,\Sigma_0,I_{n+1},...,I_1,\mu)
\propto 
P(\delta_{n+1},...,\delta_1, \Sigma_0, I_{n+1},...,I_1,\mu)
\end{equation}
with
\begin{align}
  P&(\delta_m,...,\delta_1, \Sigma_0, I_m,...,I_1,\mu) \\
   &= \int d\lambda\, d\eta\, dr_m\cdots dr_0\, P(\delta_m,...,\delta_1,\Sigma_0,r_m,...,r_0,I_m,...,I_1,\lambda,\eta,\mu) \nonumber\\
   &= \int d\lambda\, d\eta\, P(I_m|r_m,r_{m-1},\eta,\mu)\cdots P(I_1|r_1,r_0,\eta,\mu)
     P(\delta_m|r_{m-1},\lambda,\mu) \cdots P(\delta_1|r_0,\lambda,\mu) P_0(\lambda,\eta,\Sigma_0,\mu) \nonumber
\end{align}
and, when setting $I_1=...=I_{n+1}=1$, it reduces to
\begin{align}
  P&(\delta_m,...,\delta_1,\Sigma_0,I_m=1,...,I_1=1,\mu) \\ \nonumber
  &= \int d\lambda\, d\eta\, \theta(\eta r_{m-1}(\mu)-r_m(\mu))\cdots
    \theta(\eta r_0(\mu)-r_1(\mu)) P(\delta_m|r_{m-1},\lambda,\mu) \cdots P(\delta_1|r_0,\lambda,\mu) P_0(\lambda,\eta,\Sigma_0,\mu),
\end{align}
having used the identity $\delta_{1,\theta(x)}=\theta(x)$.
Note that this result could have been obtained in the ``standard'' approach (without introducing auxiliary variables)
if the likelihood Eq.~\eqref{eq:SClik2} depended on $\lambda$ and $\eta$ as
\begin{equation}\label{eq:SClik3}
P(\delta_k|\delta_{k-1},...,\delta_1,\Sigma_0,\lambda,\eta,\mu) \propto
\frac1\lambda \theta\(\lambda r_{k-1}(\mu)-\abs{\delta_k(\mu)}\) \theta\(\eta r_{k-1}(\mu)-r_k(\mu)\).
\end{equation}
Note that this cannot be the case, as it's clear that if $\lambda$ is large enough the likelihood should
become independent of it, because the other constraint is more stringent.
This means that the implementation of the constraint on the scale dependence through the procedure
proposed here and depicted in the network Fig.~\ref{fig:SC2net} is not entirely equivalent to
the ``standard'' approach based on the likelihood Eq.~\eqref{eq:SClik2}.
Nevertheless, it provides an easy and legitimate way of realizing the same constraint.

\begin{figure}[t]
  \centering
  \includegraphics[width=0.6\textwidth,page=1]{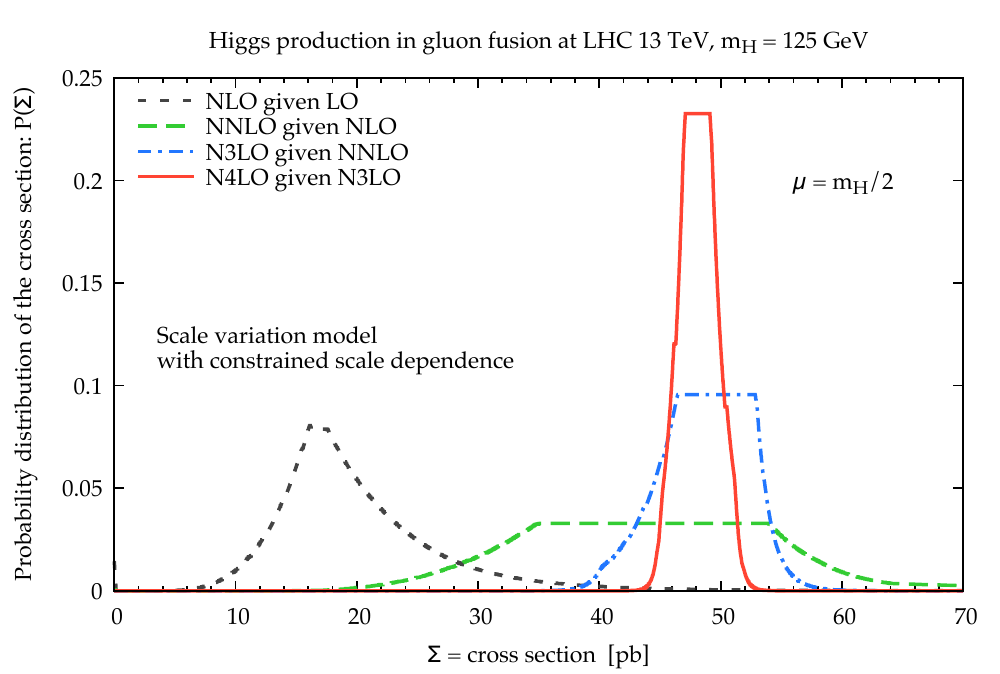}
  \caption{Probability distributions for the observable at the next order
    for the Higgs production process within the scale variation model with constrained scale dependence.}
  \label{fig:SCmod2}
\end{figure}

In order to implement the model in practice, we need to specify the prior for $\eta$.
Of course, the smaller $\eta$, the better the convergence.
We could force $\eta$ to lie in a finite range, but this would provide an upper bound on the scale variation
that seems a bit too restrictive.
Therefore, we just use a prior that exponentially suppresses large values of $\eta$, namely $P_0(\eta)=\exp(-\eta)$.
The results are shown in Fig.~\ref{fig:SCmod2} for the usual $ggH$ process.
In comparison with the result of the default scale variation model, Sect.~\ref{sec:SCresults},
the distributions are now narrower,
and asymmetric, given that the dependence of $r_k$ on $\delta_k$ is not linear.
The tails die way faster, and effectively exclude the regions where $\delta_{n+1}$
generates a too large scale dependence $r_{n+1}$.
This results in more localized distributions, corresponding to smaller uncertainties.
This model thus seems a rather powerful improvement over the default scale variation model,
the only caveat being the slow numerical evaluation due to the inability of computing inference analytically.

\subsection{Mixing models}
\label{sec:mix}

So far we have seen two main models, characterized by the ability to compute most integrations analytically
and thus leading to a fast numerical implementation, and a number of potentially improved versions of each.
We now consider the possibility of combining them.

The simplest but also the least performing option consists in considering the assumptions of two
(or more) models as \emph{alternative hypothesis}, each with its own prior, and just ``average''
the outcome of each model with a weight given by the prior.
In this way, each model works independently from the other(s), and the combination is done at the end.
This approach is similar to the combination of results from different experiments.
However, there is a difference: the experiments perform a measurement,
while the output of the two methods is directly the probability distribution for the next orders.
So, for the experiments one can apply Bayes theorem using the likelihood of the measurment
given the measurand, while in our case there is no such likelihood
(it should be the likelihood of the model given the prediction, which is difficult to define in general).
Therefore, for each model we just use its prior, so that the result of mixing any number of models is
\begin{equation}\label{eq:mix1}
P(\Sigma|\delta_n,...,\delta_1,\Sigma_0)
= \sum_i P(\Sigma|\delta_n,...,\delta_1,\Sigma_0,H_i) P_0(H_i),
\end{equation}
where $H_i$ represents the hypothesis of each model (now written explicitly in the probability distribution).
Obviously, $\sum_iP_0(H_i)=1$. If one has no preference on a specific model, it is natural to use equal
probability for the priors of the models.

This approach, though very simple, is not necessarily the best.
Indeed, suppose two models look at different characteristics of the perturbative expansions,
and the user trusts both models. In this case, the models are not \emph{alternative}, but \emph{complementary}.
In such cases, the user wants both conditions to be satisfied simultaneously.
This means, for instance, that if one model excludes (probability zero)
or at least makes very unlikely (probability almost zero)
a region of the observable $\Sigma$, after mixing the models this region should remain excluded.
However, in the average process of Eq.~\eqref{eq:mix1} this does not happen.
Indeed, Eq.~\eqref{eq:mix1} represents the condition in which the users trusts one model \emph{or} the other,
while here we want to consider the case in which the user trusts one model \emph{and} the other.

We stress that the two families of models considered so far rely on different assumptions: one looks at the behaviour of the
perturbative expansion, and the other looks at the scale dependence.\footnote
{Note that the assumption considered in Sect.~\ref{sec:SCv2} is in principle independent from the
model of Sect.~\ref{sec:ModelScaleVar}, and could therefore be considered as another family.}
These assumptions are unrelated, and therefore one is allowed to assume that both are satisfied at the same time.
In this case we need to act at the core of the model, and modify the likelihood directly.
For definiteness, we consider the mixing of the two basic models of Sect.~\ref{sec:GB} and Sect.~\ref{sec:ModelScaleVar}.
Thus, assuming that both conditions Eq.~\eqref{eq:GBhyp} and Eq.~\eqref{eq:SChyp} are satisfied,
the likelihood becomes
\begin{equation}
P(\delta_k|r_{k-1},c,a,\lambda,\mu)
\propto \theta\(ca^k-\abs{\delta_k}\) \theta\(\lambda r_{k-1}-\abs{\delta_k}\).
\end{equation}
The two theta functions can be combined in a single one, from which the normalization
is also immediate to compute. The result is
\begin{equation}\label{eq:Mixlik}
P(\delta_k|r_{k-1},c,a,\lambda,\mu)
=\frac1{2\min\[ca^k, \lambda r_{k-1}\]} \theta\(\min\[ca^k, \lambda r_{k-1}\]-\abs{\delta_k}\).
\end{equation}
The presence of the min function unavoidably entangles the two original approches,
and makes the computation of the inference difficult from an analytical point of view.
Using the custom model feature of the \texttt{THunc} code, Sect.~\ref{sec:custom}, we can easily implement this model numerically.

\begin{figure}[t]
  \centering
  \includegraphics[width=0.6\textwidth,page=1]{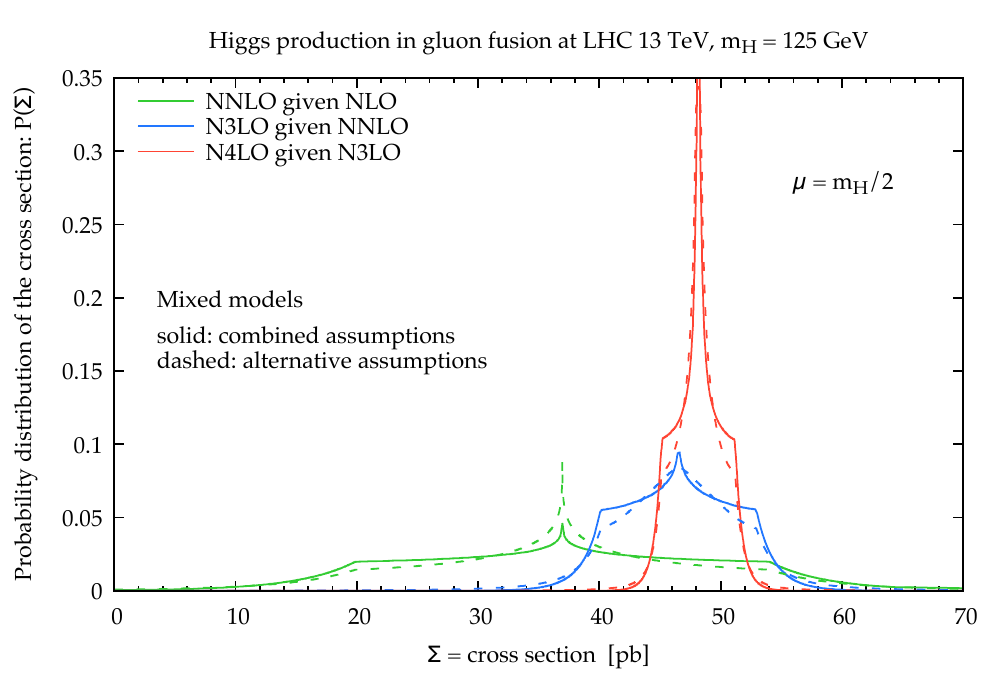}
  \caption{Probability distributions for the observable at the next order
    for the Higgs production process mixing the default geometric behaviour and scale variation models either
    as alternative hypotheses (dashed lines) or as combined hypotheses (solid lines).}
  \label{fig:Mix}
\end{figure}

In order to appreciate the way this mixed model work, we show the distributions obtained
with it for the usual example of the $ggH$ process.
This is depicted in Fig.~\ref{fig:Mix} (solid lines).
For comparison, we also show in the same plot the result obtained by averaging the results according
to Eq.~\eqref{eq:mix1}, using equal priors for each model (dashed lines).
We see that, with respect to each individual model, these distributions are somewhat more precise:
the tails die faster than in the geometric behaviour model, and the central region is more peaked
than in the scale variation model.
In the dashed curves, corresponding to the combination Eq.~\eqref{eq:mix1}, the tails are higher
due to the component coming from the geometric behaviour model.
Instead, in the solid curves, corresponding to the combined likelihood Eq.~\eqref{eq:Mixlik},
the tails are those of the scale variation model (up to a normalization factor), and thus exponentially suppressed.
Therefore, as expected, mixing the models assuming both conditions Eq.~\eqref{eq:GBhyp} and Eq.~\eqref{eq:SChyp}
at the same time gives stronger constraints leading to smaller uncertainties.
We stress that, while in the plot of Fig.~\ref{fig:Mix} the two combinations look overall very similar,
with different models, e.g.\ including the constraint on the scale dependence as in Sect.~\ref{sec:SCv2}
where the probability in some regions is strictly zero, the two ways of combining the models may lead to
results that differ not only quantitatively but also qualitatively.

\subsection{Custom model}
\label{sec:custom}

From the discussion so far it is clear that it is always possible to invent new models
to improve some aspects or to adapt them to the behaviour of the specific perturbative expansion under consideration.
Since it is not possible to predict and thus to implement all possible models,
we provide in the public code \texttt{THunc} a so-called ``custom model''.
This custom model is defined through the following inputs:
\begin{itemize}
\item the number of hidden parameters $\vec p$;
\item the prior of the hidden parameters $P_0(\vec p)$;
\item the likelihood of the model $P(\delta_k|\delta_{k-1},...,\delta_1,\Sigma_0,\vec p)$.
\end{itemize}
These ingredients are sufficient to define the model, and to allow the code to perform the inference
and compute the sought distributions.

The likelihood has explicit access to the derived quantities $r_k$,
to simplify an implementation based on these objects (as the scale behaviour model).
Note that the likelihood has to be properly normalized, namely
\begin{equation}
\int d\delta_k\, P(\delta_k|\delta_{k-1},...,\delta_1,\Sigma_0,\vec p) = 1.
\end{equation}
If this is not possible, the normalization can be restored at the end by normalizing the final distribution.
However, for the inference procedure to work properly, the $\vec p$ dependence of the likelihood must be complete.
In other words, the likelihood must be correct up to a normalization factor that can be either purely numerical
or dependent on just the previous orders.

The inference is made by performing numerical integrations.
By default, the parameters $\vec p$ are assumed to range between 0 and 1.
Therefore, for different ranges, the parameters must be rescaled and the jacobian
properly included in the prior.
Because of the numerical nature of the integration, using the custom model results
in a longer running time of the code.
The model variants considered in this appendix have been implemented in the code through this custom model feature.

\subsection{Adding information on the sign of each order}
\label{sec:sign}

So far all the models we have proposed consider only the size
of the various orders, but not their sign.
Therefore, at fixed scale, higher orders will be predicted with positive and negative sign with equal likelihood.\footnote
{There is one exception, namely the model with controlled scale variation of Sect.~\ref{sec:SCv2},
where the additional condition on the value of $r_k$ makes the likelihood for $\delta_k$ asymmetric.}
As a consequence, the probability distribution for the observable $\Sigma$ at fixed scale
will be symmetric and centered on the highest known order.
This prediction may not be satisfactory in many cases.
For instance, if the first known orders are all positive, one is tempted to
expect with a greater degree of belief that the next order(s) will also follow the same pattern.
Moreover, in asymptotically free theories like QCD,
at higher scales the coupling is smaller and the various orders are consequently smaller,
and a same-sign pattern is generated even if at smaller scales the various orders
have apparently random signs.
Keeping track of the sign in these cases where a pattern is present would allow to obtain
more realistic and precise predictions.

We propose a way to include information on the sign which is simply based on looking for a pattern.
We assume that a given pattern corresponds to an hypothesis, and we make inference on that hypothesis
by checking the sing pattern of the known coefficients of the expansion.
Specifically, we will only consider three hypotheses:
\begin{itemize}
\item $H_+$, assuming all positive signs;
\item $H_-$, assuming alternating signs;
\item $H_0$, assuming a ``random sign pattern'' (i.e., none of the above).
\end{itemize}
Remember that using the normalized coefficients $\delta_k$ we always have $\delta_0=1$,
so the hypothesis ``all negative signs'' is not possible
(or better, it would correspond to $H_+$ as an overall minus sign would be factored out in $\Sigma_0$).
For the same reason, the hypothesis $H_-$ implies that the sign of $\delta_k$ is $(-1)^k$.
Of course, there are infinitely many other possible sign patterns, but we
believe that only the two considered here, $H_\pm$, can be useful,
while the hypothesis $H_0$ can cover efficiently all other cases.\footnote
{There is a simple variation of the aforementioned patterns that could occur,
in which the pattern is present but it starts from the $\delta_1$.
This is the case if all $\delta_k$ are negative except $\delta_0$ (variant of $H_+$)
or if the signs are alternating as $\sign\delta_k=(-1)^{k+1}$ with the exception of $\delta_0$ (variant of $H_-$).
We have seen an example of each: the toy convergent series at large scale, Sect.~\ref{sec:convser},
and the anharmonic oscillator, Sect.~\ref{sec:ana}, respectively.
We will not cover this case explicitly, but testing these additional hypotheses is straightforward.}
After all, testing a more complicated pattern would require a significant number of known orders,
which is never the case.

The assumptions of each hypothesis can be expressed in terms of the likelihood
of the sign of the generic order $k$ given one of the hypotheses.
These probabilities are given by
\begin{align}
P(\sign\delta_k=+|H_+) &= 1, &
P(\sign\delta_k=+|H_-) &= \frac{1+(-1)^k}2, &
P(\sign\delta_k=+|H_0) &= \frac12, \\
P(\sign\delta_k=-|H_+) &= 0, &
P(\sign\delta_k=-|H_-) &= \frac{1-(-1)^k}2, &
P(\sign\delta_k=-|H_0) &= \frac12.
\end{align}
In the case of $H_-$, the probability of each sign is either 1 or 0 depending on whether $k$ is even or odd.
Note that for $H_0$ the formula is strictly speaking valid only for $k>0$,
as for $k=0$ we have $\delta_0=1$ and thus the probability of $\delta_0>0$ has to be 1 for all hypotheses.
At first sight, it may seem that for a generic order each of the three hypotheses are to be considered
(in practice, only for $k$ even all three probabilities are non-zero).
However, we will always consider a sequence of consecutive orders, so that what we really test
is not just the sign at a given order, but the sign pattern of the sequence.
The sequences generated by $H_+$ and $H_-$ are clearly incompatible ($++++...$ vs $+-+-...$),
so in practice only one of the two has to be considered, and compared with $H_0$
(which is compatible with any possible sequence).
Of course, the sequence may be incompatible with both $H_+$ and $H_-$, and in this case only the $H_0$ hypothesis survives.

Let's construct the inference process in practice step by step.
With the only knowledge of the LO, one cannot test any sign hypothesis, as $\delta_0=1$ is always positive.
The first non-trivial information comes from the NLO.
The sign of $\delta_1$ already allows to exclude one hypothesis:
if $\delta_1>0$, then $H_-$ is excluded, viceversa if $\delta_1<0$, then $H_+$ is excluded.\footnote
{If $\delta_1=0$, then one cannot exclude any pattern.
  In this case, one needs to check the next order, $\delta_2$, before excluding something.
  Since it is very unlikely that both $\delta_1$ and $\delta_2$ are zero simultaneously,
  one can always exclude one of the patterns at this order.}
We are thus left with two hypotheses, as anticipated.

Let us assume that $\delta_1>0$, so that $H_-$ is excluded and only $H_+$ and $H_0$ survive
(the complementary case with $\delta_1<0$ proceeds identically and we will not repeat it).
The probability of this sequence is given by
\begin{equation}
P(++|H_+) = 1, \qquad
P(++|H_0) = \frac12,
\end{equation}
where we have used a shorthand notation $++$ to indicate the signs of $\delta_0$ and $\delta_1$,
and we have used the fact that $P(\sign\delta_0=+|H_0)=1$ and not $1/2$.
At this point, if the next order has a minus sign, the probability of that sequence is 0 in the $H_+$ hypothesis,
leaving only $H_0$:
\begin{equation}
P(++-|H_+) = 0 \quad\Rightarrow\quad
P(H_+|++-) = 0 \quad\Rightarrow\quad
P(H_0|++-) = 1.
\end{equation}
It follows that the only sequence that gives non-trivial results is the one with all plus signs.
At this order it gives
\begin{equation}
P(+++|H_+) = 1, \qquad
P(+++|H_0) = \frac1{2^2}.
\end{equation}
Now let us assume that we know the first $n+1$ orders, up to $\delta_n$, and that all signs are positive.
We have
\begin{equation}
P(+^{n+1}|H_+) = 1, \qquad
P(+^{n+1}|H_0) = \frac1{2^n},
\end{equation}
where the meaning of the notation $+^{n+1}$ is obvious.
From this, we can infer the probabilities of each hypothesis. Using Bayes theorem, we find
\begin{align}
  P(H_+|+^{n+1})
  &= \frac{P(+^{n+1}|H_+)P_0(H_+)}{P(+^{n+1}|H_+)P_0(H_+)+P(+^{n+1}|H_0)P_0(H_0)} \nonumber\\
  &= \frac{P_0(H_+)}{P_0(H_+)+\frac1{2^n}P_0(H_0)} \nonumber\\
  &= \frac{1}{1+\frac1{2^n}\frac{P_0(H_0)}{P_0(H_+)}}, \nonumber\\
  P(H_0|+^{n+1}) &= 1-  P(H_+|+^{n+1}) \nonumber\\
  &= \frac1{1+2^n\frac{P_0(H_+)}{P_0(H_0)}},
\end{align}
where $P_0(H_{0,+})$ are the priors for $H_{0,+}$.
The posterior probability of $H_+$ increases as the number of known orders increases,
provided they are all positive, while the one for $H_0$ dies exponentially.
Analogous results hold for the comparison of $H_-$ with $H_0$, provided the pattern is replaced
with the alternating sign one.
Note that this inference procedure is exactly the same that occurs when tossing a coin
an testing if it's regular ($H_0$) or if it has ``head'' on both faces ($H_+$).

The result obviously depends on the prior probability of each hypothesis.
There is no reason to strongly favour one of the sign hypotheses a priori.
If a given sign pattern is present, the $H_0$ hypothesis becomes very unlikely after a few orders
due to the exponential suppression, and if no sign pattern is present,
the posterior probability for $H_0$ is 1 anyway.
Therefore, an equal prior for each hypothesis makes perfect sense.

Let us now show how this information on the sign pattern is used in the models for the uncertainty
from missing higher orders.
First of all, for each hypothesis we need to generate a different variant of the model,
accounting for the assumptions on the sign pattern.
This is expressed in a different likelihood for each hypothesis.
For example, in the geometric behaviour model, the likelihoods are
\begin{align}
P(\delta_k| c, a, \mu, H_0) &= \frac{1}{2ca^k}\theta\(ca^k - \abs{\delta_k}\), \nonumber\\
P(\delta_k| c, a, \mu, H_+) &= \frac{1}{ca^k}\theta\(ca^k - \delta_k\) \theta\(\delta_k\), \nonumber\\
P(\delta_k| c, a, \mu, H_-) &= \frac{1}{ca^k}\theta\(ca^k - \abs{\delta_k}\) \theta\((-1)^k\delta_k\).
\end{align}
The likelihood for the $H_0$ hypothesis corresponds to the original one, Eq.~\eqref{eq:GBlik}.
Similar changes also hold for the scale variation model.
The use of these modified likelihoods have a rather simple effect on the probability for
unknown orders given the known ones.
We can indeed write
\begin{align}\label{eq:PdeltajH+0}
P(\delta_{n+j},...,\delta_{n+1}|\delta_n,...,\delta_1,\Sigma_0,\mu,H_+) &=
2^j\theta(\delta_{n+j})\cdots\theta(\delta_1)\, P(\delta_{n+j},...,\delta_{n+1}|\delta_n,...,\delta_1,\Sigma_0,\mu,H_0), \nonumber\\
P(\delta_{n+j},...,\delta_{n+1}|\delta_n,...,\delta_1,\Sigma_0,\mu,H_-) &=
2^j\theta((-1)^{n+j}\delta_{n+j})\cdots\theta(\delta_2)\theta(-\delta_1)\nonumber\\
&\qquad\times P(\delta_{n+j},...,\delta_{n+1}|\delta_n,...,\delta_1,\Sigma_0,\mu,H_0),
\end{align}
which is written in terms of the same distribution in the $H_0$ hypothesis, which we know already.
This result is rather general, and it holds for all models for which the $H_0$ hypothesis
leads to a symmetric likelihood (thus, it would not work for the model of Sect.~\ref{sec:SCv2}).
In these cases, the numerical implementation of this procedure is very easy.
The last step consists in combining the results from the various hypotheses.
For the full observable $\Sigma$ we simply have
\begin{align}
P(\Sigma|\delta_n,...,\delta_1,\Sigma_0,\mu)
&= P(\Sigma|\delta_n,...,\delta_1,\Sigma_0,\mu,H_+) P(H_+|\sign\delta_n,...,\sign\delta_1)\nonumber\\
&+ P(\Sigma|\delta_n,...,\delta_1,\Sigma_0,\mu,H_-) P(H_-|\sign\delta_n,...,\sign\delta_1)\nonumber\\
&+ P(\Sigma|\delta_n,...,\delta_1,\Sigma_0,\mu,H_0) P(H_0|\sign\delta_n,...,\sign\delta_1).
\end{align}
The same structure also holds for the probability of anything given the first $n+1$ orders.
Thus, for example, the probability of the first missing higher order, using Eq.~\eqref{eq:PdeltajH+0},
is given by
\begin{align}
P(\delta_{n+1}|\delta_n,...,\delta_1,\Sigma_0,\mu)
&= P(\delta_{n+1}|\delta_n,...,\delta_1,\Sigma_0,\mu,H_0) \nonumber\\
&\qquad\times  \begin{cases}
    \dfrac{2\theta(\delta_{n+1})+\frac1{2^n}\frac{P_0(H_0)}{P_0(H_+)}}{1+\frac1{2^n}\frac{P_0(H_0)}{P_0(H_+)}} & \text{if all signs are $+$'s}\\
    \dfrac{2\theta((-1)^{n+1}\delta_{n+1})+\frac1{2^n}\frac{P_0(H_0)}{P_0(H_-)}}{1+\frac1{2^n}\frac{P_0(H_0)}{P_0(H_-)}} & \text{if signs alternate}\\
    1 & \text{otherwise}
  \end{cases}
\end{align}
which is given in terms of the $H_0$ results and a modifying function
that depends on the priors of each sign pattern.

\begin{figure}[t]
  \centering
  \includegraphics[width=0.495\textwidth,page=5]{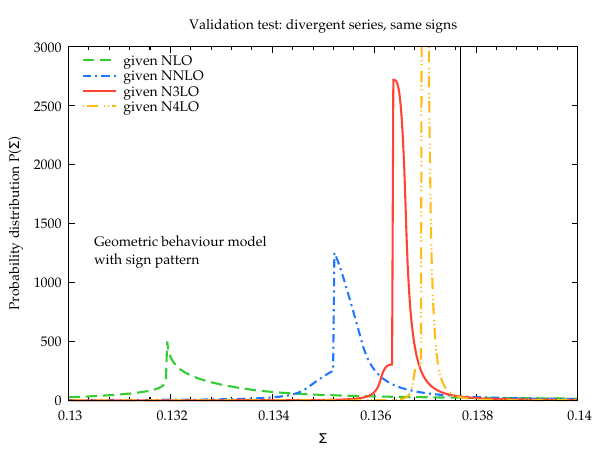}
  \includegraphics[width=0.495\textwidth,page=5]{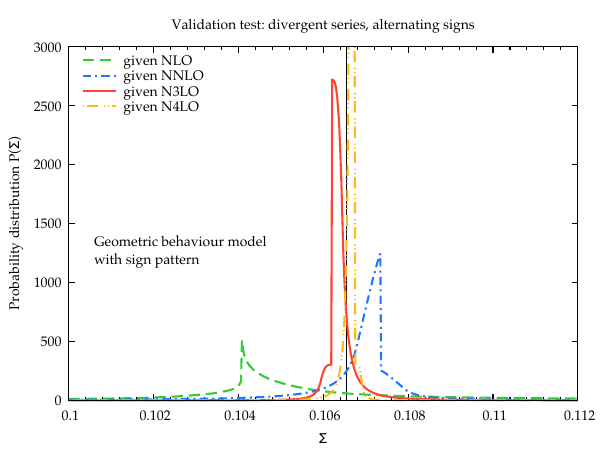}
  \caption{Probability distributions in the geometric behaviour model for the observable at the next order
    for the same sign (left) and alternating sign (right) divergent series
    discussed in Sect.~\ref{sec:Div}, taking into account the sign pattern.}
  \label{fig:sign}
\end{figure}

To visualize the effect of this procedure, we apply it to the divergent series
introduced in Sect.~\ref{sec:Div}, one of which has all coefficients with the same sign
and the other with alternating signs.
Therefore, the sign pattern is present and recognised by the algorithm,
thus favouring the hypothesis $H_\pm$ over $H_0$ more and more by increasing the order.
The distributions for the observable are shown in Fig.~\ref{fig:sign},
for the geometric behaviour model (similar results hold for the scale variation model).
The distributions are clearly asymmetric, due to the fact the hypotheses $H_\pm$
predict zero for one of the sign of the next order.\footnote
{In the case of the hypothesis $H_-$, using more than one order for approximating the observable
would result in a (small) contribution in the region where the prediction based only on the first missing
higher order is zero. However, the effect is very mild, especially
in comparison with the component from the $H_0$ hypothesis in the same region.}
In both cases, the asymmetry favours the half-region where the exact result (thin vertical line) lies,
thus showing that accounting for the sign pattern does indeed improve the model prediction.
Note that, for a real observable in quantum field theory, after marginalizing over the scale the distributions
become smooth (without discontinuities).

\phantomsection
\addcontentsline{toc}{section}{References}

\bibliographystyle{jhep}
\bibliography{references}

\end{document}